\documentclass[%
  fontsize=12pt,
  twoside,
  paper=letter,
  abstract=false,
  pagesize=auto,
  version=last,
  DIV=calc
  ]{scrartcl}
\usepackage{scrpage2}

\newcommand{\helv}{\fontfamily{phv}\fontseries{b}\fontsize{9}{11}\selectfont}
\setkomafont{pagehead}{\normalfont\footnotesize\sffamily\bfseries}
\setkomafont{pagenumber}{\helv}
\deftripstyle{CPI}[0pt][0.5pt]{\rightmark}{}{\thepage}{}{}{}
\typearea[current]{calc}
\areaset[current]{\textwidth}{\textheight}
\usepackage[english]{babel}
\usepackage{cmap}
\usepackage[margin=1in]{geometry}
\usepackage{fix-cm}
\usepackage{ae,aecompl,aeguill}
\usepackage{lmodern}
\usepackage[T1]{fontenc}
\usepackage[charter]{mathdesign}

\usepackage[protrusion=true,expansion=false]{microtype}
\usepackage[raise]{engord}
\usepackage{ccicons}    
\usepackage{ucs}
\usepackage[utf8x]{inputenc}
\usepackage{xspace}
\usepackage{setspace}
\usepackage[x11names]{xcolor}
\usepackage{graphicx}
\DeclareGraphicsExtensions{.pdf,.gif,.png,.jpg}
\usepackage{tikz}
\usetikzlibrary{calc,positioning}
\usepackage[colorlinks=true,urlcolor=RoyalBlue3,linkcolor=OrangeRed3,citecolor=SpringGreen3,linktocpage=true]{hyperref}
\hypersetup{
  pdftitle    = {Complex Path Integral and the Space of Theories},
  pdfauthor   = {D.D. Ferrante <danieldf@het.brown.edu>, G.S. Guralnik <gerry@het.brown.edu>,
    Z. Guralnik <zack@het.brown.edu>, C. Pehlevan <cengiz@het.brown.edu>},
  pdfsubject  = {High Energy Physics - Theory},
  pdfcreator  = {LaTeX2e with hyperref package},
  pdfproducer = {pdflatex},
  pdfkeywords = {Non-Perturbative, Quantum Mechanics, Supersymmetry,
    Quantum Field Theory, Gauge Theory, Path Integral, Partition Function,
    Complexification, Analytic Continuation, Langlands Duality,
    Langevin, Stochastic Quantization, Coherent State Quantization,
    Heat semi-group, Witten Index, Witten Complex, Linear Canonical
    Transform, Phase Space, Symplectic Geometry, Geometric Topology,
    Topological Field Theory, Topological String Theory, Holomorphic Quantization,
    Geometric Quantization, Deformation Quantization, Fractional
    Fourier Transform, Fractal Derivative,
    Quantum Algebra, Geometric Algebra, High Energy Theory},
  pdfview = {FitH},
  pdflang = {en_US}
}   
\usepackage{bookmark}
\usepackage{cite}
\usepackage{xfrac,xkeyval}
\usepackage{amsmath,mathtools,dsfont,amstext,eucal,amsbsy}
\DeclareMathAlphabet{\mathpzc}{T1}{pzc}{m}{it}   
\usepackage{pifont}
\usepackage{textcomp}
\usepackage{wasysym}
\makeatletter
\def\equalsfill{$\m@th\mathord=\mkern-7mu
  \cleaders\hbox{$\!\mathord=\!$}\hfill
  \mkern-7mu\mathord=$}
\makeatother
\def\hksqrt{\mathpalette\DHLhksqrt}
\def\DHLhksqrt#1#2{\setbox0=\hbox{$#1\sqrt{#2\,}$}\dimen0=\ht0
  \advance\dimen0-0.2\ht0
  \setbox2=\hbox{\vrule height\ht0 depth -\dimen0}%
{\box0\lower0.4pt\box2}}

\newcommand{\bbC}{\ensuremath{\mathpzc{C}}\xspace}
\newcommand{\bbQ}{\ensuremath{\mathbb{Q}}\xspace}
\newcommand{\bbP}{\ensuremath{\mathbb{P}}\xspace}

\DeclareMathOperator{\diag}{diag}
\DeclareMathOperator{\Ai}{Ai}
\DeclareMathOperator{\Bi}{Bi}
\DeclareMathOperator{\He}{He}

\DeclareMathOperator{\tr}{tr}

\def\blfootnote{\xdef\@thefnmark{}\@footnotetext}
\long\def\symbolfootnote[#1]#2{\begingroup%
  \def\thefootnote{\fnsymbol{footnote}}\footnote[#1]{#2}\endgroup}
\begin{document}
\begin{titlepage}
  \vspace*{-2cm}
  \begin{flushright}
    {\small\textsf{BROWN-HET-1611}}
  \end{flushright}

  \begin{center}
    {\Large \textbf{\textsf{$\boldsymbol{\mathbb{C}}$omplex Path Integrals and the Space of Theories}}} \newline

    \bigskip
    {D.~D. Ferrante, G.~S. Guralnik, Z. Guralnik, C. Pehlevan}
  \end{center}

  \begin{abstract}
    {\sffamily\noindent The Feynman Path Integral is extended in order
      to capture all solutions of a quantum field theory. This is done
      via a choice of appropriate integration cycles, parametrized by
      $\mathds{M}\in\mathsf{SL}(2,\mathbb{C})$, i.e., the space of
      allowed integration cycles is related to certain $Dp$-branes and
      their properties, which can be further understood in terms of
      the ``physical states'' of another theory. We also look into
      representations of the Feynman Path Integral in terms of a
      Mellin--Barnes transform, bringing the singularity structure of
      the theory to the foreground. This implies that, as a sum over
      paths, we should consider more generic paths than just Brownian
      ones. Finally, we are able to study the Space of Theories
      through our examples in terms of their Quantum Phases and
      associated Stokes' Phenomena (wall-crossing).
  }
  \end{abstract}

    \noindent\rule{\textwidth}{1pt}
    \tableofcontents
    \smallskip
    \noindent\rule{\linewidth}{1pt}
\end{titlepage}
\section{Introduction}\label{sec:intro}
This paper has the goal to extend our previous work, \cite{msols}, bringing to
the foreground some issues of Quantum Field Theory that have not quite found
their place yet. In particular, it aims to introduce \emph{quantum phases} as the
analogous object to chamber walls (wall-crossing and Stokes' phenomena
\cite{devilsinvention}), in order to study the ``Space of Theories'' via
suitable generalizations of the [Feynman] Path Integral.

We will start by considering the quantization procedure as defined by
the path integral and its respective algebra of observables,
\begin{align*}
  \mathscr{Z}^{\Sigma}_{\tau}[J] &= \varint e^{i\,S_{\tau}^{\Sigma}[\varphi]}\, e^{i\, J\, \varphi}\,\mathcal{D}\varphi\;;\\
  &= \mathcal{F}[e^{i\,S^{\Sigma}_{\tau}[\varphi]}] [J] \;;\\
  \langle\mathcal{O}_1\dotsm\mathcal{O}_n\rangle^{\Sigma}_{\tau} &=
    \varint \mathcal{O}_1\dotsm\mathcal{O}_n\,
    e^{i\,S_{\tau}^{\Sigma}[\varphi]}\,
    e^{i\, J\,\varphi}\,\mathcal{D}\varphi\;; \\
  &= \mathcal{F}[\mathcal{O}_1\dotsm\mathcal{O}_n\, e^{i\,S^{\Sigma}_{\tau}[\varphi]}] [J] \;;
\end{align*}
where $S^{\Sigma}_{\tau}$ is the Action of the theory, with $\Sigma$
representing a possible surface term needed in order to establish the
self-adjointness of the Laplace--Beltrami operator in the kinetic
term of the Action, interpolating among the possible boundary conditions
suitable for the problem at hand, and $\tau$ represents the possible coupling
constants of the theory; $\mathcal{O}_i = \mathcal{O}_i[\varphi], i=1,\dotsc,n$ are
operators forming an algebra in an appropriate Hilbert space; and
$\mathcal{F}$ represents the Fourier Transform (with respect to the
source $J$), and the path integral and the trace of the operators are
cast as suitable Fourier transforms of the Feynman weight, $e^{i\,
  S_{\tau}^{\Sigma}[\varphi]}$.

In order for the Fourier transforms above to be well defined, the
integrand and the kernel have to be single-valued. To that end, we can
consider their $\mathbb{C}$omplexification and look for suitable
integration cycles labeled by $\tau$ and $\Sigma$,
$\mathcal{C}_{\tau}^{\Sigma}$, i.e., Riemann sheets rendering our
problem well defined. If $\mathcal{C}_{\tau}^{\Sigma}$ is ultimately
$\mathbb{R}$, we recover the above definitions in the sense of Fourier
transforms. However, for non-trivial cycles we end up with a more general
integral transform, one that preserves the symplectic structure of the
problem at hand. In fact, we can ask ourselves what is the most
general integral transform that preserves the symplectic structure of
a problem. And, as we will see in sections \ref{sec:lct} and
\ref{sec:pilct}, this is accomplished by the Linear Canonical
Transform (a linear map of the phase space preserving the
symplectic form).

In this sense, it is clear that $\tau$ and $\Sigma$ will determine the
analytic structure of our quantization procedure: in fact, the integration cycle
$\mathcal{C}_{\tau}^{\Sigma}$ ultimately determines the fall-off properties of
the fields in our theory, thus relating the decaying properties of the fields
with the analyticity of the integral transform at hand,
\cite[in particular, the Paley--Wiener theorem (and its various
generalizations), describing functions through their analytic extensions in the
complex domain]{complexsaddlepoint}. Actually, making this relation explicit
will motivate us, in sections \ref{sec:ghiTransforms} and
\ref{sec:PIghiTransform}, to cast the Path Integral as a Mellin--Barnes
transform, bringing the role of the analytic structure of our theory to the
foreground.

%
%
%

Thus, in a sense that will become clearer later, the central concept will be
the analytical structure of our quantization procedure via the Feynman Path
Integral, including its source term, $e^{i\, J\, \phi}$: both, the integrand and
the source term, have to live in the \emph{same} Riemann sheet, otherwise we get
an ill-defined theory.
Therefore, realizing that the source term reflects the analytic structure of
the Feynman weight (after all, $J$ and $\varphi$ are related by a Legendre
transform), we can already infer that $\varphi\mapsto -i\, \partial_J$ is not
going to express the most general case: this will only be true in the situation
where $\mathcal{C}_{\tau}^{\Sigma}$ defines a Fourier transform; otherwise, the
Legendre-dual relation between $J$ and $\varphi$ will typically be more
elaborate.

As a side remark, we can note that in order for the path integral quantization
to be consistent with its respective Schwinger--Dyson equations, the field,
$\varphi$, has to be determined by its respective quantum equations of
motion. Therefore, given that $\tau$ and $\Sigma$ determine the properties
(monodromy, Stokes' phenomena, etc) of the quantum equations of motion, i.e., of
the Schwinger--Dyson equations, the integrand and $e^{i\, J\, \varphi}$ have to
live in the same Riemann sheet. This is a trivial observation if the quantum
equations of motion are of \engordnumber{1} degree. However, they are typically
of higher degree, meaning that there are multiple solutions, which we can label
through $\tau$ and $\Sigma$, thinking that each solution is associated to a
quantum equation of motion with suitable values of $\tau$ and $\Sigma$.

Thus, in order to bring these issues to the foreground, we will generalize the
path integral (thought as an integral transform in a general sense, or as a
Fourier transform in a more specific one) in two different ways: as a
\emph{Linear Canonical Transform} (sections \ref{sec:lct} and \ref{sec:pilct}),
and as a Mellin--Barnes Transform (sections \ref{sec:ghiTransforms} and
\ref{sec:PIghiTransform}). These are particularly useful, since the former aims
to be invariant by canonical transformations, while the latter is relevant to
make explicit the analytic structure of the theory, e.g., the modern subject of
scattering amplitudes in $\mathcal{N} = 4$ Super Yang--Mills. In this way, we
will be able to associate certain theories formulated as Linear Canonical
Transforms with their respective Mellin--Barnes transforms, making clear how one
framework is related to the other, i.e., we will be able to relate certain
orbits of the $\mathsf{SL}(2,\mathbb{C})$ parametrizing the Linear Canonical
Transform (as seen below) with appropriate Mellin--Barnes transforms, connecting
the symmetries described by these orbits to the singularity structure made
explicit in the Mellin--Barnes transforms.

This brings us to the question of what kinds of paths are allowed within
the context of the path integral. After all, taking the analytic structure of the
quantization process, of the path integral, seriously, how do we know that more
exotic paths (woven across several allowed and distinct Riemann sheets) are not
contributing, despite cancellations due to the highly oscillatory nature of the
integrand?

In quantum mechanics, the usual story says that all paths are allowed,
although the equivalence between the integral formulation (path integral) and
the differential one (Schrödinger equation) is only established at the level of
Brownian paths, where we think of Schrödinger's equation as a diffusion
equation, analytically continued. We are interested in this connection because
of its possible implications in Quantum Field Theory, where we deal, instead,
with Feynman's Path Integral and Schwinger--Dyson's equations as the two sides
of the quantization procedure. We want to study the connection between the
dimension of the solution space of the Schwinger--Dyson equations and the number
of vacuum states present in a physical theory: in quantum mechanics, the
dimension of the solution space of Schrödinger's equation determines the number
of ground states a certain model has --- as is the case with all differential
equations and the dimension of their respective solution spaces: the dimension
of the solution space (vector space of analytic functions) of a
$n$\textsuperscript{th} order linear homogeneous differential equation with
analytic coefficients is $n$. How does the equivalent problem work in
Quantum Field Theory? Does Feynman's Path Integral include only Brownian paths
or does it include more generic paths? It is to make sense of questions such as
these that we want to study the connections between the integral and
differential formulations of Quantum Field Theories.

The first extension of this result was done in \cite{frqm} and
can be achieved if one consider slightly more generic paths, called
Lévy flights. In this case, the path integral becomes a Fractional
Fourier Transform. The gist of this approach is that Schrödinger's
equation becomes \emph{fractional} (i.e., rather than differential
operators one uses fractional differential operators), while the path
integral now includes paths of fractal dimension. More to the point, while
Brownian motions have a scaling relation of the form $x^2 \propto t$ and fractal
dimension $d_{\text{fractal}}^{\text{Brownian}} = 2$, Lévy flights scale as
$x^{\alpha} \propto t$ and have fractal dimension
$d_{\text{fractal}}^{\text{Lévy}} = \alpha$, with $0 < \alpha \leqslant 2$. This
means that, for example, in the case of a nonrelativistic particle, its
dispersion relation is given by $E = D_{\alpha}\, p^{\alpha}$, where
$D_{\alpha}$ is the generalized fractional quantum diffusion coefficient. If
$\alpha=2$, we have that $D_2 = 1/2\, m$ and $E = p^2/2\,m$, which is the case
for Brownian paths.

The connection between this analysis in terms of scaling and dispersion
relations, and our previous argument, is through the fall-off properties of the
field content of our theory. Roughly speaking, the integration cycle
$\mathcal{C}_{\tau}^{\Sigma}$ ultimately determines these decaying
properties (Stokes' phenomena, monodromy of the quantum equations of motion,
etc). And, as such, $\mathcal{C}_{\tau}^{\Sigma}$ is related to the scaling and
dispersion relations above, determining whether they are isotropic or
anisotropic (and the respective scaling factors). Finally, the subject of
anisotropic scaling relations has been studied before in Hořava--Lifshitz
gravity, where the scaling goes as $x \propto t^z$, meaning that $\alpha \sim
1/z$, and considering the Lorentz symmetry as emergent when the theory flows to
some value of $z \sim 1/\alpha$ (e.g., $z=3=1/\alpha$),
\cite{horava}. Unfortunately, the relation between anisotropic scaling,
fractality, and integration cycles of the path integral does not seem to have
been analyzed before.

However, this is not the most general case one can consider. In fact, this is
just one among several possibilities. The bottom line is that the path
integral can be parametrized by $\mathsf{SL}(2,\mathbb{C})$, where
different orbits of the group correspond to distinct possible
transformations: from a Fourier and Fractional Fourier, to a Linear
Canonical Transform.

Furthermore, we will use the expression of the path integral as a Linear
Canonical Transform in order to introduce a representation of the path integral
as a Mellin--Barnes transform, given by Fox's $H$-function. This is possible, as
we will see later, because of a result stating that all stable probability
densities can be represented in terms of Fox's $H$-function,
\cite{stabledistros}. The advantage of such a construction is that we
make explicit the analytic structure of our theory: the monodromy of the
integration cycles defining the $H$-function representing the path integral will determine
the ramification of the theory at hand. The study of ramifications is
generally described by Picard--Lefschetz theory.

Thus, we tie in with our previous construction in different
ways. Firstly, this approach has been used before, namely in
\cite{frqm}, in order to express the fundamental solutions for a
fractional diffusion process in terms of an appropriate
$H$-function. Secondly, the properties of the
$\mathsf{SL}(2,\mathbb{C})$ parametrization of the Linear Canonical Transform
are reflected in the properties of the $H$-function: given that we explicitly have
the analytic structure of the theory, it is not difficult to see the
results of certain analytic continuations. Finally, different
$H$-functions, representing different path integrals, will be associated to
different stable probability distributions and, as such, with different allowed
paths in the path integral, giving further evidence that we need to include more
exotic paths than just Brownian ones.

At this point, a note is in order: the above results are not
dependent on the particulars of our example(s). For instance, $\varphi$ could be
$\mathbb{R}$- or $\mathbb{C}$-valued, it could also be $\mathds{M}$atrix-valued
(or, more generally, $\mathds{T}$ensor-like in character), or even
$\mathscr{L}$ie algebra-valued  (including graded $\mathscr{L}$ie algebras,
i.e., SUSY): the exact same results would be true (with minor modifications,
e.g., tracing out the matrix degrees-of-freedom, or defining the Fourier
Transform over the appropriate group, etc).

The outline of this paper is as follows: in section \ref{sec:lct} we will
introduce the Linear Canonical Transform, and in section
\ref{sec:pilct} we will describe the path integral as an LCT; in section
\ref{sec:ghiTransforms} we will introduce Fox's $H$-function and some of its
generalizations, while in section \ref{sec:PIghiTransform} we will describe the
path integral as a Mellin--Barnes transform via the use of the $H$-function; in section
\ref{sec:examples} we will develop several examples supporting our previous
constructions, and in section \ref{sec:ext} we will suggest how this
approach can be used to deal with higher-dimensional systems. Finally, in
section \ref{sec:summary} we will make a synopsis and hint at possible further
developments.
\section{Linear Canonical Transform}\label{sec:lct}
Canonical Transforms are those transformations which map the
operators \bbQ and \bbP into linear combinations of themselves. The \emph{Linear
Canonical Transformation} is the linear map which preserves the symplectic form:
it is understood as a family of integral transforms, parametrized by
$\mathsf{SL}(2,\mathbb{R})$, generalizing several classical transforms, such as
the Fourier, the fractional Fourier, the Laplace, the Gauss--Weierstrass, the
Bargmann, and the Fresnel transforms.

One of the Fourier Transform's main properties is that the operator \bbQ
(multiplication by the argument: $(\bbQ\,f)(q) = q\, f(q)$) is mapped into the
operator \bbP (differentiation: $(\bbP\,f)(q) = -i\, \tfrac{\mathrm{d} f(q)}{\mathrm{d} q}$), and
vice-versa with a minus sign. Such behavior is analogous of a rotation by
$\pi/2$ in the \bbQ--\bbP plane, the Phase Space. However, we will look for
linear operators \bbC which map \bbQ and \bbP into linear combinations of each
other,
\begin{align*}
  \bbQ' &= \bbC\, \bbQ\, \bbC^{-1} = d\, \bbQ - b\, \bbP\;;\\
  \bbP' &= \bbC\, \bbP\, \bbC^{-1}  = -c\, \bbQ + a\,\bbP\;;
\end{align*}
where $a, b, c, d\in\mathbb{R}$ for now, with the caveat that $[\bbQ', \bbP'] =
\bbC\, [\bbQ, \bbP]\, \bbC^{-1} = i\,\mathds{1}$, implying that $a\, d - b\, c =
1$.

We readily see that the Fourier Transform corresponds to $a = 0 = d$, $b = 1 =
-c$, and that the identity transformation corresponds to $a = 1 = d$, $b = 0 =
c$. So, we will label the transform operator as $\bbC_{\mathds{M}}$ with the
unimodular matrix
\begin{equation*}
  \mathds{M} =
  \begin{pmatrix}
    a & b \\
    c & d
  \end{pmatrix}\; ,
  \quad\det\mathds{M} = 1\;.
\end{equation*}

The operator $\bbC_{\mathds{M}}$ is linear and has the following properties:
\begin{align*}
  (\bbC_{\mathds{M}} f)(q) &= f^{\mathds{M}}(q)\; ;\\
  (\bbC_{\mathds{M}}\, \bbQ f)(q) &= (\bbC_{\mathds{M}} \bbQ \bbC_{\mathds{M}}^{-1}\, \bbC_{\mathds{M}} f)(q)\; ;\\
  &= d\, \bigl(q f^{\mathds{M}}(q)\bigr) + i\, b\, \biggl(\frac{\mathrm{d} f^{\mathds{M}}(q)}{\mathrm{d}q}\biggr)\; ;\\
  (\bbC_{\mathds{M}}\, \bbP f)(q) &= (\bbC_{\mathds{M}} \bbP \bbC_{\mathds{M}}^{-1}\, \bbC_{\mathds{M}} f)(q)\; ;\\
  &= -c\, \bigl(q f^{\mathds{M}}(q)\bigr) - i\, a\, \biggl(\frac{\mathrm{d} f^{\mathds{M}}(q)}{\mathrm{d}q}\biggr)\; .
\end{align*}

\paragraph{Integral Transform and Kernel} To concretely realize the operator
$\bbC_{\mathds{M}}$ let us use an \emph{integral transform} with the kernel
$\bbC_{\mathds{M}}(q', q)$,
\begin{equation}
  \label{eq:lctinttransf}
  f^{\mathds{M}}(q') = (\bbC_{\mathds{M}} f)(q') = \int f(q)\,\bbC_{\mathds{M}}(q', q)\,\mathrm{d}q \; ;
\end{equation}

Combined with the properties previously derived, we have the following
expression for the kernel $\bbC_{\mathds{M}}$,

\begin{equation}
  \label{eq:lctkernel}
  \bbC_{\mathds{M}}(q', q) =
    \frac{e^{-i\,\pi/4}}{\hksqrt{2\,\pi\,b}}\, e^{i\,(a\,q^2 - 2\,q\,q' + d\, q^{\prime\, 2})/2\,b}\;.
\end{equation}
For $\mathbb{R}$eal parameters, \eqref{eq:lctkernel} oscillates
strongly in the limits where $|q|\rightarrow\infty$ and
$|q'|\rightarrow\infty$ but with a fixed modulus at
$1/\hksqrt{2\,\pi\,|b|}$.

The inverse of the $\bbC_{\mathds{M}}$ transform
\eqref{eq:lctinttransf} can be shown to be given by,
\begin{equation}
  \label{eq:lctinttransfinv}
  f(q) = (\bbC_{\mathds{M}}^{-1} f^{\mathds{M}})(q) = \int f^{\mathds{M}}(q')\,\bbC_{\mathds{M}}^{*}(q,q')\,\mathrm{d}q'\;.
\end{equation}

The composition of two such transforms, $\bbC_{\mathds{M}_1}$ and
$\bbC_{\mathds{M}_2}$, is given by the transform whose matrix
parameter is the product of the constituent transforms, i.e.,
$\bbC_{\mathds{M}_2\,\mathds{M}_1}$. In Group Theory, the
$\bbC_{\mathds{M}}$ are a ray representation of
$\mathsf{SL}(2,\mathbb{R})$ ($2\times 2$ unimodular matrices), also
known as the metaplectic representation (Weil).

The identity $\bbC_{\mathds{M}}$ transform corresponds to $\mathds{M}
= \mathds{1}$, as expected: $\lim_{\mathds{M}\rightarrow\mathds{1}} \bbC_{\mathds{M}}(q', q) = \delta(q - q')$.
From this result, it is possible to show that all lower-triangular
transformations $\mathds{M}$, where $b=0$, have the form,
\begin{equation*}
  (\bbC_{\mathds{M}(b=0)} f)(q) = \frac{e^{i\, c\, q^2/2\,a}}{\hksqrt{a}}\, f(q/a)\;.
\end{equation*}
These are called \emph{geometric} transformations, once they consist
of dilations by $a$ and/or multiplication by an oscillating Gaussian.

As for the inverse canonical transform, $\bbC_{\mathds{M}^{-1}}(q',q)
= \bbC_{\mathds{M}}^{*}(q,q')$, we just have to note that,
\begin{equation*}
  \mathds{M}^{-1} =
  \begin{pmatrix}
    d            & e^{-i\,\pi}\,b \\
    e^{i\,\pi}\, c & a
  \end{pmatrix}
  \;;
\end{equation*}
where we have to watch out for the appropriate sheet in the complex
$b$-plane.

Thus, we have showed that the set of LCTs $\bbC_{\mathds{M}}$ forms a
\emph{group} of unitary transformations, $\mathsf{SL}(2,\mathbb{R})$;
i.e., LCTs are a family of integral transforms parametrized by a
3-manifold that can be understood as the action of
$\mathsf{SL}(2,\mathbb{R})$ on Phase Space, the field-source
plane. The name comes from the map preserving the symplectic (or
multiplectic, for the case of multi-dimensional transforms)
structure, canonical transformations, once $\mathsf{SL}(2,\mathbb{R})$
can be understood as the symplectic group $\mathsf{Sp}_2$. Thus, the
LCTs are linear maps in the coordinate--momentum domain
\bbQ--\bbP (i.e., phase space), preserving the symplectic form.

The LCT, parameterized by $\mathds{M}\in \mathsf{SL}(2,\mathbb{R})$,
$\mathds{M} = (\begin{smallmatrix} a & b\\ c & d\end{smallmatrix})$
with $a\, d - b\, c = 1$, can also be defined more straightforwardly
as follows,

\begin{align*}
  \mathscr{F}_{\mathds{M}}[J] &= \hksqrt{-i{\,}}\, e^{i\,\pi\,J^2\,d/b}\,\int_{-\infty}^{\infty} F[\varphi]\, e^{i\,\pi\,\varphi^2\,a/b}\, e^{-2\,\pi\,i\,\varphi\,J/b}\,\mathrm{d}\varphi \;; &b\neq 0\;;\\
  \mathscr{F}_{\mathds{M}'}[J] &= \hksqrt{d{\,}}\, e^{i\,\pi\,J^2\,c\,d}\, F[J\, d\,] \;; &b = 0\;;
\end{align*}
where the prime denotes those elements where $b=0$:
$\mathds{M}' = (\begin{smallmatrix} a & 0\\ c & d\end{smallmatrix})\in
\mathsf{SL}(2,\mathbb{R})$.

Let us describe some of the special cases generalized by the LCT,
i.e., physically relevant cases characterized by symmetries described
by orbits of $\mathsf{SL}(2,\mathbb{C})$:

\begin{description}
\item[Parabolic Subgroup or Dilation Transform] Corresponds to a dilation by a factor
  $\Delta$ (and are a family of parabolic elements, represented by $a
  = e^{-\beta/2}\,,\; \beta\in\mathbb{R}$),
  \begin{equation*}
    \mathds{M}_{\text{\textsf{DT}}} = \begin{pmatrix} a & b\\ c & d\end{pmatrix} = \begin{pmatrix} \Delta & 0\\ 0 & \Delta^{-1}\end{pmatrix}\;.
  \end{equation*}
\item[Fresnel or Gauss--Weierstrass Transform] The Fresnel Transform
  corresponds to shearing by $\Lambda$ (cutoff), while the
  Gauss--Weierstrass ``diffusive'' transform is its analytical
  continuation, $\Lambda \mapsto \Lambda = -i\, b$:
  \begin{equation*}
    \mathds{M}_{\text{\textsf{GWT}}} = \begin{pmatrix} a & b\\ c & d\end{pmatrix} = \begin{pmatrix} 1 & \Lambda\\ 0 & 1\end{pmatrix}\;.
  \end{equation*}
\item[Gaussian Transform] Multiplication by a Gaussian of width $-i\,w$,
  \begin{equation*}
    \mathds{M}_{\text{\textsf{GT}}} = \begin{pmatrix} a & b\\ c & d\end{pmatrix} = \begin{pmatrix} 1 & 0\\ w^{-1} & 1\end{pmatrix}\;.
  \end{equation*}
\item[Elliptic Subgroup or Fractional Fourier Transform] Corresponds to a rotation by an arbitrary
  angle $\omega$; they are the elliptic elements of $\mathsf{SL}(2,\mathbb{R})$, represented by:
  \begin{equation*}
    \mathds{M}_{\text{\textsf{FrFT}}} = \begin{pmatrix} a & b\\ c & d\end{pmatrix} = \begin{pmatrix} \cos\omega & \sin\omega\\ -\sin\omega & \cos\omega\end{pmatrix}\;;
  \end{equation*}
\item[Fourier Transform] Corresponds to a rotation by 90°, which is a
  special case of the Fractional Fourier Transform for
  $\mathds{M}_{\text{\textsf{FrFT}}}(\omega=\pi/2)$, with the corresponding integral
  kernel given by $\bbC_{\mathds{M}_{\text{\textsf{\textsf{FT}}}}} = e^{-i\,\pi/4}\, \mathds{M}_{\text{\textsf{FT}}}$, where
  \begin{equation*}
    \mathds{M}_{\text{\textsf{FT}}} = \begin{pmatrix} a & b\\ c & d\end{pmatrix} = \begin{pmatrix} 0 & 1\\ -1 & 0\end{pmatrix}\;;
  \end{equation*}
\item[Hyperbolic Subgroup] 
  \begin{equation*}
    \mathds{M}_{\text{\textsf{HS}}} = \begin{pmatrix} a & b\\ c & d\end{pmatrix} = \begin{pmatrix} \cosh\omega & -\sinh\omega\\ -\sinh\omega & \cosh\omega\end{pmatrix}\;;
  \end{equation*}
\item[Laplace Transform] The bilateral Laplace Transform can be obtained once we extend into the $\mathbb{C}$omplex domain, i.e., $\mathsf{SL}(2,\mathbb{C})$,
  \begin{equation*}
    \mathds{M}_{\text{\textsf{LT}}} = \begin{pmatrix} a & b\\ c & d\end{pmatrix} = \begin{pmatrix} 0 & i\\ i & 0\end{pmatrix}\;;
  \end{equation*}
\item[Bargmann Transform] This is a particular case of the
  $\mathbb{C}$omplex LCTs, analogous to the Fourier Transform for $\mathbb{R}$eal LCTs, where we use $-i = e^{-i\,\pi/2}$
  \begin{equation*}
    \mathds{M}_{\text{\textsf{BT}}} = \begin{pmatrix} a & b\\ c & d\end{pmatrix} = \frac{1}{\hksqrt{2}}\,\begin{pmatrix} 1 & -i\\ -i & 1\end{pmatrix}\;.
  \end{equation*}
\end{description}

For transformations with $a\neq 0$ (i.e., excluding \textsf{FTs} and special
cases of the \textsf{FrFTs}), we can show that every $\bbC_{\mathds{M}}\in\mathsf{SL}(2,\mathbb{R})$
can be written as the product of a \textsf{GT}, a \textsf{DT} and a \textsf{GWT}, as follows:
\begin{equation*}
  \mathds{M} =
  \begin{pmatrix}
    a & b \\
    c & d
  \end{pmatrix}
  =
  \begin{pmatrix}
    1   & 0 \\
    c/a & 1
  \end{pmatrix}
  \,
  \begin{pmatrix}
    a & 0 \\
    0 & 1/a
  \end{pmatrix}
  \,
  \begin{pmatrix}
    1 & b/a \\
    0 & 1
  \end{pmatrix}
  = \mathds{M}_{\text{\textsf{GT}}}(c/a)\,\mathds{M}_{\text{\textsf{DT}}}(a)\,\mathds{M}_{\text{\textsf{GWT}}}(b/a)\;.
\end{equation*}
When involving the Fourier Transform, we have that,
\begin{equation*}
  \mathds{M} =
  \begin{pmatrix}
    a & b \\
    c & d
  \end{pmatrix}
  =
  \begin{pmatrix}
    b & 0 \\
    d & 1/b
  \end{pmatrix}
  \,
  \begin{pmatrix}
    0  & 1 \\
    -1 & 0
  \end{pmatrix}
  \,
  \begin{pmatrix}
    1   & 0 \\
    a/b & 1
  \end{pmatrix}
  = \mathds{M}_{\text{\textsf{DT}}}(b)\,\mathds{M}_{\text{\textsf{FT}}}\,\mathds{M}_{\text{\textsf{GT}}}(a/b)\;.
\end{equation*}
As expected, the generalized uncertainty relation we
get from $\Delta_f$ and $\Delta_{f^{\mathds{M}}}$, respectively the
dispersion of $f$ and $f^{\mathds{M}}$, is given by:
\begin{equation*}
  \Delta_f \, \Delta_{f^{\mathds{M}}} \geqslant \frac{b^2}{4} \; .
\end{equation*}

Finally, we notice that the transform kernel,
$\mathpzc{C}_{\mathds{M}}$, plays the role of the Dirac $\delta$, in
the sense of closing the relation of completeness. This is called the
Coherent-State basis.
\paragraph{Hyperdifferential Operator Realization}\label{subsec:hyperdiff}
We want to find a differential operator in the following context:
given a 1-parameter family of integral transforms,
\begin{equation}
  \label{eq:intransffamily}
  (\mathpzc{C}_{\mathds{M}(\tau)} f)(q') = \int \mathpzc{C}_{\mathds{M}(\tau)}(q',q)\,f(q)\,\mathrm{d}q = f(q', \tau)\;;
\end{equation}
including the identity for $\tau = 0$, i.e., $\mathds{M}(0) =
\mathds{1}$ and $f(q',0) = f(q')$, we want a differential operator
$\mathpzc{H}$ that can be written as,
\begin{equation*}
  (\mathpzc{C}_{\mathds{M}(\tau)} f)(q) = \exp(i\, \tau\,\mathpzc{H})\, f(q) = \sum_{n=0}^{\infty} \frac{(i\, \tau\, \mathpzc{H})^n}{n!}\,f(q) \; .
\end{equation*}
Operator of this kind appear mainly in connection with the time
evolution of the wave and diffusion equations, translations and
dilatations.

Formally, we can write
\begin{equation*}
  \mathpzc{H} f(q') = -i\,\frac{\partial}{\partial\tau}\, \int \mathpzc{C}_{\mathds{M}(\tau)}(q',q)\,f(q)\,\mathrm{d}q\,\bigg|_{\tau=0}\;.
\end{equation*}
The operator $\mathpzc{H}$ generates the 1-parameter integral
transform family \eqref{eq:intransffamily}.

Of particular interest are the following 1-parameter subgroups of
$\mathsf{SL}(2,\mathbb{R})$:

\begin{description}
\item[Parabolic Subgroup or Dilatation Transforms] $\mathpzc{H}^{\text{\textsf{D}}} = \tfrac{1}{2}\,(\mathbb{Q}\,\mathbb{P} + \mathbb{P}\,\mathbb{Q})$,
  \begin{align*}
    \mathds{M}^{\text{\textsf{DT}}}(e^{-\tau}) &=
      \begin{pmatrix}
        e^{-\tau} & 0 \\
        0       & e^{\tau}
      \end{pmatrix}
      \; ;\\
    \frac{\partial}{\partial\tau}\mathpzc{C}_{\mathds{M}^{\text{\textsf{DT}}}}(q',q)\,\bigg|_{\tau=0} &= \frac{1}{2}\,\delta(q-q') - q'\,\frac{\partial}{\partial q}\delta(q-q') \;; \\
    &= \biggl(\frac{1}{2} - q\,\frac{\partial}{\partial q}\biggr)\,\mathpzc{C}_{\mathds{M}^{\text{\textsf{DT}}}}(q',q)\,\bigg|_{\tau=0}\;;
  \end{align*}
\item[Fresnel or Gauss--Weierstrass Transforms] $\mathpzc{H}^{\text{\textsf{GW}}} = \tfrac{1}{2}\,\mathbb{P}^2$,
  \begin{align*}
    \mathds{M}^{\text{\textsf{GWT}}}(\tau) &=
      \begin{pmatrix}
        1 & -\tau \\
        0 & 1
      \end{pmatrix}
      \; ;\\
    \frac{\partial}{\partial\tau}\mathpzc{C}_{\mathds{M}^{\text{\textsf{GWT}}}}(q',q)\,\bigg|_{\tau=0} &= -\frac{i}{2}\,\frac{\partial^2}{\partial q^2}\mathpzc{C}_{\mathds{M}^{\text{\textsf{GWT}}}}(q',q)\,\bigg|_{\tau=0}\; ;
  \end{align*}
\item[Gaussian Transform] $\mathpzc{H}^{\text{\textsf{G}}} = \tfrac{1}{2}\,\mathbb{Q}^2$,
  \begin{align*}
    \mathds{M}^{\text{\textsf{GT}}}(\tau) &=
      \begin{pmatrix}
        1     & 0 \\
        \tau  & 1
      \end{pmatrix}
      \; ;\\
    \frac{\partial}{\partial\tau}\mathpzc{C}_{\mathds{M}^{\text{\textsf{GT}}}}(q',q)\,\bigg|_{\tau=0} &= \frac{i}{2}\,q^2\,\mathpzc{C}_{\mathds{M}^{\text{\textsf{GT}}}}(q',q)\,\bigg|_{\tau=0}\; ;
  \end{align*}
\item[Hyperbolic Subgroup] $\mathpzc{H}^{\text{\textsf{HS}}} = \tfrac{1}{2}\,(\mathbb{P}^2 - \mathbb{Q}^2)$,
  \begin{align*}
    \mathds{M}^{\text{\textsf{HS}}}(\tau) &=
      \begin{pmatrix}
        \cosh\tau  & -\sinh\tau \\
        -\sinh\tau & \cosh\tau
      \end{pmatrix}
      \; ;\\
    \frac{\partial}{\partial\tau}\mathpzc{C}_{\mathds{M}^{\text{\textsf{HS}}}}(q',q)\,\bigg|_{\tau=0}
      &= \frac{i}{2}\,\biggl(-\frac{\partial^2}{\partial q^2}-q^2\biggr)\,\mathpzc{C}_{\mathds{M}^{\text{\textsf{HS}}}}(q',q)\,\bigg|_{\tau=0}\;;
  \end{align*}
\item[Elliptic Subgroup or Fractional Fourier Transform] $\mathpzc{H}^{\text{\textsf{FrF}}} = \tfrac{1}{2}\,(\mathbb{P}^2 + \mathbb{Q}^2)$,
  \begin{align*}
    \mathds{M}^{\text{\textsf{FrFT}}}(\tau) &=
      \begin{pmatrix}
        \cos\tau & -\sin\tau \\
        \sin\tau & \cos\tau
      \end{pmatrix}
      \; ;\\
    \frac{\partial}{\partial\tau}\mathpzc{C}_{\mathds{M}^{\text{\textsf{FrFT}}}}(q',q)\,\bigg|_{\tau=0}
      &= \frac{i}{2}\,\biggl(-\frac{\partial^2}{\partial q^2}+q^2\biggr)\,\mathpzc{C}_{\mathds{M}^{\text{\textsf{FrFT}}}}(q',q)\,\bigg|_{\tau=0}\;; 
  \end{align*}
\end{description}
Again, if we allow for $\mathbb{C}$omplex parameters in the
expressions above, we obtain the group $\mathsf{SL}(2,\mathbb{C})$ of
$\mathbb{C}$omplex 2-dimensional special (unimodular) linear
transformations.

We now see that linear canonical transformations are generated by all
operators constructed out of quadratic expressions in $\mathbb{Q}$ and
$\mathbb{P}$. Further, we have the following objects,
\begin{enumerate}
\item Integral Transforms $\mathpzc{C}_{\mathds{M}}$;
\item Hyperdifferential Operators $\exp(i\,\tau\mathpzc{H})$; \&
\item $2\times 2$ matrices $\mathds{M}$.
\end{enumerate}
For every element in one there are corresponding elements in the other
two, and this correspondence is preserved under composition, sum and
multiplication.
\section{Path Integral as a Linear Canonical Transform}\label{sec:pilct}
One of the more common quantization techniques
is via the Feynman Path Integral formulation, generalizing
the Action Principle. Furthermore, the Path Integral allows us to easily change
coordinates between different canonically conjugated descriptions of
the same quantum system. However, it is rather unfortunate that most
treatments of the Path Integral do not do it full justice, e.g., \textit{a
  priori} assuming that its measure is $\mathbb{R}$-valued, and thus
missing on possible extra solutions to the quantum equations of motion and
extensions of the theory (c.f.,
\cite{reedsimon,devilsinvention,jextension,complexsaddlepoint}, roughly
summarized in appendix \ref{sec:extop}).

%
%

Historically, defining the Path Integral's measure, in any way (ranging from
more mathematically rigorous approaches all the way to the more formal
and heuristic ones), has always been the Achilles’ heel of the Path Integral
quantization scheme. What we will show below is a generalization that
brings to the foreground the Legendre-dual relation between the field,
$\varphi$, and its source, $J$. Once again, note that $\varphi$ can be quite
general: $\mathbb{R}$- or $\mathbb{C}$-valued,
$\mathds{V}$ector- or $\mathds{M}$atrix- or $\mathds{T}$ensor-valued, and even
$\mathscr{L}$ie- or graded $\mathscr{L}$ie Algebra-valued (SUSY): the necessary
modifications, e.g., tracing matrix degrees-of-freedom, do not affect our
discussion: there is no loss of generality in the results below dealing with
scalar fields (spin 0 bosonic fields) or $D0$-branes.

The idea of generalizing the Path Integral not only as an integral transform,
but as a general Linear Canonical transform, explicitly bringing forth the
symplectic structure of the theory, is motivated by several reasons. In what
follows, we can consider a ``finite'' (discrete) version of the Path Integral in several
different but analogous ways: we can think of our theory as having an IR- and an
UV-cutoff (including defining the theory via an appropriate Vertex Operator
Algebra that implements the desired IR- and UV-cutoffs), or we can consider
different discretization schemes (discrete differential forms, finite exterior
calculus, $\mathds{M}$atrix models --- \textsf{BFSS}/\textsf{IKKT}, infinite momentum frame:
$N\times N$ matrices describing $N$ Dirichlet particles ---,
naïve lattice discretizations, etc). In these cases the Path Integral as well as
the questions and motivations below are well-defined and tractable.

\begin{description}
\item[Nonlinear Stokes' Phenomena \& Path Integral Convergence:] We want the
  Path Integral to be well-defined in a certain sense, even if it is
  mathematically loose (and only satisfies the physicist's level of rigour). One
  of the ways to achieve this is through studying the asymptotics of the
  oscillatory Riemann--Hilbert problem associated to the nonlinear
  Schwinger--Dyson equation describing the system at hand: this means
  that the fall-off behavior of the fields of our theory will ultimately
  determine their algebra and quantization procedure. One of the
  methods used to attack such problems is the \emph{steepest descents}
  (also called `saddle-points' or `stationary phase'), which
  identifies the contribution from the dominant saddle-point: once the
  integration contour is appropriately deformed, the original integral
  is extended (by a change of variables) to the entire $\mathbb{R}$eal
  $\varphi$-axis. In this sense, we will consider the Path Integral
  over a $\mathbb{C}$omplex integration cycle (a multi-dimensional
  integration contour), that we deform in order to obtain a Path
  Integral over the $\mathbb{R}$eal $\varphi$-axis (which is the usual
  way they are presented to us). Moreover, we want to associate to
  each integration cycle a $\mathds{M}$atrix label/parameter whose job
  is to `lift' the Feynman Path Integral to an integral transform
  called \emph{Linear Canonical Transform}, \cite{wolfLCT}.

  A previous generalization of the Path Integral along these lines has already
  been done, albeit only using the Elliptic subgroup of
  $\mathsf{SL(2,\mathbb{R})}$: it re-interprets the Path Integral as a
  Fractional Fourier Transform, \cite{frqm}. Unfortunately, this is but
  a subset of all that can be done, although we can already see some of the
  features of the general case: the Elliptic subgroup of
  $\mathsf{SL(2,\mathbb{R})}$ consists of rotations parametrized by an
  arbitrary angle,
  \begin{equation*}
    \mathds{M}_{\text{\textsf{FrFT}}} =
    \begin{pmatrix}
      \cos\omega  & \sin\omega \\
      -\sin\omega & \cos\omega
    \end{pmatrix}
    \;;
  \end{equation*}
  which means that we are performing a rotation of $\omega$ in Phase Space
  ($\varphi$--$J$ plane). This means that the Fractional Fourier
  Transform, i.e., the Path Integral, has real and imaginary parts, and its
  imaginary part only vanishes for special values of
  $\omega$.
  
  However, this rotation is limited by the integration
  cycle defining the Fractional Fourier Transform (viz. Path Integral): we
  cannot cross the poles of the integrand being transformed, given
  that the integration cycle delimits the allowed regions. The
  boundary between two allowed regions is subjected to Stokes'
  phenomena, also known as wall-crossing phenomena in this setting,
  \cite{stokeswall}.

  These Stokes' lines are curves in the $\mathbb{C}$omplex
  $\varphi$-plane, where the integrand of the Path Integral decays without
  oscillations; analogously, anti-Stokes' lines are curves where the integrand
  oscillates without change in amplitude. Upon crossing the Stokes' lines, the
  integrand jumps discontinuously; equivalently, the integrand becomes more and
  more wildly oscillatory as it approaches an anti-Stokes' line. The steepest descent
  path is relevant for topological reasons (more about this on the next item below),
  i.e., we need to know which saddle-point lies on the path. Lastly, we should
  remind ourselves between the connection of Stokes' phenomena and the study of
  hyperasymptotics: the analysis of divergent asymptotic series, where key parts
  of the answer lie `beyond all orders', because the desired features are
  exponentially small in the perturbation parameter,
  \cite{devilsinvention}. This should gives us a clear connection between the
  perturbative series for our model, Stokes' phenomena, and wall-crossing.

  Along these lines, we can define each ``branch'' (wall chamber) of the theory
  determined by $S_{\tau}[\varphi]$, as bounded by appropriate Stokes'
  lines (codimension one manifold). Thus, we can translate the quantities
  computed in one ``branch'' (wall chamber) into another using the discontinuous
  jump determined by crossing Stokes' lines: this enables the computation of the
  discontinuous change of quantities such as integer invariants, space of BPS
  states, etc, across walls of marginal stability. In this sense, maybe we can
  borrow the definition of \emph{quantum phases}, which originally means to
  indicate different quantum states of matter at zero temperature: the idea is
  that varying parameters of the theory, quantum fluctuations can drive phase
  transitions. In our current case, the integration cycle and associated Stokes'
  phenomena essentially determine the allowed range of continuous values of
  the coupling constants. It is at the point where the coupling constants suffer
  a discontinuous leap (along a Stokes' line) that a \emph{quantum phase
    transition} happens. Thus, a given theory $S_{\tau}[\varphi]$ may have more
  than one quantum phase associated to it, provided we find more than one
  integration cycle rendering the Path Integral well-defined.

  \begin{center}
    \begin{tikzpicture}
      %
      \fill[black,opacity=0.25, minimum size=0pt, inner sep=0pt] (0,0.5) circle (1.5cm);
      \fill[Cyan4, minimum size=0pt, inner sep=0pt, outer sep=0pt] (0,1.1) circle (0.5cm);
      \node[minimum size=0pt, inner sep=0pt, outer sep=0pt] (cyan) at (0.4,1.5) {};
      \fill[Magenta4, minimum size=0pt, inner sep=0pt, outer sep=0pt] (-0.6,0.1) circle (0.5cm);
      \node[minimum size=0pt, inner sep=0pt, outer sep=0pt] (magenta) at (-1.0,-0.3) {};
      \fill[Yellow4, minimum size=0pt, inner sep=0pt, outer sep=0pt] (0.6,0.1) circle (0.5cm);
      \node[minimum size=0pt, inner sep=0pt, outer sep=0pt] (yellow) at (1.0,-0.3) {};
      %
      \node[minimum size=0pt] (cqp) at (0,2.5)
        {$\overbrace{\mbox{\hspace{5.5cm}}}^{\displaystyle S_{\tau}[\varphi]}$};
      \node[minimum size=0pt] (cqp) at (2.4,1.5) {$S_{\tau_1}$};
      \node[minimum size=0pt] (yqp) at (2.4,-0.3) {$S_{\tau_2}$};
      \node[minimum size=0pt] (mqp) at (-2.4,-0.3) {$S_{\tau_3}$};
      \path[->, very thick, Red3] (cqp.west) edge [bend right] (cyan);
      \path[->, very thick, Red3] (yqp.west) edge [bend left] (yellow);
      \path[->, very thick, Red3] (mqp.east) edge [bend right] (magenta);
      %
      \draw (3.75,1) node[right, text width=6.75cm] { \small{ %
        \noindent Here we depict a ``single theory'',
        $S_{\tau}[\varphi]$, that has three quantum phases for
        different values of the couplings: $\tau = \tau_1,\, \tau_2,\,
        \tau_3$. Separating each set of allowed couplings are Stokes'
        lines (codimension one manifolds). In this sense, we can say
        that this theory has three quantum phases (wall chambers),
        such that crossing a Stokes' line (wall-crossing) is analogous
        to a quantum phase transition.
    } };
    \end{tikzpicture}
  \end{center}

  Different Quantum Phases are associated with distinct integration
  cycles that are bounded by the appropriate Stokes' lines. This
  yields distinct values for the matrix parameter of the Linear
  Canonical Transform, $\mathds{M}\in\mathsf{SL}(2,\mathbb{C})$, as
  shown in the examples below. Further, because each quantum phase is
  delimited by Stokes' lines, crossing from one quantum phase to
  another implies the picking up of a Stokes phase factor, i.e., a
  finite contribution that accounts for the discontinuity between the
  two quantum phases, and is related to the monodromy of the
  Schwinger--Dyson equations.

  Finally, given the nature of the Linear Canonical Transform as
  generalizing several different kinds of integral transforms, it also
  represents an extension in the allowed classes of paths present in the Path
  Integral (viewed as a sum over paths): on top of Brownian paths, one should
  also consider more generic cases, such as Lévy flights, and so on, \cite{frqm}.
\item[Symplectic Structure \& Complex Path Integral:] We can consider the above
  as the problem of quantizing a symplectic manifold, namely the Phase Space of
  the particular theory we have at hands: we choose a \emph{Lagrangian
  splitting} (a maximal isotropic subspace, a choice of coordinates and momenta,
  i.e., Phase Space $\mathds{P}\simeq \mathds{L}\otimes \mathds{L}'$ where
  $\mathds{L}$ and $\mathds{L}'$ are isotropic subspaces and
  $\dim_{\mathbb{R}}\mathds{L} = \dim_{\mathbb{R}}\mathds{L}' =
  \text{\textonehalf}\,\dim_{\mathbb{R}}\mathds{P}$\/: this is tantamount to
  choosing a \emph{polarization} for the phase space at hand), and the
  integration cycle for the Path Integral is the extra information provided
  towards integrability.

  In this sense, this is very reminiscent of the Bargmann--Segal representation
  (cf. Bargmann--Fock or holomorphic representation)
  equipped with a Kähler polarization and compatible affine and $\mathbb{C}$omplex
  structures, \cite{geometricquantization}. This is reasonable given that the
  Linear Canonical Transform also generalizes the Bargmann transform, as we saw
  in the previous section, with the matrix parameter being
  $\mathds{M} _{\text{\textsf{BT}}} = \tfrac{1}{\hksqrt{2}}\,\bigl(\begin{smallmatrix}
    1 & -i\\ -i & 1
  \end{smallmatrix}\bigr)$ and $-i = e^{-i\,\pi/2}$.

  The relation between Phase Space quantization (also known as Weyl
  or deformation quantization: focus on the multiplicative structure of quantum
  observables), geometric quantization (focus on the representation of observables),
  and the $A$-model in string theory is discussed
  in \cite{geometricquantization,branesquantization}. Essentially, the basic idea is to
  start with a symplectic manifold and choose a suitable
  complexification for it. There are three points of view to this
  subject:
  \begin{enumerate}
  \item Complexify the phase space (i.e., the original symplectic
    manifold) to a complex symplectic manifold that has a good
    $A$-model with respect to the symplectic structure given by the
    imaginary part of the complexified one;
  \item Start with a complex symplectic manifold and pick a suitable
    coisotropic brane, assuming it to be an $A$-brane with respect to
    the symplectic structure given by the imaginary part of the
    original complex one; \&
  \item This is the more natural one from the viewpoint of
    2-dimensional TFTs, bringing to the foreground the $A$-model of
    the complexified phase space: there may be many inequivalent
    choices of $A$-branes, corresponding to different choices of a
    complex structure. Thus, the same $A$-model can lead to
    quantization of the phase space in different symplectic
    structures.
  \end{enumerate}

  The point is to choose a suitable behavior at infinity for the functions that
  we can quantize. Generally, the condition chosen is that only functions of
  polynomial growth are allowed. However, the integration
  cycle that we choose for our Path Integral will ultimately determine this
  fall-off behavior, \cite{complexsaddlepoint}: understanding the Path
  Integral as a Fourier Transform, the idea is to relate the decay
  properties at infinity of its integrand with the analiticity
  properties of the Path Integral (in a loose analogy with the
  Paley--Wiener theorem, \cite{complexsaddlepoint}). This very same observation can
  be made through the study of the [global] boundary conditions of the
  Schwinger--Dyson operator (viz. Bethe--Salpeter equation,
  Ward--Takahashi identities, and anomalies). Thinking in terms of differential
  equations (local systems and Higgs bundles, \cite{delignesimpson}) makes the
  discussion about boundary conditions more straightforward, where different
  boundary conditions are related to distinct \emph{extensions} of the
  respective differential operators, \cite{jextension} (c.f., appendix
  \ref{sec:extop}).

  \begin{center}
    \begin{tikzpicture}[scale=2,cap=round]
      \colorlet{rotaxis}{orange!80!black}
      \colorlet{anglecolor}{green!50!black}
      \tikzstyle{axes}=[]
      \tikzstyle{important line}=[thick]
      \tikzstyle{information text}=[rounded corners,fill=red!10,inner sep=1ex]
      
      \draw[style=help lines,step=0.5cm] (-1.4,-1.4) grid (1.4,1.4);
      \draw[->,style=important line] (-1.5,0) -- (1.5,0) node[right] {\scriptsize{Base manifold ($M$)}};
      \draw[->,style=important line] (0,-1.5) -- (0,1.5) node[above] {\scriptsize{Typical fiber ($\mathsf{G}$)}};
            
      \draw (-0.95,0.65) node[left,green!50!black] {\footnotesize{Section}};
      \draw[<-,style=thick,green!50!black] (-1.4,0.5) -- (0,0.5);
      \draw[->,style=thick,green!50!black] (0,0.25) -- (1.4,0.25);
      \node[green!50!black] (SD1) at (1.45,0.35) {$\scriptstyle \nabla|_{\Sigma_1}$};
      \node[green!50!black] (SD2) at (-1.25,0.3) {$\scriptstyle \nabla|_{\Sigma_2}$};

      \draw (0.75,1.4) node[right] {$\scriptstyle\Sigma_1$};
      \draw[->,style=thick,orange!80!black] (0.785,-1.4) -- (0.785,1.4);
      
      \draw (-0.75,1.4) node[left] {$\scriptstyle\Sigma_2$};
      \draw[->,style=thick,orange!40!black] (-0.785,-1.4) -- (-0.785,1.4);

      \node[circle,opacity=0.65,fill=red!70!black,scale=0.6] (b3) at (0,0.25) {};
      \node[] (dummy3) at (0,0.2) {};
      \node[circle,opacity=0.65,fill=red!70!black,scale=0.6] (b4) at (0.785,0.25) {};
      \node[] (dummy4) at (0.785,0.2) {};
      \node[circle,opacity=0.5,fill=red!80!black,scale=0.6] (b2) at (0,0.5) {};
      \node[] (dummy2) at (0,0.45) {};
      \node[circle,opacity=0.5,fill=red!80!black,scale=0.6] (b1) at (-0.785,0.5) {};
      \node[] (dummy1) at (-0.785,0.45) {};

      \node[] (dummy25) at (0,0.35) {};
  
    \node[red!75!black,inner sep=-1.2cm] (BC) at (1.85,-0.65) {\footnotesize{Boundary Condition}};
      \node[left=of BC] {}
        edge[->,very thick,red!75!black,bend left=-13] (dummy4.south east)
        edge[->,very thick,red!75!black,bend left=-27] (dummy25.east)
        edge[->,very thick,red!75!black,bend left=45] (dummy1.south east);
      %
  
      \draw (2.8,0.1) node[right,text width=6.75cm] { \small{ %
        \noindent In the general case, we have a $\mathsf{G}$-bundle $\pi\,:\,
        P\rightarrow M$ with fiber $\mathsf{G}$. The different boundary
        conditions for the differential equations (local systems) are given by
        $\Sigma_1$ and $\Sigma_2$. The fields of the theory are sections of the
        bundle. The red dots represent the points where the boundary conditions
        are satisfied. The Schwinger--Dyson operators $(\nabla)$ restricted to
        the global boundary conditions are $\nabla|_{\Sigma_1}$ and
        $\nabla|_{\Sigma_2}$: these are two possible \emph{extensions} of
        $\nabla$, \cite{jextension}.
    } };
  \end{tikzpicture}
  \end{center}

  Historically, this study can be traced back to some early ideas by
  V.~I. Arnold, regarding Hermitian operators and
  its ground states, as well as the role played by
  solitons/in\-stant\-ons/mon\-o\-poles (loosely referred to as
  ``quasimodes''), \cite{arnolds}. Later, these were extended to
  symmetric operators with multiple ground states (c.f., appendix
  \ref{sec:extop}, and \cite{jextension}), and also to
  $\mathbb{C}$omplex structures on moduli spaces, \cite{arnolds}.

  Finally, the above can also be understood in terms of the Symplectic
  Field Theory developed by Y.~Eliashberg,
  \cite{symplecticfieldtheory}, including its connection to Floer
  Homology. (Seiberg--Witten--Floer homology yields knot invariants
  formally similar to combinatorially defined Khovanov homology: these
  are closely related to Donaldson and Seiberg invariants of
  4-manifolds, as well as the Gromov invariant of symplectic
  4-manifolds --- solutions to the relevant differential equations,
  that we take to be the Schwinger--Dyson equations.)

\item[Effective Action \& Schwinger's Action Principle:] Expanding the
  motivations for this proposed generalization of the Path
  Integral, we now focus on the construction of the Legendre
  conjugation between $\varphi$ and $J$, and explore its implications:

  \begin{align*}
   \text{if}\quad \mathscr{W}[J] &= -i\, \log\mathscr{Z}[J] \;; &
      \text{and}\quad \Gamma[\varphi] &=  \mathscr{W}[J] - J\,\varphi\;; &\\
    \text{then}\quad \varphi &\mapsto -i\,\tfrac{\delta}{\delta J} \mathscr{W}[J]\;; &
      \text{and}\quad J &\mapsto -i\,\tfrac{\delta}{\delta {\varphi}} \Gamma[\varphi]\;; &\\
    & \text{yielding}\quad &\Gamma[\varphi] = S[\varphi] +
      \tfrac{i}{2}\, \log\det\Bigl(\tfrac{\delta^2}{\delta\varphi^2} S[\varphi]\Bigr) \;.& &
  \end{align*}

  As long as the relation established by the Legendre transform
  between $J$ and $\varphi$ is at most quadratic (in the expression
  defining the Effective Action, $\Gamma$, above), the Linear
  Canonical Transform is the most general integral transform
  preserving the symplectic (resp. metaplectic) form at hand, i.e.,
  preserving the structure defined by the Legendre conjugation of
  $\varphi$ and $J$. This implies that, in fact, we can even extend
  this construction from 1PI to 2PI, \cite{cjt}, and the Canonical
  Linear Transform can again be used to construct the Path Integral of
  the system, given that it still preserves the symplectic structure
  of the theory (once the 2PI construction saturates the quadratic
  relation between $\varphi$ and $J$, \cite{cjt}). Speaking in terms of
  Deformation Quantization, the Linear Canonical Transform preserves
  the structure of the theory up to $\mathcal{O}(\hslash^3)$
  deformations.
  
  Thus, considering the symplectic structure of the plane
  $\varphi$--$J$, i.e., the theory's Phase Space, we can think of the case where $\varphi,
  J\in\mathbb{C}\cup\{\infty\}$, i.e., the field and source are valued
  in the Riemann Sphere: in such a scenario the symplectic structure is then preserved
  by the Möbius group, $\mathsf{SL(2,\mathbb{C})}$, which is the
  automorphism group of the Riemann Sphere. In the special case where
  the coefficients of a Möbius transformation are integers, we obtain
  the \emph{Modular Group}, $\mathsf{SL(2,\mathbb{Z})}$, which is
  relevant for ${S}$- and ${T}$-dualities.

  Finally, with a bit of hindsight from \eqref{eq:fpilctM} and
  \eqref{eq:fpilct0}, we can discuss the Legendre Transform construction of the
  Effective Action based on the Linear Canonical Transform expression for the
  Path Integral, in analogy to what is usually done when we treat the
  Path Integral as a Fourier Transform. Let us assume that $b=-2\,\pi$ in
  the Linear Canonical Transform (cf. \eqref{eq:fpilctM}), for if $b=0$ we only
  have trivial [free] dynamics (cf. \eqref{eq:fpilct0}). Now, we have to discern
  between two cases: $d=0$ and $d \neq 0$. If $d$ vanishes, we have that
  $\varphi \mapsto -i\,\delta_J$, meaning that $\langle\varphi^n\rangle =
  \delta_J^n \mathscr{Z}_{\mathds{M}}/(-2\pi\, i)^n$, implying
  $\langle\varphi\rangle= 0$. However, if $d$ does not vanish, we get
  $\varphi\mapsto \tfrac{b^2}{4\,\pi^2\,d}\, \tfrac{1}{J}\delta_J$, and
  $\langle\varphi\rangle\neq 0$. This gives us a direct way to account for
  symmetry breaking straight from the parametrization of the Path Integral given
  by $\mathds{M}=(\begin{smallmatrix} a & b\\ c & d\end{smallmatrix})\in \mathsf{SL}(2,\mathbb{C})$.

  In this sense, thinking of the Path Integral as a ``sum over paths'', we see
  once again that we are required to include more exotic paths than just
  Brownian ones, such as Lévy flights, and so on: a generic Linear
  Canonical Transform, parametrized by a generic $\mathds{M}\in
  \mathsf{SL}(2,\mathbb{C})$, will be characterized by
  different moments, $\langle\varphi^n\rangle$, depending on the physical system
  at hand, ultimately describing distinct [stable] probability densities,
  \cite{stabledistros}, that are associated to their respective paths and
  diffusion processes. Further, in anticipation of Section
  \ref{sec:PIghiTransform}, it is worth noting that all stable probability
  distributions can be associated with a certain Fox $H$-function,
  \cite{stabledistros}.

  Lastly, we can use the connection between stochastic processes and orthogonal
  polynomials in order to try and establish a Wiener--Askey polynomial chaos
  expansion of the Path Integral, \cite{wieneraskey}, loosely along the
  following lines,

  \begin{align*}
    \mathscr{Z}[J] &= \sum_{i=0}^{\infty} c_i\, Z_i[\zeta]\;;
  \end{align*}
  where $\{Z_i\}$ is the complete orthogonal polynomial basis of the
  Wiener--Askey chaos expansion, and $\zeta$ is a random variable chosen
  according to the type of the random distribution at hand, e.g., free
  fields, i.e., Gaussian random variables, yield Hermite-chaos,
  \cite[cf. table 4.1]{wieneraskey}.

  The idea is to use appropriate Fox $H$-functions for the
  $Z_i$'s, thus developing an ``$H$-chaos expansion'' for the Path
  Integral. This should become more clear in Section \ref{sec:PIghiTransform},
  when we venture into writing the Path Integral in terms of Fox's
  $H$-functions.

\item[Penrose Transform:] Finishing the list of motivations for our generalization of the
  Path Integral, we will borrow an analogy with the Penrose and the Penrose--Ward
  transform. The insight is that we are also trying to define some sort of
  $\mathbb{C}$omplex analogue of the Radon transform, where the straight lines
  we are looking for are not the light-cone as done in the Penrose Transform
  (massless fields, as in twistor theory), but the lines delimiting the Stokes
  wedge where a certain integration cycle is defined, i.e., the Stokes' lines of
  the integrand. In this sense, we would like to establish a parametrization of
  the space of integration cycles by the $\mathds{M}$atrix index, $\mathds{M}\in
  \mathsf{SL}(2,\mathbb{C})$, labeling the Linear Canonical Transform, i.e., we
  would like to parametrize a certain space of compact $\mathbb{C}$omplex
  manifolds (the integration cycles) by a particular $3$-dimensional submanifold of
  $\mathsf{SL}(2,\mathbb{C})$ determined by the allowed values of
  $\mathds{M}$.

  For massless fields, we already know that the Penrose
  Transform establishes a correspondence between two spaces, where one
  parametrizes certain compact complex submanifolds (cycles) of the
  other. Further, its non-linear extension, the Penrose--Ward
  transform, relates certain holomorphic vector bundles (on
  $\mathbb{CP}^3$) with solutions of the self-dual Yang--Mills
  equations (on $S^4$).

  The idea is to construct a map between integration cycles
  (compact $\mathbb{C}$omplex manifolds) and solutions to the Schwinger--Dyson
  equations, given by the Linear Canonical Transform evaluated at a particular
  $\mathds{M}\in \mathsf{SL}(2,\mathbb{C})$.
\end{description}
Therefore, given all of the above, we will define our extended Feynman Path
Integral to be given by the expressions below:

\begin{align}
  \label{eq:fpilctM}
  \mathscr{Z}_{\mathds{M}}[J] &= \hksqrt{-i{\,}}\, e^{i\,\pi\,J^2\,d/b}\,\varint \mathcal{O}[\varphi]\,e^{i\,\pi\,\varphi^2\,a/b}\,e^{-2\,\pi\,i\,\varphi\,J/b}\,\mathcal{D}\varphi\;; &b\neq 0\;;\\
  \label{eq:fpilct0}
  \mathscr{Z}_{\mathds{M}'}[J] &= \hksqrt{d{\,}}\, e^{i\,\pi\,J^2\,c\,d}\, \mathcal{O}[J\, d\,] \;; &b = 0\;;
\end{align}
where $\mathds{M}=(\begin{smallmatrix} a & b\\ c & d\end{smallmatrix})\in \mathsf{SL}(2,\mathbb{C})$
parametrizes our Path Integral, the prime denotes those elements where $b=0$, i.e.,
$\mathds{M}' = (\begin{smallmatrix} a & 0\\ c & d\end{smallmatrix})\in \mathsf{SL}(2,\mathbb{C})$,
and $\mathcal{O}[\varphi]$ is a certain functional of the fields.
%
%
\section{Fox's \emph{H}-, and Generalized \emph{H}-function Transforms}\label{sec:ghiTransforms}
The history of Mellin--Barnes transforms is intimately related to that of
differential and difference equations, as well as the development of
hypergeometric functions: the robust understanding of the asymptotic
properties of functions, hyperasymptotics, Stokes' phenomena, monodromy, and so
on, \cite{devilsinvention}.

Fox's $H$-function was introduced to better the asymptotic control and to be the
most generalized symmetrical Fourier kernel (a generalized Fredholm operator,
\cite{genfredholm}) involving Mellin--Barnes integrals. The $H$-function
includes and generalizes several elementary, higher, and special functions,
ranging from the exponential, Airy, Bessel, special polynomials, hypergeometric,
Riemann Zeta function, up to Meijer's $G$-function, multi-index Mittag--Leffler
function, polylogarithms, etc. This means that we can give a Mellin--Barnes
representation to a wide range of functions.

In particular, the $H$-function has recently found applications in a large variety of
situations, ranging from diffusion all the way up to fractional differential and
integral equations, and so on. It is worth noting that the $H$-function can be
extended to have $\mathds{M}$atrix arguments ($\mathbb{R}$- or
$\mathbb{C}$-valued), to include polylogarithms, and so on,
\cite{Ifunction,foxHfunction}.

The notation we will use for the $H$-function is as follows,

\begin{align}
  \nonumber
  H_{p,q}^{m,n} (z) &= H_{p,q}^{m,n} [\vec{a}_p , \vec{A}_p\, ; \vec{b}_q , \vec{B}_q\, ;\, z] =
    H_{p,q}^{m,n} \biggl[
    \begin{matrix}
       (a_p, A_p) \\
      (b_q, B_q)
    \end{matrix}
    \,\Big|\, z\biggr] =
    H_{p,q}^{m,n} \biggl[
    \begin{matrix}
       (a_1, A_1), \dotsc, (a_p, A_p) \\
       (b_1, B_1), \dotsc,  (b_q, B_q)
    \end{matrix}
    \,\Big|\, z\biggr] \;;\\
    \nonumber
    & \\
    \label{eq:FoxH}
    &= \frac{1}{2\,\pi\,i}\, \int_{\mathcal{L}}\,\frac{\prod_{j=1}^{m} \Gamma(b_j + B_j\, s)\, \prod_{j=1}^{n} \Gamma(1 - a_j - A_j\, s)}
    {\prod_{j=m+1}^{q}\Gamma(1 - b_j - B_j\, s)\, \prod_{j=n+1}^{p} \Gamma(a_j + A_j\, s)} \, z^{-s}\,\mathrm{d}s\;;
\end{align}
where $i = \hksqrt{-1}$, $z\neq 0$, $z^{-s} = \exp\bigl(-s\, (\log|z| + i\, \arg
z )\bigr)$, and $\arg z$ is not necessarily the principal value, and an empty
product is always taken as unity. There are three different integration cycles
$\mathcal{L}$, schematically depicted below, where red ticks are singularities
of $\Gamma(1 - a_j - A_j\, s)$ and green ones are singularities of $\Gamma(b_j +
B_j\, s)$:

\begin{center}
  \begin{tikzpicture}[scale=2,cap=round]
    \colorlet{rotaxis}{orange!80!black}
    \colorlet{anglecolor}{green!50!black}
    \tikzstyle{axes}=[]
    \tikzstyle{important line}=[thick]
    \tikzstyle{information text}=[rounded corners,fill=red!10,inner sep=1ex]
    
    \draw[style=help lines,step=0.5cm] (-1.4,-1.4) grid (1.4,1.4);
    \draw[->,style=important line] (-1.7,0) -- (2,0) node[right] {$\boldsymbol\Re$};
    \draw[->,style=important line] (0,-1.5) -- (0,1.5) node[above] {$\boldsymbol\Im$};

    \foreach  \x in {-1.5, -1.0, -0.5, 0, 0.5, 1.0}
      \draw[very thick,red] (\x,1pt) -- (\x,-1pt);
    \foreach  \x in {0.25, 0.75, 1.25, 1.75}
      \draw[very thick,green] (\x,1pt) -- (\x,-1pt);

    \draw (1.125,1.4) node[right] {\textcolor{orange!80!black}{$\boldsymbol{\mathcal{L}}_{\mathbf{1}}$}};
    \draw[->,style=important line,orange!80!black] (1.125,-1.5) -- (1.125,-0.25) -- (0.125,-0.25) --
      (0.125,0.25) -- (0.375,0.25) -- (0.375,-0.13) -- (0.625,-0.13) --
      (0.625,0.25) -- (0.875,0.25) -- (0.875,-0.13) -- (1.125,-0.13) -- (1.125,1.5);

    \draw (1.9,-0.35) node[above] {\textcolor{blue!80!black}{$\boldsymbol{\mathcal{L}}_{\mathbf{2}}$}};
    \draw[->,style=important line,blue!80!black] (1.9,-0.35) -- (1.1,-0.35) -- (1.1,0.17) --
      (0.9,0.17) -- (0.9,-0.35) -- (0.6,-0.35) -- (0.6,0.17) -- (0.4,0.17) --
      (0.4,-0.35) -- (0.1,-0.35) -- (0.1,0.35) -- (1.9,0.35);

    \draw (-1.9,-0.35) node[below] {\textcolor{LightGoldenrod1!60!black}{$\boldsymbol{\mathcal{L}}_{\mathbf{3}}$}};
    \draw[->,style=important line,LightGoldenrod1!60!black] (-1.9,-0.35) -- (0.05,-0.35) --
      (0.05,0.35) -- (-1.9,0.35);

    \draw (2.1,0.2) node[right,text width=8cm] { %
      \begin{description}
      \item[$\boldsymbol{\mathcal{L}}_{\mathbf{1}}$:] goes from $r -
        i\,\infty$ to $r + i\,\infty$, $r\in\mathbb{R}$, so that all
        sin\-gu\-lar\-i\-ties of $\Gamma(b_j + B_j\, s)$ lie to the
        right of $\mathcal{L}_1$, while all singularities of
        $\Gamma(1 - a_j - A_j\, s)$ lie to the left of it.
      \item[$\boldsymbol{\mathcal{L}}_{\mathbf{2}}$:] loop beginning
        and ending at $+\infty$ and encircling all the singularities
        of $\Gamma(b_j + B_j\, s)$ once in the clockwise
        direction, but none of the singularities of $\Gamma(1 - a_j - A_j\, s)$.
      \item[$\boldsymbol{\mathcal{L}}_{\mathbf{3}}$:] loop beginning
        and ending at $-\infty$ and encircling all the singularities
        of $\Gamma(1 - a_j - A_j\, s)$ once in the
        anti-\- -clock\-wise direction, but none of the singularities of
        $\Gamma(b_j + B_j\, s)$.
      \end{description}
    };
  \end{tikzpicture}
\end{center}

The $H$-function is, in general, multi-valued, due to the presence of the
$z^{-s}$ factor in the integrand. However, it is one-valued on the Riemann
surface of $\log(z)$.

For example, here are some functions expressed in terms of the
$H$-function, i.e., the Mellin--Barnes representations of some
functions:

\begin{align*}
  H^{1,0}_{0,1} \biggl[
    \begin{matrix}
       \\
      (b, B)
    \end{matrix}
    \,\Big|\, z\biggr] &= B^{-1}\, z^{b/B}\,\exp(-z^{1/B}) \;;\\
  & \\
  H^{1,1}_{1,1} \biggl[
    \begin{matrix}
       (1-\nu, 1)\\
      (0, 1)
    \end{matrix}
    \,\Big|\, z\biggr] &= \Gamma(\nu)\, (1 + z)^{-\nu} 
    = \Gamma(\nu)\; {}_1F_0(\nu; \;; -z)\;;\\
  & \\
  H^{1,1}_{1,2} \biggl[
    \begin{matrix}
      (1-\gamma, 1)\\
      (0, 1),\, (1-\beta, \alpha)
    \end{matrix}
    \,\Big|\, -z\biggr] &= \Gamma(\gamma)\, E^{\gamma}_{\alpha,\, \beta}(z)\;;\\
  & \\
  H^{m,n}_{p,q} \biggl[
    \begin{matrix}
      (a_p, A_p=1)\\
      (b_q, B_q=1)
    \end{matrix}
    \,\Big|\, z\biggr] &= G^{m,n}_{p,q}(\vec{a}_p;\, \vec{b}_q;\, z)\;;\\
  & \\
  G^{1,\, s+1}_{s+1,\, s+1} \biggl[
    \begin{matrix}
      0, 1-a, \dotsc, 1-a\\
      0, -a, \dotsc, -a
    \end{matrix}
    \,\Big|\, -1\biggr] &= \zeta(s,a)\;;
\end{align*}
where $\Gamma(z)$ is the Gamma function, ${}_pF_q(\vec{a}_p;\,
\vec{b}_q;\, z)$ is the generalized hypergeometric function,
$E^{\gamma}_{\alpha,\beta}(z)$ is the generalized Mittag--Leffler function,
$G^{m,n}_{p,q}(\vec{a}_p;\, \vec{b}_q;\, z)$ is Meijer's $G$-function, and
$\zeta(s,a)$ is Hurwitz's zeta function (that yields Riemann's zeta function as
$\zeta(s,1)$).

The $H$-function can be further generalized along similar lines as the
construction done above, in order to include some functions which are not
particular cases of Fox's $H$-function, e.g., Polylogarithms, including Polylogs
of complex order, \cite[Appendix A.5]{foxHfunction}, and some further examples
from Physics, \cite{Ifunction}.
 
For Polylogarithms and generalizations of the Riemann-zeta function, among
others, we need to use one certain generalization of the $H$-function, called
the $\bar{H}$-function, $\bar{H}^{m, n}_{p, q}(z)$. The $\bar{H}$-function is
defined as,

\begin{align}
  \label{eq:Hbarfunction}
    \bar{\mathrm{H}}^{m, n}_{p, q} &
    \biggl[
    \begin{matrix}
      (a_j,\alpha_j,A_j)_{1,n} \, , (a_j,\alpha_j)_{n+1, p} \\
      (b_j,\beta_j) _{1,m} \, , (b_j,\beta_j,B_j)_{m+1, q}
    \end{matrix}
    \,\Big|\, z\biggr]
    = \frac{1}{2\,\pi\,i}\, \int_{\mathcal{L}_{i\, r\, \infty}} \chi(s)\; z^{-s}\, \mathrm{d}s \;;\\
  \intertext{where $\chi(s)$ is given by,}
  \label{eq:HbarfunctionDef}
  \chi(s) &= \frac{\prod_{j=1}^{m} \Gamma(b_j + \beta_j\, s)\; \prod_{j=1}^{n} \Gamma^{A_j}(1 - a_j - \alpha_j\, s)}
    {\prod_{j=m+1}^{q}\Gamma^{B_j}(1 - b_j - \beta_j\, s)\; \prod_{j=n+1}^{p} \Gamma(a_j + \alpha_j\, s)} \;;
\end{align}
where the contour $\mathcal{L}_{i\, r\, \infty}$ starts at $r -
i\,\infty$ and goes to $r + i\,\infty$, with $r\in\mathbb{R}$, separating all
the singularities of $\Gamma(b_j - \beta_j\, s)$ from those of $\Gamma(1 - a_j +
\alpha_j\, s)$. For non-integer $A_j$ or $B_j$ the poles of the Gamma functions
in \eqref{eq:HbarfunctionDef} become branch points, whose branch
cuts can be chosen so that the integration cycle can be distorted for each of
the three $\mathcal{L}$'s shown above as long as there is no coincidence of
poles from any pair(s) of $\Gamma$'s.

Then, the polylog of $\mathbb{C}$omplex order $\nu$, $L^{\nu}(z)$, is represented as
follows:
\begin{equation}
  \label{eq:hbarpolylog}
  L^{\nu}(z) = 
  \bar{H}^{1, 2}_{2, 2}
  \biggl[
    \begin{matrix}
      (0,1,1), (1,1,\nu) \\
      (0,1), (0,1,\nu-1)
    \end{matrix}
  \,\Big|\, -z\biggr] \;;
\end{equation}
and a generalization of Hurtwitz's zeta function is given by,
\begin{equation}
  \label{eq:hbarzeta}
  \phi(z,\, q,\, \eta) = 
  \bar{H}^{1, 2}_{2, 2}
  \biggl[
    \begin{matrix}
      (0,1,1), (1-\eta,1,q) \\
      (0,1), (-\eta,1,q)
    \end{matrix}
  \,\Big|\, -z\biggr] = \sum_{k=0}^{\infty}\frac{z^k}{(\eta + k)^q}\;.
\end{equation}

The reason for our studying of the $H$-function is because it shows up when one
generalizes from derivative operators to Fractional Differentiation,
\cite{frqm,foxHfunction,stabledistros}. By extending the Path Integral to be
represented by a Linear Canonical Transform, we have brought into play not only
the elliptic subgroup of $\mathsf{SL}(2,\mathbb{C})$ --- related to fractional
differential operators ---, but also the parabolic,
the Fresnel and Gauss--Weierstrass, the hyperbolic, the Bargmann, etc, subgroups
of $\mathsf{SL}(2,\mathbb{C})$, as already explained in previous
sections. Moreover, as mentioned in the beginning of this section, the
$H$-function is intimately related to the asymptotic behavior of the system it
is describing, as well as its Stokes' phenomena and monodromy properties. In
this fashion, if we can express the Path Integral as an $H$-function, we can
condense all this information in the analytic structure of the
$H$-function. This is the subject of the next section.
\section{Path Integral as a Fox \emph{H}-, or Generalized \emph{H}-function Transform}\label{sec:PIghiTransform}
Throughout the paper, we will also try and express the calculated Path
Integral in terms of the $H$-function and its generalizations,
\cite{Ifunction}. The reason for this stems from the fact that the $H$-function
is a Mellin--Barnes transform, which shows up in generalizations from derivative
operators to Fractional Differentiation, etc, \cite{frqm,foxHfunction,stabledistros}.

As such, we look for a representation of our theory in terms of a Mellin--Barnes transform,
because of its relation to asymptotic expansions (i.e., Stokes Phenomena and
wall-crossing), and its closeness to Fourier Transforms: in this sense we would like to
capture the properties of the theory straight from its singularities. We will
consider the Path Integral as a hypergeometric integral transform, or simply an
$H$-transform, \cite{integraldiscriminants,frqm}. This view in terms of an
$H$-transform is of particular significance given Fox's $H$-function's
definition in terms of a Mellin--Barnes transform, that clearly brings the
singularity structure of the particular $H$-function being used to the
foreground. This shows that the Path Integral considered as an $H$-transform
encodes all of the interesting and relevant singularity structure, which can be
as simple as that of Polylogarithms, or far more intricate as we will show in
the examples below.

In the particular case of Polylogs, represented by the $\bar{H}$-function, their
recent importance for the computation of scattering amplitudes (maximally
helicity violating amplitudes, MHV, and Wilson loops in $\mathcal{N}=4$ SYM),
their relation with Grassmannians and Cluster Varieties (viz. Stokes Phenomena),
their relation to the Bloch--Wigner--Ramakrishnan--Zagier functions and volumes of
polytopes in AdS and mixed Tate motives, makes them a very relevant case to be
studied. In this way, by expressing the Path Integral in terms of a generic
$H$-function, we can more easily see what is induced by this structure.

As an example of how the singularity structure of the $H$-function determines a
theory, we can look at \cite{frqm}, where the fundamental solution
(Green's function) for a fractional diffusion process is given by
a generalization of the Airy function, expressed in terms of Fox's
$H$-function: $\mathrm{H}^{1,0}_{1,1}[(1-\beta/2,\beta/2);
(0,1); \varphi]$, where the limit $\beta\rightarrow 1$ recovers the
Gaussian density $\mathrm{H}^{1,0}_{1,1}[(1/2,1/2);
(0,1); \varphi] = \exp(-\varphi²/4)/2\,\hksqrt{\pi}$, and the
limit $\beta\rightarrow 2/3$ recovers 
the Airy function, $\mathrm{H}^{1,0}_{1,1}[(1/3,2/3);
(0,1); \varphi] = 3^{2/3}\, \Ai(\varphi/3^{1/3})$. In fact,
a stronger result holds: all \emph{stable} probability densities
can be represented in terms of Fox $H$-functions \cite{stabledistros}.
(A random variable is
called \emph{stable} (and has a stable distribution) if a linear
combination of two independent copies of itself has the same
distribution, up to translations and scalings. Examples of
stable distributions include the Gaussian, Cauchy and Lévy
distributions.)

    %
    %
%
\section{${\mathbf{D0}}$-brane Examples}\label{sec:examples}
We perform the analysis suggested above in three different examples, progressively
showing more and more relevant details and properties of our extension for the
Path Integral.

The models we study include the free field, the cubic, and quartic interactions,
revealing their different structure dependent on $\mathds{M}\in\mathsf{SL}(2,\mathbb{C})$,
i.e., using a 3-dimensional manifold as a parameter space.

We would like to re-interpret our construction along the lines of a
BFSS/IKKT matrix-model and consider its large-$N$ limit, which is conjectured to be
equivalent to $M$-theory. As such, rather than considering the problem from a
bottom-up, constructionist, viewpoint, we would like to think about it using a
top-down strategy: rather than solving a $D0$-brane model and considering
infinitely many copies of it in order to build an extended model, we can think
in ``reversed'' terms, where certain extended structures of the theory determine
its properties. This has been named Brane Quantization,
\cite{branesquantization}. Our plan is to use our $D0$-brane examples
in this section, considering the implications of the
large-$N$ limit of these models to certain structures in each one of
them, e.g., the convergence of the Path Integral defining the model
or, analogously, the boundary conditions of the Schwinger--Dyson
equation each model obeys. For example, let us consider a
$\mathds{M}$atrix-valued $D0$-brane model with a generic polynomial
for an Action. Its Feynman Path Integral (FPI) and Schwinger--Dyson
Equation (SDE) are given by,

\begin{align}
  \label{eq:MatrixD0braneFPI}
  \mathscr{Z}_{\tau}[\mathds{J}] = \varint_{\mathcal{C}}
    e^{i\, \tr\sum_{k=0}^{n} \mathds{A}_k\,\boldsymbol{\varphi}^k}\, 
    e^{i\, \mathds{J}\, \boldsymbol{\varphi}}\, \mathcal{D}\boldsymbol{\varphi} &< \infty\;;\\
  \label{eq:MatrixD0braneSDE}
  \tr\Bigl(\sum_{k=1}^{n} k\,\mathds{A}_k\, (-i\, \partial_{\mathds{J}})^{k-1} - \mathds{J}\Bigr)\, \mathscr{Z}_{\tau}[\mathds{J}]&= 0\;;
\end{align}
where $\boldsymbol{\varphi}\in\mathbb{C}^{n\times n}$ is our
$\mathds{M}$atrix-valued field and $\mathds{J}\in\mathbb{C}^{n\times
 n}$ is its source, $\mathcal{D}\boldsymbol{\varphi}$ is
the canonical integration measure in $\mathbb{C}^{n\times n}$,
$\mathcal{C}$ is an integration cycle rendering
\eqref{eq:MatrixD0braneFPI} finite, and $\tau$ is a label to remind us
of the connection between $\mathcal{C}$ and
$\{\mathds{A}_k\}_{k=0}^{n}$ (this will be further explored
later), i.e., different sets of coupling constants will yield Actions
that will be convergent under distinct cycles $\mathcal{C}$, and
$\tau$'s role is to perform this bookkeeping.

We could as well have written the above for a graded $\mathscr{L}$ie
Algebra-valued field, i.e., for a superfield
$\boldsymbol{\Phi}$ with superpotential $W[\boldsymbol{\Phi}]$,

\begin{align}
  \label{eq:SUSYD0braneFPI}
  \mathscr{Z}_{\tau}[\mathfrak{J}, \bar{\mathfrak{J}}] &= \displaystyle\varint_{\mathcal{C}}
    e^{i\, (L^a\,\boldsymbol{\Phi}_a + \tfrac{M^{a\, b}}{2}\,\boldsymbol{\Phi}_a\,\boldsymbol{\Phi}_b + \tfrac{y^{a\, b\, c}}{6}\,\boldsymbol{\Phi}_a\,\boldsymbol{\Phi}_b\,\boldsymbol{\Phi}_c + \text{ c.c.})}\, 
    e^{i\, \mathfrak{J}^a\, \boldsymbol{\Phi}_a}\, e^{i\, \bar{\mathfrak{J}}^a\, \bar{\boldsymbol{\Phi}}_a}\, \mathcal{D}\boldsymbol{\Phi}\, \mathcal{D}\bar{\boldsymbol{\Phi}} < \infty\;;\\
  \nonumber
  &\\
  \label{eq:SUSYD0braneSDE}
  &\quad\bigl(M^{a\, b}\, \partial_{\mathfrak{J}^b} - i\, y^{a\, b\, c}\, \partial_{\mathfrak{J}^b}\, \partial_{\mathfrak{J}^c}
    - \tilde{\mathfrak{J}}^a\bigr)\, \mathscr{Z}_{\tau}[\mathfrak{J}]= 0\;;\\
  \label{eq:SUSYD0braneCCSDE}
  &\quad\bigl(\bar{M}^{a\, b}\, \partial_{\bar{\mathfrak{J}}^b} - i\, \bar{y}^{a\, b\, c}\, \partial_{\bar{\mathfrak{J}}^b}\, \partial_{\bar{\mathfrak{J}}^c}
    - \tilde{\bar{\mathfrak{J}}}^a\bigr)\, \mathscr{Z}_{\tau}[\bar{\mathfrak{J}}]= 0\;;
\end{align}
where $\tilde{\mathfrak{J}}^a = \mathfrak{J}^a + L^a$, i.e., the source is shifted by the
linear term (\emph{ditto} for its complex conjugate term). This motivates our treatment of the
cubic Action on Section \ref{subsec:airy}, i.e., cubic potentials are relevant to SUSY models,
as well as to [non-Abelian] Chern-Simons theory.

At this point we have some observations to make regarding the coupling constants
of the model at hand, that analogously to the fields themselves, can be valued
in $\mathbb{R}$ or $\mathbb{C}$, or be $\mathds{V}$ector or $\mathds{M}$atrix
valued, or $\mathscr{L}$ie algebra or graded $\mathscr{L}$ie algebra valued.

The couplings can be understood as background fields, meaning that the
renormalized, effective Action is constrained by symmetries (superselection
rules), by its holomorphic properties (Lee--Yang and Fisher zeros), and by its
different properties in different limits (weak and strong coupling
limits). Moreover, a holomorphic function (i.e., a section in an appropriate
bundle) is determined by its asymptotic behavior and singularities, connecting
us to he formalism developed earlier, where the Path Integral is represented in
terms of the appropriate $H$-function encoding its asymptotic behavior and
singularities.

Furthermore, we can implement desired boundary conditions for the
Schwinger--Dyson equations in terms of a VEV(s) for the field(s)
(at the boundary), $\langle\varphi\rangle|_{\partial M} = \text{const}$ (flux
quantization), or via explicit terms in the action. See Appendix \ref{sec:extop}
and \cite{jextension,branesquantization}.

So, considering the couplings as parametrizing a family of functions between two
Morse functions (namely, the Hamiltonians described by the initial and final
values of the couplings in a certain range), its degeneracies involve a birth
(branching out) or death (merging together) transition of critical points. This
study can be done via Cerf theory, which tackles stratified spaces labelled by
the co-dimension of the strata, \cite{arnolds}. As such, in the examples below
we will show plots to ilustrate these transitions at particular values of the
couplings.
\subsection{Free Field}\label{subsec:gaussian}
In order to get our feet wet and highlight some of the properties that will be
relevant later on, let us start with the simplest possible example: the free
field.

This $D0$-brane model has an Action given by $S[\varphi] = \mu\,\varphi^2/2$,
yielding the following FPI and SDE:

\begin{align}
  \label{eq:freeLCT}
  \mathscr{Z}[J] &= \varint_{\mathcal{C}} e^{i\,\mu\,\varphi^2/2}\,e^{i\, \varphi\, J}\,\mathcal{D}\varphi < \infty\;;\\
  \label{eq:freeSDE}
  (-i\,\mu\,\partial_J &- J)\, \mathscr{Z}[J] = 0\Rightarrow \mathscr{Z}[J] = a\,e^{i\,J^2/2\mu}\;;
\end{align}
where $a$ is an arbitrary constant, and $\mathcal{C}$ is an integration cycle.

In order to cast it as an LCT, we will follow \eqref{eq:fpilctM} and choose
$\mathcal{O}[\varphi] = \mathds{1}$ and $\mathds{M} = \bigl(\begin{smallmatrix} -\mu & -2\,\pi\\ \frac{1}{2\,\pi} & 0\end{smallmatrix}\bigr)$.
This choice of $\mathds{M}\in\mathsf{SL}(2,\mathbb{C})$ gives us the rational
function determining the analytic structure of this transformation,

\begin{equation}
  \label{eq:analstruct}
  \mathds{M}(\varphi) = f_{\mu}(\varphi) = \frac{-2\,\pi\, (\mu\,\varphi + 2\,\pi)}{\varphi}\;.
\end{equation}
These are plots of $f_{\mu}(\varphi)$ for different values of $\mu$:

\begin{center}
  %
  \hspace*{-2.9cm}\mbox{%
    \raisebox{7mm}{%
      \includegraphics[scale=0.65]{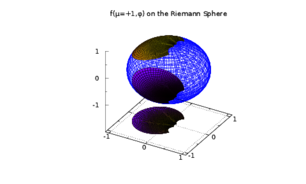} %
      \hspace*{-1.5cm} %
      \includegraphics[scale=0.65]{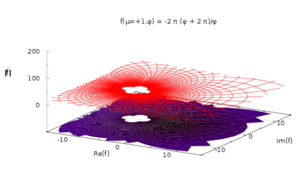}} %
    \raisebox{1.3cm}{%
      \includegraphics[scale=0.25]{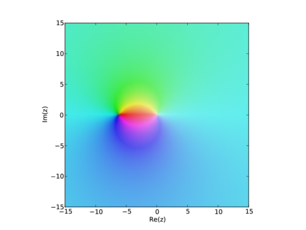}}   %
  }
  \hspace*{-2.9cm}\mbox{%
    \raisebox{7mm}{%
      \includegraphics[scale=0.65]{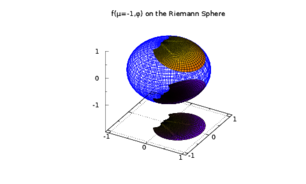} %
      \hspace*{-1.5cm} %
      \includegraphics[scale=0.65]{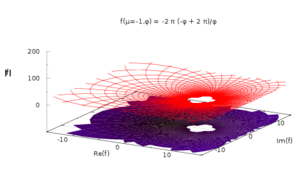}} %
    \raisebox{1.3cm}{%
      \includegraphics[scale=0.25]{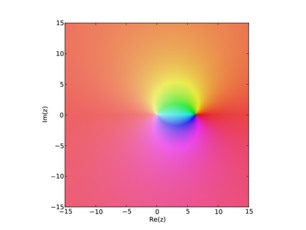}}   %
  }
  \hspace*{-2.9cm}\mbox{%
  \raisebox{7mm}{%
    \includegraphics[scale=0.65]{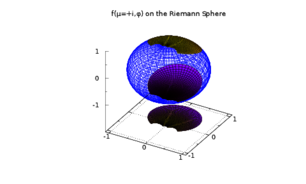} %
    \hspace*{-1.5cm} %
    \includegraphics[scale=0.65]{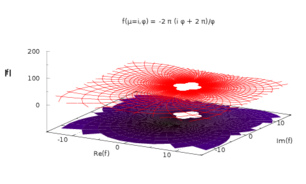}} %
  \raisebox{1.3cm}{%
    \includegraphics[scale=0.25]{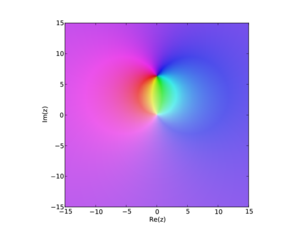}}   %
  }
  \hspace*{-2.9cm}\mbox{%
    \raisebox{7mm}{%
      \includegraphics[scale=0.65]{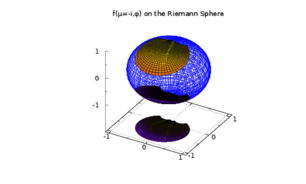} %
      \hspace*{-1.5cm} %
      \includegraphics[scale=0.65]{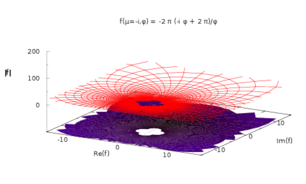}} %
    \raisebox{1.3cm}{%
      \includegraphics[scale=0.25]{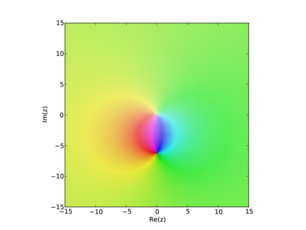}}   %
  }%
  \vspace{-5mm}
   
  \begin{minipage}[c]{0.85\linewidth}
    {\small\textsf{%
        \noindent Plots of $f_{\mu}(\varphi)=\tfrac{-2\,\pi\, \mu\,\varphi -
          4\,\pi^2}{\varphi}$, for $\mu=1, -1, i, -i$, respectively.
        The plots on the right
        have $\Re(f_{\mu=\pm 1, \pm i})$ as
        the $x$-axis and $\Im(f_{\mu=\pm 1, \pm i})$ as the $y$-axis, with variations of
        color (phase $\theta$) and brightness (magnitude $\rho$) to
        indicate $z=\rho\,e^{i\,\theta}=f_{\mu=\pm 1, \pm i}$; plots on the left are simply their
        wrapping over the Riemann Sphere.
      }%
    }
  \end{minipage}
\end{center}

It is not difficult to see that these graphs are related to each other by a
Möbius transform of the Complex $\varphi$-plane.

For the special case where $\mu = i$ we recover a well known identity about
Fourier Transforms: the Hermite polynomials are its eigenfunctions. This can be
seen in \eqref{eq:freeLCT} simply by noting that the 0\textsuperscript{th} order
Hermite polynomial is given by $\mathrm{He}_0(\varphi) = 1$, as defined in
\eqref{eq:fteigen} and \eqref{eq:hermite}, and
$\mathscr{F}[e^{-\varphi^2/2}\,\mathrm{He}_n(\varphi)] =
(i)^n\,e^{-J^2/2}\,\mathrm{He}_n(J)$, taking $n=0$ for this special case of
ours. This result can be extended to other values of $\mu$ with a change
of variables given by, $\varphi\mapsto \hksqrt{i\,\mu}\,\varphi$ and $J\mapsto
J/\hksqrt{i\,\mu}$, provided we stay within the same Stokes wedge defining the
original integral transform. This is in complete analogy with the following
extension of Hermite polynomials: $\mathrm{He}_n^{[\alpha]}(\varphi) =
\alpha^{-n/2}\, \mathrm{He}_n^{[1]}(\varphi/\hksqrt{\alpha}) = e^{-\alpha\, \partial^2_{\varphi}/2}\,\varphi^n$;
meaning that we can choose $\alpha = i\, \mu$ and talk in terms of
$\mathrm{He}_n^{[i\,\mu]}(\varphi)$.

As mentioned before, we want to represent these results in terms of Fox's
$H$-function. In order to do so, we will use the following conversion table:

\begin{align*}
  \mathrm{He}_{2\,n}(\varphi) &= (-1)^n\,\frac{(2\,n)!}{n!}\,{}_{1}\mathrm{F}_{1}(-n ; \tfrac{1}{2} ; \varphi^2) \;;\\
  &= (-1)^n\,\frac{(2\,n)!}{n!}\, \frac{\Gamma(\tfrac{1}{2})}{\Gamma(-n)}\,\mathrm{H}^{1,1}_{1,2}[(1+n,1); (0,\tfrac{1}{2},1); \varphi^2]\;;\\
  \mathrm{He}_{2\,n+1}(\varphi) &= (-1)^n\,\frac{(2\,n+1)!}{n!}\,{}_{1}\mathrm{F}_{1}(-n ; \tfrac{3}{2} ; \varphi^2) \;;\\
  &= (-1)^n\,\frac{(2\,n+1)!}{n!}\, \frac{\Gamma(\tfrac{3}{2})}{\Gamma(-n)}\,\mathrm{H}^{1,1}_{1,2}[(1+n,1); (0,-\tfrac{1}{2},1); \varphi^2]\;;\\
\end{align*}
where ${}_{1}\mathrm{F}_{1}$ is Kummer's confluent hypergeometric function,
$\Gamma$ is the Gamma function, and $\Gamma(\tfrac{1}{2}) = \hksqrt{\pi}$ and $\Gamma(\tfrac{3}{2}) = \hksqrt{\pi}/2$.

With this in mind, it is not difficult to write the FPI as an
$H$-transform,
\begin{align}
  \label{eq:freeFPISDE}
  \mathscr{Z}_{\mathds{M}}[J] &= \overbrace{\varint e^{i\,\mu\,\varphi^2/2}\, e^{i\, \varphi\, J}\, \mathcal{D}\varphi}^{\text{FPI}}
    = \overbrace{a\, e^{i\, J^2/2\mu}}^{\text{SDE}} \;;\; \text{where }
    \mathds{M} = \bigl(\begin{smallmatrix} -\mu & -2\,\pi \\ 1/(2\pi) & 0 \end{smallmatrix}\bigr)\;;\\
  \label{eq:freeHtransf}
  &= a\, \hksqrt{\pi}\,\mathrm{H}^{1,0}_{1,1}[(\tfrac{1}{2},\tfrac{1}{2}) ; (0,1) ; \hksqrt{\tfrac{-2\,i}{\mu}}\,J]\;.
\end{align}
In this particular case, it is possible to reinterpret
\eqref{eq:freeFPISDE} in terms of an LCT with $b=0$, following
\eqref{eq:fpilct0}. All we need to do is choose $\mathcal{O}[J\, d] =
\mathds{1}$ and,

\begin{equation*}
  \mathds{M}' =
  \begin{pmatrix}
    1 & 0 \\
    \omega^{-1}  & 1 \\
  \end{pmatrix}
  \;;
\end{equation*}
where $\omega = i\, 2\,\pi\,\mu$. This means that
$\mathscr{Z}_{\mathds{M}'}[J]$ is a Gaussian Transform of width $-i\,
\omega$.

Now, we can use \eqref{eq:freeLCT} parametrized by \eqref{eq:analstruct} in
order to construct Coherent States $\Upsilon_{\varphi_0}^{\mu}[J]$, parametrized
by $\mu$ and $\varphi_0 = \langle\varphi\rangle$.

The general expression for $\Upsilon_{\varphi_0}[J]$ is given in terms of the
matrix $\mathds{A} = (\begin{smallmatrix} a & b\\ c & d\end{smallmatrix})$
diagonalizing $\mathds{M}$, $\mathds{A}^{-1}\, \mathds{M}\, \mathds{A} = \diag(\lambda_1,\lambda_2)$, in the following fashion:

\begin{equation}
  \label{eq:LCTcs}
  \Upsilon_{\varphi_0}(J) = (\mathpzc{C}_{\mathds{A}}\,\delta_{\varphi_0}) (J)
    = \hksqrt{-i{\,}}\, e^{i\,\pi\,J^2\,d/b}\,\varint_{-\infty}^{\infty} \delta(\varphi-\varphi_0)\,e^{i\,\pi\,\varphi^2\,a/b}\,e^{-2\,\pi\,i\,\varphi\,J/b}\,\mathrm{d}\varphi\;;
\end{equation}
where the notation is the same as in \eqref{eq:lctinttransf},
\eqref{eq:lctkernel}, \eqref{eq:lctinttransfinv}: that is, the coherent states
are the Linear Canonical Transforms of the Dirac-delta function parametrized by
the diagonalization transform, i.e., localized states defining each quantum
phase (representing each integration cycle and Stokes' wedge).

For this case of ours, the coherent state is given by,
\begin{equation}
  \label{eq:freeCS}
  \Upsilon_{\varphi_0}^{\mu}[J] = \hksqrt{-i}\, e^{i\, J^2/2\,\lambda_{\mu}}\, e^{i\,\pi\,\varphi_0^2}\,e^{-2\,\pi\,i\,\varphi_0\,J}\;;
\end{equation}
where $\lambda_{\mu} = \tfrac{-\mu - \hksqrt{\mu^2-4}}{2}$, and
$\lambda_{\mu=\pm 2} = \mp 1$, $\lambda_{\mu=\pm 2\,i} = \mp i - i\,\hksqrt{2}$.

The plots below depict $\Upsilon_{\varphi_0}^{\mu}[J]$ for $\mu = \pm
2$ and $\mu = \pm 2\, i$: the graphs on the left are the wrapping over
the Riemann Sphere of the ones on the right.

\begin{center}
  %
  \includegraphics[scale=0.77]{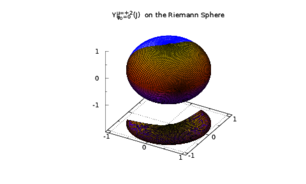}
  \raisebox{0.7cm}{\includegraphics[width=0.4\linewidth,keepaspectratio]{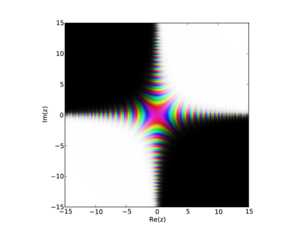}}

  \includegraphics[scale=0.77]{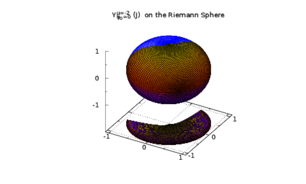}
  \raisebox{0.7cm}{\includegraphics[width=0.4\linewidth,keepaspectratio]{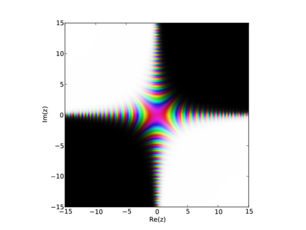}}

  \includegraphics[scale=0.77]{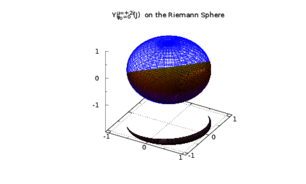}
  \raisebox{0.7cm}{\includegraphics[width=0.4\linewidth,keepaspectratio]{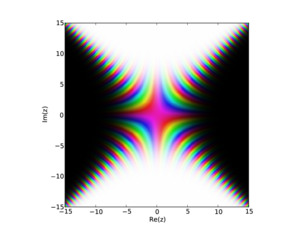}}

  \includegraphics[scale=0.77]{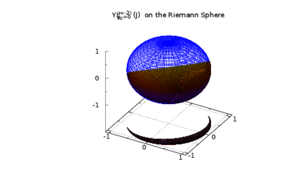}
  \raisebox{0.7cm}{\includegraphics[width=0.4\linewidth,keepaspectratio]{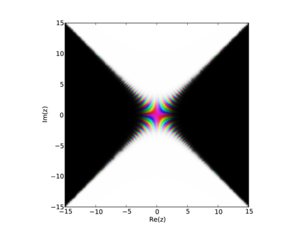}}
  %
  %
  %
  %

  \begin{minipage}[c]{0.85\linewidth}
    {\small\textsf{%
        \noindent Coherent State plots for different combinations of
        the parameters $\mu$ and $\varphi_0$. The plots on the right
        have $\Re(\Upsilon^{\mu=\pm 2, \pm 2\,i}_{\varphi_0=0}[J])$ as
        the $x$-axis and $\Im(\Upsilon^{\mu=\pm 2, \pm
          2\,i}_{\varphi_0=0}[J])$ as the $y$-axis, with variations of
        color (phase $\theta$) and brightness (magnitude $\rho$) to
        indicate $z=\rho\,e^{i\,\theta}=\Upsilon^{\mu=\pm 2, \pm
          2\,i}_{\varphi_0=0}[J]$; plots on the left are simply their
        wrapping over the Riemann Sphere.
      }%
    }
  \end{minipage}
\end{center}
\subsection{Cubic Action}\label{subsec:airy}
At this point we can try and study a slightly non-trivial model, given
by a $D0$-brane with $S[\varphi] = \varphi^3/3$. It's Feynman Path Integral and
Schwinger--Dyson equation are shown below:

\begin{align}
  \label{eq:airyFPI}
  \mathscr{Z}[J] &= \varint_{\mathcal{C}} e^{i\, \varphi^3/3}\, 
    e^{i\, J\, \varphi}\, \mathcal{D}\varphi < \infty\Rightarrow \mathcal{C}:\, \mathrm{Arg}(\varphi) = \{0,\,\pm 2\,\pi/3\}\;;\\
  \nonumber  & \\
  \label{eq:airySDE}
  \bigl(\partial_J^2 &- J\bigr)\, \mathscr{Z}[J] = 0\Rightarrow \mathscr{Z}[J] = a\,\Ai[J] + b\,\Bi[J]\;;
\end{align}
where the contour $\mathcal{C}$ over which the Path Integral is well defined is
along the lines of $\varphi$ such that its arguments are either $0$ or
$\pm2\,\pi/3$, and $\Ai$ and $\Bi$ are the Airy functions, that can
also be expressed in terms of Fox's $H$-function as
$3^{2/3}\Ai(\varphi/3^{1/3}) =
\mathrm{H}^{1,0}_{1,1}\bigl((\tfrac{2}{3},\tfrac{1}{3}); (0,1);
\varphi/3^{1/3}\bigr)$.

The $\Bi$ function can be expressed in terms of the $\Ai$ function via a
combination of different integration cycles: $\Bi[J] = e^{-i\,\pi/6}\,
\Ai[J\, e^{-2\,\pi\,i/3}] + e^{i\,\pi/6}\,\Ai[J\, e^{2\,\pi\,i/3}]$. That is, we
scale and rotate the argument. The combination of the two
integration cycles can be understood as follows: the pre-factor $e^{\pm i\,\pi/6}$
is a critical value for the scaling, and $g = e^{\pm 2\,\pi\,i/3}$ is the
appropriate Stokes' factor associated to crossing each cycle (upon which we
cross a cycle and pick up another Stokes' factor).

Now, we want to choose an appropriate $\mathds{M}$ and
$\mathcal{O}[\varphi]$ in order to parametrize our Path Integral along the lines
of \eqref{eq:fpilctM}. Let us choose
$\mathds{M}_{\Ai} = \bigl(\begin{smallmatrix} 0 & -2\,\pi\\ 1/2\pi & 0\end{smallmatrix}\bigr)$
and $\mathcal{O}[\varphi] = e^{i\, \varphi^3/3}$.

\begin{center}
  \begin{tikzpicture}[scale=2,cap=round]
    \colorlet{rotaxis}{orange!80!black}
    \colorlet{anglecolor}{green!50!black}
    \tikzstyle{axes}=[]
    \tikzstyle{important line}=[thick]
    \tikzstyle{information text}=[rounded corners,fill=red!10,inner sep=1ex]
    
    \draw[style=help lines,step=0.5cm] (-1.4,-1.4) grid (1.4,1.4);
    \draw[->,style=important line] (-1.5,0) -- (1.5,0) node[right] {$\boldsymbol\Re$};
    \draw[->,style=important line] (0,-1.5) -- (0,1.5) node[above] {$\boldsymbol\Im$};

    \draw (1.3,1.2) node {$\mathcal{C}_1 = \Ai[J\,e^{+2\,\pi\,i/3}]$};
    \draw[style=important line,orange!80!black] (0,0) -- (1.2,0) arc(0:120:1.2cm) -- (0,0);

    \draw (1.3,-1.2) node {$\mathcal{C}_2 = \Ai[J\,e^{-2\,\pi\,i/3}]$};
    \draw[style=important line,orange!40!black] (0,0) -- (1.2,0) arc(0:-120:1.2cm) -- (0,0);

    \filldraw[fill=green!20,draw=anglecolor,opacity=0.45] (0,0) -- (1.126,0.65) arc(30:-30:1.3cm) -- (0,0);

    \filldraw[fill=green!20,draw=anglecolor,opacity=0.45] (0,0) -- (0,1.3) arc(90:150:1.3cm) -- (0,0);
    \filldraw[fill=green!20,draw=anglecolor,opacity=0.45] (0,0) -- (0.,-1.3) arc(-90:-150:1.3cm) -- (0,0);

    \draw (2.5,0) node[right,text width=8cm] { %
      The shaded regions are Stokes' wedges and represent zones of convergence:
      cycles beginning and ending within these regions will yield meaningful
      integrals. $\mathcal{C}_1$ and $\mathcal{C}_2$ are particular cycles
      giving rise to the two independent solutions indicated with their
      respective Stokes' factor.
    };
  \end{tikzpicture}
\end{center}

As we mentioned before, $\mathds{M}_{\Ai}$ can be decomposed in the
following fashion: $\mathds{M}_{\Ai} =
\mathds{M}_{\text{DT}}(-2\,\pi)$ $\mathds{M}_{\text{FT}}\, \mathds{M}_{\text{GT}}(0)$,
i.e., we use a composition of a Gaussian Transform with a Fourier
Transform and with a Dilation. However, because we now have an
$\mathcal{O}[\varphi] = e^{i\, \varphi^3/3}$ that is non-trivial, it
is easier to read the analytic structure of our Path Integral straight from its
representation in terms of its associated $H$-function.

As a passing comment, let us just note that \eqref{eq:airySDE} can be
easily generalized in two different ways:

\begin{align*}
  \bigl(\partial^2_J \pm g^2\,J\bigr)\, \mathscr{Z}_{\mathds{M}_{\Ai_g}}[J] &= 0 \;;\\
  \bigl(\partial^3_J -4\,J\,\partial_J -2\bigr)\, \mathscr{Z}_{\mathds{M}^2_{\Ai}}[J] &= 0 \;;
\end{align*}
that are finite at the origin, and where the first generalization can
be understood in terms of a scaling by a factor of $g^{2/3}$, while the
second generalization can be interpreted in terms of completing the
squares --- as expected, these solutions can easily be expressed in
terms of $H$-functions (via dilations of its variable or by taking powers of it),
and are also parametrizable by distinct $\mathds{M}$'s. That is, completing the
polynomial potential such that we have an Action along the lines of $S[\varphi] = 
a\, \varphi^2/2 + b\,\varphi^3/3$, the coefficient of the highest power
(``top coupling'') determines the analytical structure. As $b\rightarrow 0$,
$a$ plays a more and more relevant role: only a cycle contained in both
regions would yield convergent solutions for both powers --- a cycle in
non-matching convergence regions yields a divergent $\mathscr{Z}$, while an
integration cycle that can be deformed from one convergence region to
another yields a vanishing $\mathscr{Z}$ (analogous to the accumulation of
Lee--Yang zeros).

\begin{center}
  \begin{tikzpicture}[scale=2,cap=round]
    \colorlet{rotaxis}{orange!80!black}
    \colorlet{anglecolor}{green!50!black}
    \tikzstyle{axes}=[]
    \tikzstyle{important line}=[thick]
    \tikzstyle{information text}=[rounded corners,fill=red!10,inner sep=1ex]
    
    \draw[style=help lines,step=0.5cm] (-1.4,-1.4) grid (1.4,1.4);
    \draw[->,style=important line] (-1.5,0) -- (1.5,0) node[right] {$\boldsymbol\Re$};
    \draw[->,style=important line] (0,-1.5) -- (0,1.5) node[above] {$\boldsymbol\Im$};

    \draw (2.9,1.1) node[right,text width=7cm] { %
      The shaded regions represent convergence contours for the following
      terms, respectively: blue $\Leftrightarrow$
      cubic, and yellow $\Leftrightarrow$ quadratic.
    };

    \draw[->,style=important line] (-0.7,1.2124) .. controls (0,0) and (1,0.1) .. (1.4,0.09) node[above] {\hspace{7em}$\mathcal{C}_i : \Ai[J\, e^{+2\,\pi\, i/3}]$};
    \draw[->,style=important line] (-0.7,1.2124) .. controls (0,0) and (-0.1,-1) .. (-0.09,-1.4) node[below] {$\mathcal{C}_k$};
    \draw[->,style=important line] (-0.7,-1.2124) .. controls (0,0) and (1,-0.1) .. (1.4,-0.09) node[below] {\hspace{7em}$\mathcal{C}_j : \Ai[J\, e^{-2\,\pi\, i/3}]$};

    \filldraw[fill=blue!20,draw=blue!50!black,opacity=0.45] (0,0) -- (1.126,0.65) arc(30:-30:1.3cm) -- (0,0);

    \filldraw[fill=blue!20,draw=blue!50!black,opacity=0.45] (0,0) -- (0,1.3) arc(90:150:1.3cm) -- (0,0);
    \filldraw[fill=blue!20,draw=blue!50!black,opacity=0.45] (0,0) -- (0.,-1.3) arc(-90:-150:1.3cm) -- (0,0);

    \filldraw[fill=yellow!20,draw=yellow!50!black,opacity=0.55] (0,0) -- (1.2,0) arc(0:90:1.2cm) -- (0,0);
    \filldraw[fill=yellow!20,draw=yellow!50!black,opacity=0.55] (0,0) -- (-1.2,0) arc(180:270:1.2cm) -- (0,0);
  \end{tikzpicture}
\end{center}

Therefore, the quantum phases of this system are given by the $\Ai$
and $\Bi$ functions, and their integral representation corresponds to
the appropriate Path Integral associated to each quantum phase. Moreover, their
respective coherent states are given by,

\begin{equation}
  \label{eq:airyCS}
  \Upsilon_{\varphi_0}[J] = \hksqrt{-i}\, e^{-J^2/2}\, e^{-i\,\pi\,\varphi_0^2}\, e^{\varphi_0\,J}\;;
\end{equation}
where $\varphi_0$ is a constant (something like $\varphi_0 =
\langle\Ai|\varphi|\Ai\rangle$ or $\varphi_0 =
\langle\Bi|\varphi|\Bi\rangle$, for each quantum phase).

Below you see plots of \eqref{eq:airyCS} for $\varphi_0 = \pm 1$: the
plots on the right side are such that the complex argument (phase) is
shown as color (hue) and the magnitude is show as brightness; the
plots on the left side are nothing but the wrapping of the former over
the Riemann Sphere (keeping in mind that its group of automorphisms is
given by $\mathsf{SL}(2,\mathbb{C})/\{\pm\mathds{1}\}$).

\begin{center}
  %
  \includegraphics[scale=0.77]{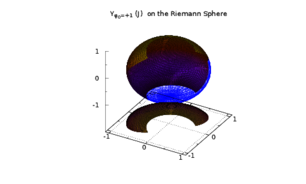}
  \raisebox{0.7cm}{\includegraphics[width=0.4\linewidth,keepaspectratio]{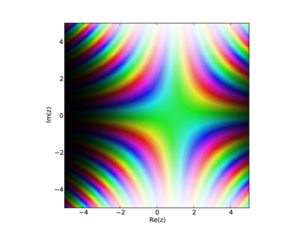}}
  
  \includegraphics[scale=0.77]{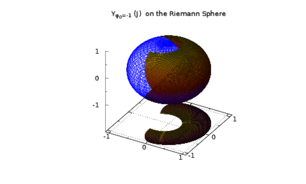}
  \raisebox{0.7cm}{\includegraphics[width=0.4\linewidth,keepaspectratio]{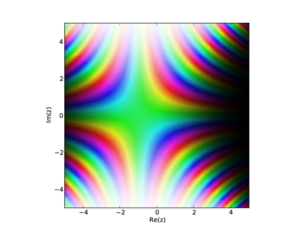}}
  %
  %

  \begin{minipage}[c]{0.85\linewidth}
    {\small\textsf{%
        \noindent Coherent State plots for each quantum phase. The
        plots on the right have $\Re(\Upsilon_{\varphi_0=\pm
          1}[J])$ as the $x$-axis and $\Im(\Upsilon_{\varphi_0=\pm
          1}[J])$ as the $y$-axis, with variations of color (phase
        $\theta$) and brightness (magnitude $\rho$) to indicate
        $z=\rho\,e^{i\,\theta}=\Upsilon_{\varphi_0=\pm 1}[J]$; plots
        on the left are simply their wrapping over the Riemann Sphere.
      }%
    }
  \end{minipage}
\end{center}
\subsection{Quartic Action}\label{subsec:quartic}
The quartic model is that of a $D0$-brane with $S[\varphi] = g\,
\varphi^2/2 + \varphi^4/4$, where $g = \mu^2/\lambda$ just for
convenience's sake, i.e., $g\rightarrow+\infty$ corresponds to the weak
coupling regime, while $g\rightarrow 0$ represents the strong
coupling one, and $g\rightarrow-\infty$ represents the ``broken
symmetric'' one. Accordingly, the Path Integral and Schwinger--Dyson equation
for this model are given by,

\begin{align}
  \label{eq:quarticFPI}
  \mathscr{Z}[J] &= \varint_{\mathcal{C}} e^{i\, (g\, \varphi^2/2 + \varphi^4/4)}\, 
    e^{i\, J\, \varphi}\,\mathcal{D}\varphi < \infty \Rightarrow\mathcal{C}:\, \mathrm{Arg}(\varphi) = \{0,\,\pm \pi/2\}\;;\\
  \label{eq:quarticSDE}
  \bigl(\partial^3_J - g\, &\partial_J - J\bigr)\, \mathscr{Z}[J] = 0\Rightarrow \mathscr{Z}[J] = a\,U[g, J] + b\,V[g, J] + c\, W[g, J]\;;
\end{align}
where the cycle $\mathcal{C}$ over which \eqref{eq:quarticFPI} is well defined is
such that the arguments of $\varphi$ are either $0$ or $\pm\pi/2$, and $U$, $V$
and $W$ are the Parabolic Cylinder Functions, that can also be cast in terms of
Fox's $H$-function as follows:

\begin{align*}
  y_1(g; \varphi) = \mathrm{PCyl}_{\text{even}}(g; \varphi)
  &= e^{-\varphi^2/4}\,\frac{\Gamma(\tfrac{1}{2})}{\Gamma(\tfrac{2\,g+1}{4})}\,
    \mathrm{H}^{1,1}_{1,2}\bigl((\tfrac{3-2\,g}{4},1); (0,\tfrac{1}{2},1); \tfrac{\varphi^2}{2}\bigr)\;;\\
  y_2(g; \varphi) = \mathrm{PCyl}_{\text{odd}}(g; \varphi)
  &= \varphi\,e^{-\varphi^2/4}\,\frac{\Gamma(\tfrac{3}{2})}{\Gamma(\tfrac{2\,g+3}{4})}\,
    \mathrm{H}^{1,1}_{1,2}\bigl((\tfrac{1-2\,g}{4},1); (0,-\tfrac{1}{2},1); \tfrac{\varphi^2}{2}\bigr)\;;\\
\end{align*}
where $U, V, W$ are particular combinations of
$\mathrm{PCyl}_{\text{even}}$ and $\mathrm{PCyl}_{\text{odd}}$ satisfying
the boundary conditions of \eqref{eq:quarticSDE} or, equivalently,
compatible with the integration cycle chosen in \eqref{eq:quarticFPI}
(in the sense of yielding a well-defined measure assuring the
convergence of the integral). Thus, the Path Integral of this model is given by
the integral representation of Fox's $H$-function, \eqref{eq:FoxH},
making it clear how the coupling constant $g$ determines the
singularity structure and integration cycle of the particular Path Integral in
question.

We can now appropriately parametrize \eqref{eq:quarticFPI} using
$\mathds{M}_{\text{PCyl}} = \bigl(\begin{smallmatrix} -g & -2\,\pi\\ \frac{1}{2\,\pi} & 0\end{smallmatrix}\bigr)$
and $\mathcal{O}[\varphi] = e^{i\, \varphi^4/4}$: in this way, the quantum
phases of our model are given by the Parabolic Cylinder Functions with respect
to the physical parameter $g$ --- varying $g$ will change the ground state
across Stokes' lines due to quantum fluctuations. As before,
$\mathds{M}_{\text{PCyl}}$ can be understood in terms of a composition of a
Gaussian Transform of width $\omega=g/2i\pi$ with a Fourier Transform
and with a Dilation of $\Delta=-2\,\pi$:
$\mathds{M}_{\text{PCyl}} = \mathds{M}_{\text{DT}}(\Delta=-2\,\pi)\,
\mathds{M}_{\text{FT}}\,\mathds{M}_{\text{GT}}(\omega=g/2\pi)$.

\begin{center}
  \begin{tikzpicture}[scale=2,cap=round]
    \colorlet{rotaxis}{orange!80!black}
    \colorlet{anglecolor}{green!50!black}
    \tikzstyle{axes}=[]
    \tikzstyle{important line}=[thick]
    \tikzstyle{information text}=[rounded corners,fill=red!10,inner sep=1ex]
    
    \draw[style=help lines,step=0.5cm] (-1.4,-1.4) grid (1.4,1.4);
    \draw[->,style=important line] (-1.5,0) -- (1.5,0) node[right] {$\boldsymbol\Re$};
    \draw[->,style=important line] (0,-1.5) -- (0,1.5) node[above] {$\boldsymbol\Im$};

    \draw (2.3,1) node[right,text width=9cm] { %
      The shaded regions represent convergence contours for the following
      terms, respectively: red $\Leftrightarrow$ quartic, blue $\Leftrightarrow$
      cubic, and yellow $\Leftrightarrow$ quadratic.
    };

    \draw[->,style=important line] (-0.3,1.45) node[above] {$\mathcal{C}_i$} .. controls (0,0) and (1,0.1) .. (1.47,0.09);
    \draw[->,style=important line] (-1.45,0.3) .. controls (0,0) and (-0.1,-1) .. (-0.09,-1.47) node[below] {$\mathcal{C}_j$};
    \draw[->,style=important line] (1.455,0.4) .. controls (0,0) and (0,0) .. (1.45,-0.3) node[right] {$\mathcal{C}_k$};

    \filldraw[fill=red!20,draw=red!50!black,opacity=0.4] (0,0) -- (1.293,0.535) arc(22.5:-22.5:1.4cm) -- (0,0);
    \filldraw[fill=red!20,draw=red!50!black,opacity=0.4] (0,0) -- (-1.293,0.535) arc(157.5:202.5:1.4cm) -- (0,0);
    \filldraw[fill=red!20,draw=red!50!black,opacity=0.4] (0,0) -- (0.535,1.293) arc(67.5:112.5:1.4cm) -- (0,0);
    \filldraw[fill=red!20,draw=red!50!black,opacity=0.4] (0,0) -- (-0.535,-1.293) arc(247.5:292.5:1.4cm) -- (0,0);

    \filldraw[fill=blue!20,draw=blue!50!black,opacity=0.45] (0,0) -- (1.126,0.65) arc(30:-30:1.3cm) -- (0,0);
    \filldraw[fill=blue!20,draw=blue!50!black,opacity=0.45] (0,0) -- (0,1.3) arc(90:150:1.3cm) -- (0,0);
    \filldraw[fill=blue!20,draw=blue!50!black,opacity=0.45] (0,0) -- (0.,-1.3) arc(-90:-150:1.3cm) -- (0,0);

    \filldraw[fill=yellow!20,draw=yellow!50!black,opacity=0.55] (0,0) -- (1.2,0) arc(0:90:1.2cm) -- (0,0);
    \filldraw[fill=yellow!20,draw=yellow!50!black,opacity=0.55] (0,0) -- (-1.2,0) arc(180:270:1.2cm) -- (0,0);
  \end{tikzpicture}
\end{center}

The coherent states for this model can also be computed, yielding,

\begin{equation}
  \label{eq:quarticCS}
  \Upsilon_{\varphi_0}^{g}[J] = \hksqrt{-i}\,
    \exp\Bigl\{i\,\tfrac{J^2}{\bigl(\hksqrt{g^2-4}-g\bigr)}\Bigr\}\,
    \exp\Bigl\{i\,\pi\,\varphi_0^2\,\tfrac{g+\hksqrt{g^2-4}}{g-\hksqrt{g^2-4}}\Bigr\}\,
    \exp\Bigl\{-2\,i\,\varphi_0\,\tfrac{J}{\bigl(\hksqrt{g^2-4}-g\bigr)}\Bigr\} \;;
\end{equation}
where $\varphi_0$ is a constant (à la $\varphi_0 = \langle
U|\varphi|U\rangle$, for each quantum phase: $U, V, W$), and $g$ is
the model's coupling constant.

Below you find plots of \eqref{eq:quarticCS} for $g = +2, 0, -2$ and
$\varphi_0 = +2\,i, 0, +2$, respectively. The plots on the right side
show the phase as color (hue) and the magnitude as brightness. The
plots on the left side are the wrapping of the former over the Riemann
Sphere (keeping in mind that its group of automorphisms is given by
$\mathsf{SL}(2,\mathbb{C})/\{\pm\mathds{1}\}$).

\begin{center}
   %
  \includegraphics[scale=0.77]{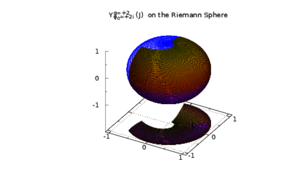}
  \raisebox{0.7cm}{\includegraphics[width=0.4\linewidth,keepaspectratio]{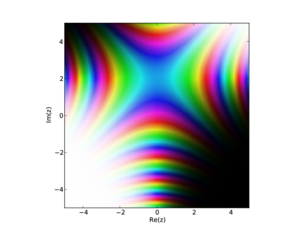}}
  
  \includegraphics[scale=0.77]{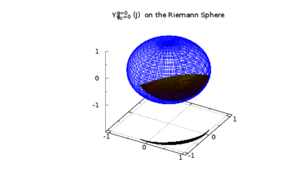}
  \raisebox{0.7cm}{\includegraphics[width=0.4\linewidth,keepaspectratio]{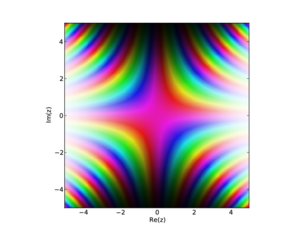}}
  
  \includegraphics[scale=0.77]{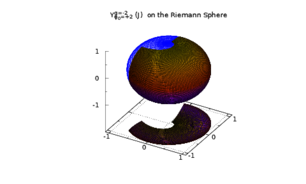}
  \raisebox{0.7cm}{\includegraphics[width=0.4\linewidth,keepaspectratio]{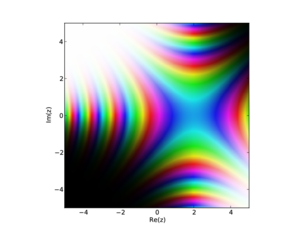}}
  %
  %
  %
  
  \begin{minipage}[c]{0.85\linewidth}
    {\small\textsf{%
        \noindent Coherent state plots for each quantum phase. The
        plots on the right have $\Re(\Upsilon_{\varphi_0=+2i,0,+2}^{g=+2,0,-2})$ as
        $x$-axis and $\Im(\Upsilon_{\varphi_0=+2i,0,+2}^{g=+2,0,-2})$ as 
        $y$-axis, with variations of color (phase $\theta$) and
        brightness (magnitude $\rho$) to indicate $z = \rho\,
        e^{i\,\theta} = \Upsilon_{\varphi_0=+2i,0,+2}^{g=+2,0,-2}$. Plots on the left
        are simply their wrapping over the Riemann Sphere. Note that
        the values for $g$ and $\varphi_0$ are obtained from their
        relation established via the Schwinger--Dyson equation: $\varphi=0$ or
        $\varphi = \pm\hksqrt{-2\, g}$, where we have chosen the positive branch
        of the square root.
      }%
    }
  \end{minipage}
\end{center}

It is very interesting to note that
$\Upsilon^{g=+2}_{\varphi_0=+2\,i}\mapsto
\Upsilon^{g=-2}_{\varphi_0=+2}$ through a rotation of $\pi/2$ ($J\mapsto i\, J$
and we also pick up a Stokes' factor, the multiplier $e^{-i\, 8\, \pi}$), which
can be interpreted in terms of the rolling of the Riemann Sphere
around the base manifold (which is just a point in this
case). Moreover, the fixed-point locus of our Action is simply given
by the set $\{0, \pm\hksqrt{-2\, g}\}$. Both of these together imply
that the Gromov--Witten invariant for this model is given by the sum
of the stationary phase approximation (viz. Atiyah--Bott fixed-point
theorem) of the Paht Integral computed at each point of the fixed-point set, for
different values of the coupling constant $g$: this brings the
asymptotic behavior of $U, V, W$ to the forefront, making the Stokes'
phenomena (wall-crossing) associated to this model quite relevant in
regards to the asymptotic behavior of each quantum phase.

The Stokes' Phenomena mentioned above is associated to understanding the
differential operator defining the Schwinger--Dyson equation \eqref{eq:quarticSDE},
$(\partial^3_J - g\,\partial_J - J)$, as a connection on the Riemann
Sphere, forming the basis of an interpretation in terms of the Riemann--Hilbert
correspondence, which  can be thought of as describing the Schwinger--Dyson equation
\eqref{eq:quarticSDE} in terms of the monodromies of its solutions. The
underlying geometry is sometimes called ``wild geometry'' (in analogy with ``wild
ramification'' in arithmetic): the [matrix of the] connection cannot be reduced to
[a matrix] having logarithmic poles, which is in turn related to the fact that
\eqref{eq:quarticSDE} is represented via Fox's $H$-function, and not some form
of polylogarithm (which is very common in models dealing with MHV amplitudes).

Both of the observations made in the two paragraphs above are intimately related
to the topic of nonlinear Fredholm theory, \cite{genfredholm}, in fact, with the
nonlinear Fredholm theory of the differential operator defining the
Schwinger--Dyson equation \eqref{eq:quarticSDE}, describing the deformations of
the given pseudoholomorphic curves (determined by the different solutions to
\eqref{eq:quarticSDE}). As such, as long as the Path Integral representation of each solution,
\eqref{eq:quarticFPI}, can be understood as a Fredholm map, these observations
are also related to the construction of Seiberg--Witten invariants.
\subsection{Non-Polynomial Actions}\label{subsec:nonpoly}

For the case of non-polynomial actions, we can perform the same construction as
for polynomial ones, just the integration cycle may be a bit more
intricate. For example, if $S = \beta\, \cos\theta$ and $\mathscr{Z}[J,\tilde{J}] =
\varint_{-\pi}^{\pi} e^{\beta\,\cos\theta}\, e^{J\, e^{i\,\theta}}\,
e^{\tilde{J}\, e^{-i\,\theta}}\, \mathcal{D}\theta$, the two relevant
integration cycles are $\mathcal{C}_1$ and $\mathcal{C}_2$, where
$\varint_{-\pi}^{\pi} = \varint_{\mathcal{C}_1} -
\varint_{\mathcal{C}_2}$, as depicted below,

\begin{center}
  \begin{tikzpicture}[scale=2,cap=round]
    \colorlet{rotaxis}{orange!80!black}
    \colorlet{anglecolor}{green!50!black}
    \tikzstyle{axes}=[]
    \tikzstyle{important line}=[thick]
    \tikzstyle{information text}=[rounded corners,fill=red!10,inner sep=1ex]
    
    \draw[style=help lines,step=0.5cm] (-1.4,-1.4) grid (1.4,1.4);
    \draw[->,style=important line] (-1.5,0) -- (1.5,0) node[right] {$\boldsymbol\Re$};
    \draw[->,style=important line] (0,-1.5) -- (0,1.5) node[above] {$\boldsymbol\Im$};

    \draw (-0.785,0) node[above] {$-\pi$};
    \draw (+0.785,0) node[above] {$+\pi$};

    \draw (0.8,1.2) node[right] {$\mathcal{C}_1$};
    \draw[->,style=very thick,orange!80!black] (-0.785,-1.3) -- (-0.785,0) -- (0.785,0) -- (0.785,1.3);

    \draw (-0.8,1.2) node[left] {$\mathcal{C}_2$};
    \draw[->,style=thick,orange!40!black] (-0.785,-1.4) -- (-0.785,1.4);
  \end{tikzpicture}
\end{center}
\section{Dimensional Extensions}\label{sec:ext}
Firstly, it is important to realize and clarify a few different facts:

\begin{description}
\item[Airy Property:] the so-called ``Airy property'', as defined by
  \cite{airyproperty}, needs to be satisfied if we want to dimensionally extend
  the models above;
\item[Topological Field Theory:] our $D0$-dimensional examples above can be
  understood as TFTs in the sense of Atiyah (the spacetime metric does not appear
  anywhere in these theories, and there is no real dynamics or propagation);
\item[$\boldsymbol{\Sigma}$-model analogy:] when we extend our field $\varphi$
  from being scalar-valued to being, e.g., $\mathds{M}$atrix-valued, we enter
  the realm of $\Sigma$-models, in the sense that the space of
  $\mathds{M}$atrices is considered the target space of our model --- the same
  argument holds true if rather than $\mathds{M}$atrix-valued we have
  $\mathds{T}$ensor-valued, $\mathscr{L}$ie or graded (SUSY) $\mathscr{L}$ie Algebra-valued fields, or even
  more generic target spaces;
\item[BFSS/IKKT analogy:] the $D0$-branes of our examples can be understood as being
  in the infinite momentum frame (light-cone frame) and, as such, as a random
  matrix model (extending the fields from scalar- to matrix-valued) where the
  $N\rightarrow\infty$ is conjectured to be equivalent to $M$-theory --- in this
  sense, matrices of finite $N$ can be thought of as a discretization scheme;
\item[Causal Dynamical Triangulation:] $\mathds{T}$ensor-valued fields are a
  generalization of $\mathds{M}$atrix models and correspond to dynamical
  triangulations, \cite{coloredtensor}, and, as such, the results previously
  obtained are relevant to the field of CDTs;
\item[Group Field Theory:] GFTs were also developed as a generalization of
  $\mathds{M}$atrix models, where the fields are valued in an appropriate
  $\mathscr{L}$ie Algebra (possibly graded, SUSY) --- once again, our results are thus relevant to this
  field as well;
\item[Boundary Conditions:] the boundary conditions of the Schwinger--Dyson equations
  (resp. integration cycles defining the Feynman Path Integrals) are required to
  not be affected in the process of taking the $N\rightarrow\infty$ limit,
  meaning that the quantum phases of the system should remain well-defined
  within their respective Stokes' wedges (topological $Dp$-brane). However, as
  $N\rightarrow\infty$, we have more and more points satisfying the boundary
  conditions, eventually yielding themselves a $Dp$-brane in their own
  right. Thus, we can revert our reasoning and say that such $Dp$-branes built
  out of the suitable boundary conditions are responsible for the quantization
  of our system, in analogy with \cite{branesquantization}.
\end{description}

Therefore, we have the following general picture. We start with a $D0$-brane model of fields
that can be either scalar-valued (e.g., $\mathbb{R}$, $\mathbb{C}$), yielding a
topological field theory, or valued in more general target spaces ($\mathds{V}$ector,
$\mathds{M}$atrix, $\mathds{T}$ensor, $\mathscr{L}$ie or graded $\mathscr{L}$ie algebra, etc), yielding a
topological $\Sigma$-model.

In any case, the quantization is done via a generalization of the Feynman Path Integral given in terms
of a Linear Canonical Transform, which is parametrized generically by a matrix $\mathds{M}
\in\mathsf{SL}(2,\mathbb{C})$ (keeping in mind that $\mathsf{SL}(2,\mathbb{C})$ is a
3-dim${}_{\mathbb{C}}$ manifold parametrizing our model). This parameter $\mathds{M}$
essentially labels the allowed integration cycles that render the Linear Canonical
Transform finite and well-defined. Furthermore, it is possible to combine different
cycles using the rule that $\mathds{M}_3 = \mathds{M}_2\, \mathds{M}_1$: if $\mathds{M}_1$
and $\mathds{M}_2$ are both labels of allowed integration cycles, so is
$\mathds{M}_3$, i.e.,

\begin{align}
  \nonumber
  \mathscr{Z}_{\mathds{M}_3 = \mathds{M}_2\,\mathds{M}_1} &= \mathscr{Z}_{\mathds{M}_2}\circ\,\mathscr{Z}_{\mathds{M}_1} \;;\\
  \label{eq:gluingrule}
  &= {-i}\, e^{i\pi\, J^2\, (d_2\,b_2^{-1} + d_1\,b_1^{-1})}\,\times\\
  \nonumber
  &\quad\times  \int_{\mathcal{C}_{\mathds{M}_{3}}=\mathcal{C}_{\mathds{M}_{2}}\circ\,\mathcal{C}_{\mathds{M}_{1}}}
    e^{-2i\pi\, J\varphi\,(b_2^{-1} + b_1^{-1})}\, e^{i\pi\,\varphi^2\, (a_2\,b_2^{-1} + a_1\,b_1^{-1})} \,
    e^{i\, S[\varphi]}\,\mathcal{D}\varphi \; .
\end{align}
This composition rule is analogous to the gluing of bordisms together, following Atiyah's
definition of a TFT, in the sense of \cite[Definition 1.1.1 and 1.1.5, Example 1.1.9]{lurieTFT}.
This is relevant for dimensional extensions for the reason that if we want to combine several
$D0$-branes into a $D1$-brane, certain relations among the boundary conditions of their respective
Schwinger--Dyson equations must be satisfied in order for this dimensional
extension to be ``stable'' in some suitable sense. These relations are more
easily encoded in terms of the algebra of the parameters $\mathds{M}$.

Thus, if we want to dimensionally extend our $0$-dimensional systems into bona fide $n$-dimensional ones,
we need to first pick an integration cycle, i.e., we need to pick an $\mathds{M}$ labeling the particular 
boundary conditions of the Schwinger--Dyson equations, determining which
solution we are about to extend. Then, we can proceed in two ways,

\begin{description}
\item[BFSS/IKKT inspired:] Interpreting the model as a $\Sigma$-model (where scalar-valued systems are understood 
  and extended to $\mathds{V}$ector-valued linear $\Sigma$-models), taking the large-$N$ limit and Lorentz 
  boosting it to a desired frame (i.e., away from the infinite momentum frame); \&
\item[Many-Body System inspired:] Constructing a discretized version of the
  Feynman Path Integral out of several copies of our system, extending the class
  of allowed paths to include not only Brownian paths, but also Lévy flights,
  and even more generic ones, using the full power of the Linear Canonical
  Transformation.
\end{description}
These two processes, in principle, should give us the same resulting dimensionally extended theory. Analogously,
we can revert the above arguments saying that
given a $D$-dimensional system, we can define its quantization through the implementation of the appropriate
boundary conditions of its respective Schwinger--Dyson equation
(resp. integration cycles of its Feynman Path Integral) in terms of suitable $Dp$-branes:
the solutions to the quantum equations of motion are only fully determined when we implement its compatible
boundary conditions, which is achieved through certain $Dp$-branes that implement them.

\section{Summary and Outlook}\label{sec:summary}
We have shown how to extend the notion of the Feynman Path Integral in a way to include more general paths than the original
formulation allows (e.g., Brownian paths, Lévy flights, etc). This extension comes labelled by a matrix
parameter that effectively determines the kind of path being used.

Furthermore, we have established the integro-differential nature of the quantization problem,
associating the boundary conditions of the Schwinger--Dyson equations to the
integration cycle of the Feynman Path Integral. In this fashion, the
matrix label previously mentioned has the job to perform the bookkeeping of all possible allowed solutions,
a fact that is reflected by the existence of more than one integration cycle guaranteeing the convergence
of the Feynman Path Integral.

A relevant property of Linear Canonical Transforms that will be important to us
is their eigenfunctions, forming a complete orthonormal set --- in analogy, for
instance, with the Fourier Transform's eigenfunctions,
\begin{align}
  \label{eq:fteigen}
  \Psi_n[\varphi] &= \frac{2^{1/4}}{\hksqrt{n!}}\, e^{-\pi\,\varphi^2}\,\He_n[2\,\hksqrt{\pi}\,\varphi] \; ;\\
  \intertext{where the Hermite polynomials are given by,}
  \label{eq:hermite}
  \He_n[\varphi] &= (-1)^n\, e^{\varphi^2/2}\, \frac{\mathrm{d}^n}{\mathrm{d}\varphi^n} e^{-\varphi^2/2}\;.
\end{align}
These will give us information about the coherent states associated to the
Linear Canonical Transform and each of its components (Fractional Fourier,
Hyperbolic, Bargmann, etc).

As a consequence, it is possible to stratify\footnote{A stratified space (or
  filtration) is one that has been decomposed into subspaces called strata, that
  are required to fit together in a certain way. Stratified spaces provide a setting
  for the study of singularities via Morse Theory on manifolds with boundary and
  manifolds with corners (e.g., orbifolds).}
the space of functions of $\varphi$ (with some suitable decaying conditions,
determined by the boundary conditions of our problem) according to the
eigenvalues of its Linear Canonical Transform. The choice of Hermite functions
in the Fourier Transform case, \eqref{eq:fteigen}, is convenient because of
their localization properties in $\varphi$ and $J$, thus allowing for a more
general transform to be introduced and used in field-source analysis (i.e.,
quantization) via Path Integrals or Schwinger--Dyson equations, namely the
Linear Canonical Transform.

For example, thinking of the Path Integral as a Sum over Paths (considering the
space of paths suitably stratified according to the problem at hand), the Path
Integral as a Fourier Transform implies these paths are Brownian; the Path
Integral as a Fractional Fourier Transform implies these paths are Lévy flights;
and so on. In this sense, we are trying to formalize the notion of a
``transform'' between two [fundamental] groupoids\footnote{A groupoid
  generalises the notion of a group in several equivalent ways, e.g., an
  oriented graph, or a category in which every morphism is invertible, etc. The
  fundamental group measures the 1-dimensional singularity structure of a
  space; for studying ``higher-dimensional singularities'', the homotopy groups
  are used. The fundamental groupoid, rather than singling out one point and
  considering the loops based at that point up to homotopy, considers all paths
  in the space (up to homotopy).},
the space of paths, $\varphi$, and the space of sources, $J$, where the Path
Integral plays the role of this transform.

On top of the explicit analysis of three different $D0$-brane examples, we also showed how to dimensionally
extend such models in two different ways. This enabled us to reverse the line of thought and think of the
quantization of a system as being determined by the $Dp$-branes that act as boundaries to it.

As for the connection with wall-crossing and crystal melting, we clearly stated the role that Stokes'
Phenomena plays in our models, showing how the asymptotic structure of each solution is relevant
in determining the allowed integration cycles of the appropriate Feynman Path Integral representing the system. Along
these lines, we used an intrinsic property of the Linear Canonical
Transformations, namely its symmetry with respect to
$\mathsf{SL}(2,\mathbb{C})$, plotting the relevant objects over the Riemann
Sphere, to make a connection with some dualities ($S$-duality, $T$-duality) and
to establish the modular character of the Feynman Path Integral representation
we used.

Finally, we have determined the appropriate representation of the Feynman Path Integral of each of our examples in
terms of Fox's $H$-function, which is a Mellin--Barnes transform. Thus, we brought to the foreground
discussions regarding the singularity structure of our models, showing that not all systems can be
described in terms of Polylogarithms (as happens in the case of MHV amplitudes): more complicated
systems require a more elaborate description, where the use of the $H$-function can be particularly
handy. This fact is associated with the nonlinear Fredholm theory of the
Schwinger--Dyson operator in question, which in turn has something to say about
the construction of Seiberg--Witten invariants in each of our systems.

%
%

To illustrate this point further, we will use some Cerf theory, which
is a way to study families of smooth functions on smooth manifolds, their
generic singularities, and the topology of the subspaces these singularities
define, as subspaces of the function space. Essentially, two Morse functions can
be approximated by one that is Morse at all but finitely many degenerate
times. The degeneracies involve a birth/death transition of critical
points. Cerf theory provides a framework for analyzing transitions between Morse
functions; it is the study of the positive co-dimensional strata of the
space of smooth functions.

For example, $f_t(x) = x^3 + t\, x$ is Morse for $t>0$ (no critical points) and
for $t<0$ (2 critical points), while $t=0$ represents a birth/death transition
of critical points, i.e., the attachment of a 1-handle connecting the manifold
$f_{t>0}$ to $f_{t<0}$. Something similar happens with $f_t(x) = x^4 + t\, x^2$,
which is Morse for $t>0$ (1 critical points) and for $t<0$ (3 critical points),
while $t=0$ indicates the birth/death transition of critical points, i.e., the
attachment of the handlebody connecting the manifolds $f_{t>0}$ and $f_{t<0}$.


The plots below highlight this construction for two relevant Potentials, the
cubic and quartic ones: $V_J(\phi) = \phi^3/3 - J\, \phi$, $V_g(\phi) = \phi^4 +
g\, \phi^2$. The birth/death transition, at $J=0$ and $g=0$ (marked as
``critical points'' in the plots), represents the addition of extra handles. We
can think of this in terms of constructing the Potential manifold via the
attachment of handlebodies. In physics parlance, we can give the fluxes that
determine the handlebody decomposition of a manifold (in particular, the vacuum
manifold, the moduli space): once the handlebodies determined by the appropriate
fluxes are known, the desired manifold is determined --- this is known as the
Heegaard diagram of a manifold. Using the help of the contour (equipotential)
lines on the plots, we can determine the set of flow lines outgoing or incoming
a certain critical point (birth/death transition), thus defining ascending
and descending manifolds (analogous to the ones computed in
\cite{complexwitten}). Finally, the intersection of these ascending and
descending manifolds marks the Stokes' lines determining the boundaries between
adjacent Stokes' wedges. These intersections, called ``handle slides'' , are the
$Dp$-branes establishing the desired bounary conditions for our model.


\begin{center}
  %
  \includegraphics[width=0.47\paperwidth,keepaspectratio]{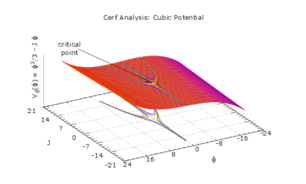}
  
  \centerline{ %
    \includegraphics[width=0.47\paperwidth,keepaspectratio]{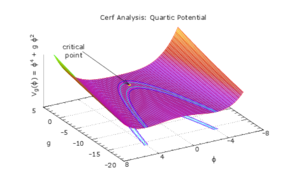} %
    \includegraphics[width=0.47\paperwidth,keepaspectratio]{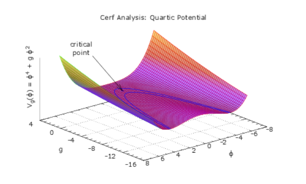}
  }
  
  \begin{minipage}[c]{0.85\linewidth}
    {\small\textsf{%
        \noindent Plots of $V_J(\phi) = \phi^3/3 - J\, \phi$ and $V_g(\phi) =
        \phi^4 + g\, \phi^2$ for ranges of the parameters $J$ and $g$, showing
        the ``critical point'' where the birth/death transition of critical
        points between two Morse functions happens, i.e., the point where the
        transition handlebody should be attached. The points where $J, g = 0$
        represent the transition between $J, g > 0$ and $J, g < 0$, where an
        extra monomial gets added to the top-coupling term.
      }%
    }
  \end{minipage}
\end{center}

 
At this point we can make some comments about similar approaches and
constructions available in the literature.

\begin{enumerate}
\item Multi-cut matrix models (which can be extended to $\mathds{T}$ensor models
  and thus Group Field Theory) have been a research topic for a couple of
  decades now, \cite{multicut}. However, their associated Stokes' phenomena, and
  their non-perturbative completion, are much more modern features. Their
  connection with our approach has only become clear in the past two years or
  so. Here is a rough summary of the approach, highlighting the common points.

  \begin{itemize}
  \item Riemann--Hilbert and Deligne--Simpson problem: given a system of
    differential equations on the Riemann Sphere, we want to reconstruct the system
    from its monodromy data, \cite{delignesimpson}, i.e., we want to solve the
    inverse monodromy problem in order to obtain the suitable Stokes'
    multipliers and factors. In our case, the system of
    differential equations is given by the Schwinger--Dyson equations, while its
    monodromy is established by the boundary conditions we choose, i.e., by the
    fall-off properties of the fields in our system. In the case of multi-cut
    boundary conditions, these turn out to be the quantum integrable
    $T$-systems; and in the case of non-critical string theory, this is
    described by the position of the cuts.
  \item Non-perturbative amplitude and corrections: the $D$-instanton chemical
    potentials are the missing information in [perturbative] string theory,
    \cite{multicut}. These are equivalent to the $\theta$-vacua in \cite{msols}
    which, in turn, are the $Dp$-branes establishing the appropriate boundary
    conditions rendering our Path Integral well defined. Further, these
    $\theta$-vacua correspond to the appropriate Stokes' multipliers and factors
    which, as seen above, are related to the Riemann--Hilbert and
    Deligne--Simpson problems via inverse monodromy, thus establishing the
    Stokes' phenomena of our system. This enables us to relate different
    perturbative vacua and study the string theory Landscape from first
    principles, where the relation between different string theory vacua in the
    Landscape happens via gluing the spectral curves non-perturbatively.
  \item Spectral Curves and perturbative string theory: the resolvent of the
    multi-cut matrix model defines the spectral curve, and thus the perturbative
    correlators, in the following way,
    \begin{align*}
      R(x) &= \biggl\langle\frac{1}{N}\,\tr\frac{1}{x-\mathds{M}}\biggr\rangle \;;\\
      \intertext{where,}
      \mathscr{Z} &= \int e^{-N\, \tr V(\mathds{M})}\, \mathrm{d}\mathds{M} \;;\\
      &= \int e^{-N\, \sum_i V(\lambda_i)}\, \prod_{i>j}(\lambda_i-\lambda_j)^2\,\mathrm{d}^N\lambda\;;\\
      \intertext{and the perturbative correlators are given by,}
      R_n(x_1,\dotsc,x_n) &= \biggl\langle\prod_{i=1}^{n}\frac{1}{N}\,\tr\frac{1}{x_i-\mathds{M}}\biggr\rangle\;;
    \end{align*}
    i.e., a $N$-body problem with potential $V$ (and eigenvalues
    $\lambda_i$). The resolvent operator allows us to determine the position of
    the eigenvalues and the eigenvalue density, \cite{multicut}. The
    non-perturbative corrections are given by the $D$-instantons, i.e., by the
    $\theta$-vacua parameters that we compute via the Stokes' factors of the
    model. The geometric meaning of these $\theta$-vacua is to determine the
    position of the ``eigenvalue cuts'', i.e., the determination of the Stokes'
    wedges.
  \item Quantum Integrability: Stokes' phenomena of the Schwinger--Dyson
    Equations satisfy the $T$-systems of quantum integrable models, linking the
    spectral analysis of differential equations and integrable models,
    establishing a Differential Equation/Integrable Model correspondence,
    \cite{multicut,odeim}.
  \end{itemize}

\item $PT$-symmetric Quantum Mechanics started as a proposal to broaden the
  usual formulation of Quantum Mechanics by relaxing the requirement of
  Hermiticity of the Hamiltonian, \cite{ptsymm}. Rather than using Hermiticity
  in order to guarantee that the Hamiltonian has a $\mathbb{R}$eal spectrum, it
  was required that the Hamiltonian satisfied the weaker condition of $PT$
  symmetry, i.e., that the Hamiltonian be invariant by parity reflection and
  time reversal. Finally, this generalization is usually studied via a
  $\mathbb{C}$omplex deformation of the harmonic oscillator, $H = p^2 + x^2\,
  (i\, x)^{\epsilon}$, where $\epsilon\in\mathbb{R}$, exhibiting two phases:
  when $\epsilon \geqslant 0$ and when $-1 < \epsilon < 0$; the phase transition
  occuring at $\epsilon = 0$.

  This scenario can be understood in terms of appropriate integration cycles for
  the Path Integral: effectively, the integration cycle is related to $\epsilon$
  in the following way. Let us choose $\epsilon=1$ to represent the
  $PT$-symmetric phase, and $\epsilon=-1$ for the broken-$PT$ one. When
  $\epsilon=0$ we have a simple harmonic oscillator (or, in QFT terms, a free
  theory). Now, we can associate a $\mathds{M}$atrix model to each one of these
  phases, the $PT$-symmetric one being given by $V_{\text{symm}}(\mathds{X}) =
  i\, \mathds{X}^3$, and the broken-$PT$ one being $V_{\text{bsymm}}(\mathds{X})
  = -i\, \mathds{X}^2\, \mathds{X}^{-1}$. Moreover, as long as the broken-$PT$
  model also has $\det \mathds{X} \neq 0$, we have $V_{\text{bsymm}}(\mathds{X})
  = -i\, \mathds{X}$. Thus, from their polynomial form alone we can already see
  that the same integration cycle will not guarantee the convergence of both
  models. In this sense, $PT$-symmetry can be associate with the suitable cycle
  guaranteeing the convergence of each respective Path Integral.

  Now, following the guidance of our examples, let us define a model without a definite
  $PT$-symmetric behavior, given by $V(\mathds{X}) = V_{\text{symm}}(\mathds{X})
  + V_{\text{bsymm}}(\mathds{X}) = i\, (u\, \mathds{X}^3 - v\, \mathds{X})$,
  where we have introduced the modular parameters $u, v\in\mathbb{C}$, and where
  the moduli space is given by $\mathcal{M}=\mathbb{CP}^1$, once only the
  ratio $u/v$ is relevant. This is in complete analogy to what was done in subsection
  \ref{subsec:airy}. As such, $u$ (i.e., the top coupling) controls the analytic
  structure of the theory, and as $u\rightarrow 0$, $v$ plays an ever more
  relevant role, and only an integration cycle contained in both Stokes' wedges
  would yield a convergent solution for both powers. As before, an integration
  cycle in a non-matching region of both Stokes' wedges yields a divergent
  $\mathscr{Z}$, and an integration cycle that can be deformed from one wedge to
  another yields a vanishing $\mathscr{Z}$, analogous to Lee--Yang zeros (see
  the plots in subsection \ref{subsec:airy}). In fact, we can recast this model as
  $V(\mathds{X}) = i\, g\, \mathds{X}^3$, where $g = -u/v\in\mathbb{CP}^1$, and the
  linear term, $i\, \mathds{X}$, is absorbed in the source term, $\mathds{J}\,
  \mathds{X}\mapsto \tilde{\mathds{J}}\,\mathds{X}$, where $\tilde{\mathds{J}} =
  \mathds{J} + \mathds{1}$. Thus, we end up with a complete parallel to the
  example of subsection \ref{subsec:airy}, scaled by a [power of the] factor
  $g$. Therefore, we can choose a particular linear combination of the cycles we
  obtained in subsection \ref{subsec:airy} in order to endow one of the
  solutions of this theory to be $PT$-symmetric, while the other solution will
  have broken-$PT$ symmetry. This will be reflected in the allowed ranges of the
  coupling $g$, just like the $\Ai$ and $\Bi$ solutions in subsection
  \ref{subsec:airy} depend on the integration cycles themselves. As such, we
  have a phase transtion at $g=0$, and symmetric and broken-symmetric phases for
  $\Re(g) > 0$ and $\Re(g) < 0$.

\item The $\mathbb{C}$omplexification of the Path Integral of Quantum Mechanics,
  the analytic continuation of Chern--Simons theory, as well as that of Liouville
  theory, have recently been shown to relate branes in a two-dimensional
  $A$-model with possibly new integration cycles for their respective Path
  Integrals, \cite{complexwitten}. The novelty introduced in these works is that
  a new integration cycle for the Path Integral is possible considering
  $\mathbb{C}$omplex-valued paths that are boundary values of pseudoholomorphic
  maps, giving a middle-dimensional cycle in the loop space of a symplectic
  manifold (classical phase space), which is one of the main ideas in Floer
  cohomology, \cite{complexwitten}.

  A pseudoholomorphic curve (or $J$-holomorphic curve, where $J$ is a
  suitable \emph{almost-$\mathbb{C}$omplex} structure, $J^2=1$) is a smooth map
  from a Riemann surface, $\Sigma$, into an almost-$\mathbb{C}$omplex manifold,
  satisfying the Cauchy--Riemann equations, \cite{donaldson}. Moreover, if
  $\Sigma$ is compact (e.g., Riemann Sphere, $\mathbb{CP}^1$), there is a
  nonlinear Fredholm theory describing the deformations of the given
  pseudoholomorphic curve, i.e., these deformations are parametrised by a
  finite-dimensional moduli space which, in turn, will be smoothly deformed by
  variations in $J$, \cite{donaldson}. This can be useful in a strategy where
  $J$-holomorphic curves are used as sigma-models, $\sigma :\,
  \Sigma\rightarrow (M,J)$, since the properties above can be used to determine
  the moduli space of the theory. Further, if $J$ is compatible with a
  symplectic structure on $M$ (think Phase Space), this provides us with an
  integrability condition extending this framework to a well-defined global
  theory (symplectic topology), \cite{donaldson}. In such a case, $I =
  \int_{\Sigma} \sigma$ is a topological invariant of $\sigma$, and controlled
  by topological data. In physics parlance, if $J$ is compatible with the
  Poisson bracket of our sigma-model's Phase Space (a fact that will depend on
  the boundary conditions of the Schwinger--Dyson equations of the model), the
  Action of the theory is a [homotopy] invariant. As such, $\sigma$ yields
  numerical invariants: the Gromov--Witten invariants, opening the door to the
  use of Floer homology, Fukaya category (mirror symmetry, Khovanov homology),
  and, in four dimensions, to Seiberg--Witten invariants, \cite{donaldson}.

  In \cite{complexwitten}, the $\sigma$-model mentioned above is a topologically
  twisted $A$-model, where the target space is the $\mathbb{C}$omplexification
  of the original classical Phase Space. The integration cycle is described by an
  $A$-brane called a coisotropic brane, establishing a relationship between the
  $A$-model and quantization. In this sense, the Path Integral localizes in the
  new integration cycle(s) of the $A$-model.

  This is not different from what we presented in this paper (particularly in
  the case of fields valued in a graded Lie algebra, so we can incorporate SUSY
  to our model and cancel eventual bosonic determinants appearing through the
  implementation of the boundary conditions via delta-functionals, on top of
  being able to use the superpotential in order to make things a bit simpler),
  provided we never cross the Stokes' lines delimiting a certain Stokes'
  wedge. Otherwise, we have to account for the Stokes' factor (hyperasymptotics)
  associated to this crossing: the bottom-line is the same as in
  \cite{complexwitten}, the single-valuedness of the integrand, i.e., the choice
  of the $n$\textsuperscript{th} root of the canonical bundle is an important
  part of the framework guaranteeing its global existence. Furthermore, the
  coherent states, $\Upsilon$, that we computed in the examples are Poincaré
  dual to their respective defining integration cycles.

  Finally, this is a good point to bring forward the similarities between the
  present paper, \cite{integraldiscriminants}, and \cite{complexwitten}. For
  example, \emph{integral discriminants} are an intrinsic property of the Action
  $S$, as shown below,
  \begin{equation}
    \label{eq:intdet}
    J_{n|r} = \int e^{-S_{i_1,\dotsc,i_r}\, \varphi_{i_1}\dotsm\varphi_{i_r}}\, \mathrm{d}^n\varphi \;;
  \end{equation}
  where the simplest cases are just powers of the algebraic discriminant, which
  determines only the singularities of $J_{n|r}$. The theory of integral
  discriminants is related to invariant theory, and can be viewed as one of the
  branches of \emph{nonlinear} algebra, \cite[and references therein]{integraldiscriminants}.

  The connections are not difficult to be seen: equation \eqref{eq:intdet} can
  be understood in terms of a $\mathds{V}\!$ector field, or we can think of the
  $\varphi_i$'s as $n$ matrices in a multi-$\mathds{M}$atrix model, or we can
  interpret the fields in terms of a $\mathds{T}$ensor model (Group Field Theory). It is very
  interesting to note that \cite{integraldiscriminants} expresses the integral
  discriminants in terms of generalized hypergeometric functions, which is in
  complete analogy with our use of Fox's $H$-function to describe the Path
  Integral, given that the $H$-function is a much broader generalization of
  hypergeometric functions, bringing the singularities of the theory to the
  foreground (in parallel to the relation between integral and algebraic
  discriminants).

  Moreover, \cite{integraldiscriminants} uses Ward identities in
  order to specify appropriate integration cycles for the integral discriminants
  at hand. This is the same role played by the Schwinger--Dyson equation(s) in
  our framework. It is interesting to see the use of a diagrammatic technique
  representing the tensor fields of our model, \cite{integraldiscriminants}:
  this gives us a graphical way to compute integral discriminants (Path
  Integrals), which we have not explored in this paper --- although it is
  possible to reconstruct these diagrams from the particular form of the
  $H$-function representing our theory.

  Further, by $\beta$-deforming our $\mathds{M}$atrix model representation
  (i.e., assuming $\beta\neq 1$ in the equation below), we
  can make connections with the Seiberg--Witten prepotentials,
  \begin{equation*}
    \mathscr{Z} = \int e^{-N\, \sum_i V(\lambda_i)}\, \prod_{i>j}(\lambda_i-\lambda_j)^{2\,\beta}\,\mathrm{d}^N\lambda\;;
  \end{equation*}
  where the Seiberg--Witten prepotential is given by $\log(\mathscr{Z})$, \cite{integraldiscriminants}.

  It is definitely a very positive reinforcement to see so many equivalences
  among such varied topics and approaches, and the framework we have constructed
  in this paper.

\item Complex Action Theory extends the position and momentum operators to be
  non-Hermitian, as well as their eigenvalues to be $\mathbb{C}$omplex, allowing
  a description of a theory with a $\mathbb{C}$omplex action or of a theory
  with a $\mathbb{R}$eal action and $\mathbb{C}$omplex saddle-points,
  \cite{holgerscat}. The basic idea is twofold:
  \begin{enumerate}
  \item Redefine the inner product of the quantum states in a physically
    reasonable way, such that the Hamiltonian's eigenstates are orthogonal with
    respect to it, i.e., the Hamiltonian has a well-defined Hermiticity under
    this new inner product --- let us call this inner product $I_Q$ and say the
    Hamiltonian is $Q$-Hermitian; \&
  \item Suppress the effect of the anti-$Q$-Hermitian part of the Hamiltonian after
    long times, thus the effect of the imaginary part of the $Q$-Hamiltonian (the
    anti-$Q$-Hermitian part) is smoothed out except for an irrelevant constant.
  \end{enumerate}

  Compared to our approach, this is the same as realizing that the integration
  cycle (i.e., the particular Stokes' wedge where the Path Integral is defined)
  determines the allowed values of the parameters of our theory (the coupling
  constants), and that gives us control over the asymptotic behavior of our
  theory: that is, ultimately the selected Stokes' wedge determines the
  parameters of the theory which, in turn, control the fall-off behavior of the
  fields in the theory and thus its asymptotic properties.

  In some sense, this is not very different from the $PT$-symmetric Quantum
  Mechanical case mentioned above, where the inner product is modified via a
  charge conjugation symmetry, $C$, such that $CPT$ is ultimately conserved in
  the theory. In this fashion, as soon as we choose a suitable integration cycle
  retaining $PT$ symmetry, it suffices to choose a compatible $I_C$ inner
  product. Thus, this charge conjugation symmetry, $C$, plays the role of the
  deformation $Q$ mentioned above, \cite{holgerscat}, defining a $C$-Hermitian
  property. This can be understood in terms of the extensions of operators, as
  done in Appendix \ref{sec:extop}, where the extension is given by $C$. In
  short, the properties of the particular chosen integration cycle select the
  parameters of the model which determine not only its asymptotic behavior but
  also the $C$-extension, clearly implying that different Stokes' wedges (i.e.,
  integration cycles) yield distinct sets of coherent states labelled
  $\mathds{1}_C = \int_{\mathcal{C}} |\Phi\rangle\langle\Phi|\,
  \mathcal{D}\Phi$, where $\mathcal{C}$ is the integration cycle associated to
  the charge conjugation symmetry, $C$.

  Complex Action Theory is then an extension of $PT$-symmetric Quantum Mechanics
  in the sense that one is free to pick any of the allowed integration cycles
  and define the charge conjugation symmetry, $C$, through it. Then, it suffices
  to define a compatible $PT$ symmetry, and build the theory from there. This
  enables the use of all Stokes' wedges, as opposed to the situation in
  $PT$-symmetric Quantum Mechanics.

\item A new framework for $\mathbb{C}$omplex-valued classical fields in the case
  of quantum field theories describing neutral particles has recently been
  proposed, generalizing Osterwalder and Schrader's construction of Euclidean
  fields via reflection positivity, studying scenarios that ensure and preserve
  it, \cite{jaffesneutral}.

  The relevant part of this framework is that the neutral fields can be either
  $\mathbb{R}$eal or $\mathbb{C}$omplex --- the usual distinction between
  neutral and charged fields does not correspond with the distinction between
  $\mathbb{R}$eal- and $\mathbb{C}$omplex-valued fields: $\mathbb{C}$omplex
  neutral fields allow for some novelty. As made quite clear in
  \cite{jaffesneutral}, the two-point function of the fields is the integral
  kernel of an operator, $D$, which need not be Hermitian. However, two requirements
  are made:
  \begin{enumerate}
  \item The Hermitian part of $D$ should have a strictly positive spectrum; \&
  \item The transformation $D$ should be reflection positive.
  \end{enumerate}

  Charged fields are also treated in an innovative fashion: distinct charged
  fields are introduced, \emph{not} related by $\mathbb{C}$omplex conjugation,
  where charge conjugation acts unitarily, $\mathbf{U}_C\, \Phi_{\pm}\,
  \mathbf{U}_C^{-1} = \Phi_{\mp}$.

  Reflection Positivity not only gives meaning to inverse Wick rotations, but is
  also used in the analytic continuation of group representations, in
  understanding the properties of the spectrum of the Transfer matrix, and in
  the theory of phase transitions, \cite{jaffesneutral}.

  The connection to our approach can be established when we are able to use the
  functional integral description of the above, which describes the decaying
  properties of the fields of the theory, i.e., their fall-off properties at the
  boundary. This is equivalent to setting boundary conditions to the
  Schwinger--Dyson equations appropriate to our model. In this sense, the
  problem of quantization is reduced to that of establishing suitable boundary
  conditions determining the desired decaying properties for the field content
  of our theory. This procedure can be accomplished in different ways, one of
  them being the construction of an appropriate Fourier Transform respecting the
  constraints of the problem, that is then understood as the partition function
  for the model (in an analogy with the Paley--Wiener theorem,
  \cite{complexsaddlepoint}).

  Essentially, this can be understood as an extension of $PT$-symmetric
  Euclidean QFT, where we can choose different combinations of $C$, $P$, and $T$
  to be held by the Hamiltonian, and the remaining symmetry ``corrects'' the
  inner product.  For example, we can choose a $CP$-symmetric Euclidean QFT
  whose inner product is corrected by $T$, etc. As such, this can be understood
  in terms of $J$-extensions of $J$-symmetric operators,
  \cite{reedsimon,devilsinvention,jextension}.

  In this sense, this discussion can be reduced to that of $PT$-symmetric
  Quantum Mechanics and of Complex Action Theory above.
\end{enumerate}

As a final remark, it is worth saying that we tried to understand the
process of quantization via the singularity structure and poles of the
path integral, $\mathscr{Z}_{\tau}^{\Sigma}[J]$, appropriately
decorated by the parameters of the theory, $\tau$, and suitably
compatible boundary conditions, $\Sigma$. In the language of
\cite{langlands}, we can loosely say that we tried to study the
functoriality of quantization via the singularity structure of the
path integral: facts on a topological setting (e.g.,
fall-off properties of the fields, asymptotic structure of the path
integral) can be translated to an algebraic setting (algebra of
observables, classification of the vacua of the theory). The
representation of the path integral as a Fox $H$-function encodes the
analytic structure of the theory in a Mellin--Barnes transform, which
can be understood as an appropriate $L$-function (viz. functoriality
via poles of $L$-functions), \cite{langlands}.

In particular, as suggested in \cite{langlands}, one can think of
Quantization $\subset$ Deformation: study the symmetries of a certain
equation of motion to be quantized and deform the Lie algebra
associated to each symmetry. These should be able to be expressed in
terms of their own path integral. Deformation and Functoriality then
can be seen as a reflection of the wave-particle duality,
respectively: functoriality $\sim$ particle $\sim$ Heisenberg picture,
deformation $\sim$ wave $\sim$ Schrödinger picture.

\section*{Acknowledgments}\label{sec:ack}
Parts of the results reported in this paper were presented at the \engordnumber{1}
Winter Workshop on Non-Perturbative Quantum Field Theory,
\href{http://wwnpqft.inln.cnrs.fr/}{WW-NPQFT} (January, 2010), and at
\href{http://cgc.physics.miami.edu/}{Miami 2010} (December, 2010) by
DF. DF wishes to thank the organizers for their hospitality, and the
participants for stimulating discussions. This work is supported in
part by funds provided by the US Department of Energy (\textsf{DOE})
under contract \textsf{DE-FG02-85ER40237}.

After this work was completed, the following references were kindly brought to our
attention, \cite{mithat, schiappa}. Also, \cite{jaffesneutral} has recently been
augmented by \cite{jaffespartial}.
\appendix
\section{Extensions of Operators}\label{sec:extop}

Another approach to the material developed above would be in terms of extensions
of symmetric operators, \cite{reedsimon}. Denoting the Schwinger--Dyson operator by
$\mathscr{O}_{\text{SD}}$, it is said to be symmetric if
\begin{equation*}
  \langle \mathscr{O}_{\text{SD}}\,\phi | \varphi\rangle = \langle\phi | \mathscr{O}_{\text{SD}}\,\varphi\rangle \; .
\end{equation*}

Its self-adjoint extensions can be understood in terms of its boundary values
(resp. $Dp$-branes), i.e., the monodromy of the
boundary conditions (resp. brane monodromy) pa\-ram\-e\-ter\-ize the possible
extensions. Analogously, we can study $\mathscr{O}_{\text{SD}}$ via its Cayley
transform, $U_{\mathscr{O}_{\text{SD}}} = (\mathscr{O}_{\text{SD}} -
i\,\mathds{1})\, (\mathscr{O}_{\text{SD}} + i\,\mathds{1})^{-1}$, where
$\mathds{1}$ is the unit operator: if $U_{\mathscr{O}_{\text{SD}}}$ is unitary
then $\mathscr{O}_{\text{SD}}$ is self-adjoint. This can be measured through the
deficiency indices of $\mathscr{O}_{\text{SD}}$, which are defined as the
dimension of the orthogonal complements of its domain and range: if both
deficiency indices are identical then $U_{\mathscr{O}_{\text{SD}}}$ is unitary.

The interesting thing to notice is that different self-adjoint extensions may
have different spectrum. That is, by varying the monodromy of the boundary
conditions (i.e., varying the brane monodromy) we obtain different operators
$\mathscr{O}_{\text{SD}}^{\prime}$, which may define different theories in terms
of their particle spectrum.

This can be further understood in terms of an extension of the resolution of the
identity, i.e., coherent state quantization à la \cite{delignesimpson},
\begin{equation*}
  \mathds{1} = \Biggl(\int_{\mathcal{C}_{1}}+\dotsb+\int_{\mathcal{C}_{n}}\Biggr)\, |\Phi\rangle\,\langle\Phi|
    = \frac{1}{\sum_{j=1}^{n} \alpha_j}\, \sum_{j=1}^{n} \alpha_j\, e^{i\, c_j}\; ;
\end{equation*}
where $\mathcal{C}_{m=1,\dotsc,n}$ labels the integration cycle associated to each monodromy
generator, and $e^{i\, c_{m=1,\dotsc,n}}$ represents the contribution from each cycle.

This can be further generalized by relaxing the condition that the extension be
self-adjoint, \cite{jextension}. Instead, we can talk in terms of
$J$-extensions, where $J$ is a \emph{conjugation}: $\langle J\,\phi, J\,\varphi\rangle =
\langle\phi,\varphi\rangle$. That is, a
$J$-extension of $\mathscr{O}_{\text{SD}}$ is such that $J\,
\mathscr{O}_{\text{SD}}\, J = \mathscr{O}_{\text{SD}}^{\dagger} \Leftrightarrow
\langle\mathscr{O}_{\text{SD}}\,\phi, J\,\varphi\rangle = \langle\phi,
J\,\mathscr{O}_{\text{SD}}\,\varphi\rangle$. For example, $\mathpzc{P\,T}$-symmetric quantum
mechanics falls in this case if we choose $J = \mathpzc{P\,T}$, where we have to modify the
inner product slightly in order for $J$ to remain a conjugation (so we can still
retain $\mathpzc{C\,P\,T}$ invariance, \cite{complexQM}).

This should not necessarily come as a surprise because of the following
observation. We can understand the boundary conditions for
$\mathscr{O}_{\text{SD}}$ via explicit boundary terms in the Action. These
explicit boundary terms are implemented via new fields that are subjected to the
actual boundary conditions we desire. The new fields are related to the
original ones: they are merely convenient objects used to express the constraints
necessary for the boundary conditions of $\mathscr{O}_{\text{SD}}$ to be
[explicitly] reflected in the Action. Finally, these boundary terms in the
Action are equivalent to $Dp$-branes described by these new fields: this is the
rationale behind brane quantization, \cite{branesquantization}.

Following this construction, the notion of $J$-extensions described above is
translated in terms of deformations of the integration cycle for the path integral,
done in such a way as to implement the particulars of the conjugation at hand. 

  {{}\symbolfootnote[0]{\href{http://www.het.brown.edu/people/danieldf/}{\copyright} \qquad
      \href{http://creativecommons.org/licenses/by-nc-sa/3.0/}{\ccbyncsa}}
  }


\begin{thebibliography}{99}
%
\bibitem{msols}
  S. Garcia, G.~S. Guralnik, Z. Guralnik,
  \emph{Theta Vacua and Boundary Conditions of the Schwinger Dyson Equations},
  ⟨\href{http://arxiv.org/abs/hep-th/9612079}{\texttt{arXiv:hep-th/9612079}}⟩.

  D.~D. Ferrante, G.~S. Guralnik,
  \emph{Mollifying Quantum Field Theory or Lattice QFT in Minkowski Spacetime and Symmetry Breaking},
  ⟨\href{http://arxiv.org/abs/hep-lat/0602013}{\texttt{arXiv:hep-lat/0602013}}⟩.

  G.~S. Guralnik, Z. Guralnik,
  \emph{Complexified Path Integrals and the Phases of Quantum Field Theory},
  ⟨\href{http://arxiv.org/abs/0710.1256}{\texttt{arXiv:0710.1256 [hep-th]}}⟩.

  D.~D. Ferrante, G.~S. Guralnik,
  \emph{Phase Transitions and Moduli Space Topology},
  ⟨\href{http://arxiv.org/abs/0809.2778}{\texttt{arXiv:0809.2778 [hep-th]}}⟩.

  D.~D. Ferrante,
  \emph{Symmetry Breaking: A New Paradigm for Non-Perturbative QFT and Topological Transitions},
  ⟨\href{http://arxiv.org/abs/0904.2205}{\texttt{arXiv:0904.2205 [hep-th]}}⟩.


\bibitem{frqm}
  Nikolai Laskin,
  \emph{Fractional quantum mechanics and Lévy path integrals},
  ⟨\href{http://dx.doi.org/10.1016/S0375-9601(00)00201-2}{\texttt{DOI: 10.1016/S0375-9601(00)00201-2}}⟩.

  Nick Laskin,
  \emph{Fractional Schrodinger equation},
  ⟨\href{http://arxiv.org/abs/quant-ph/0206098}{\texttt{arXiv:quant-ph/0206098}}⟩.

  Nick Laskin,
  \emph{Lévy Flights over Quantum Paths},
  ⟨\href{http://arxiv.org/abs/quant-ph/0504106}{\texttt{arXiv:quant-ph/0504106}}⟩.

  Nick Laskin,
  \emph{Fractional Quantum Mechanics},
  ⟨\href{http://arxiv.org/abs/0811.1769}{\texttt{arXiv:0811.1769 [math-ph]}}⟩.

  Nick Laskin,
  \emph{Principles of Fractional Quantum Mechanics},\\
  ⟨\href{http://arxiv.org/abs/1009.5533}{\texttt{arXiv:1009.5533 [math-ph]}}⟩.

  Francesco Mainardia, Gianni Pagninib, R.~K. Saxena,
  \emph{Fox $H$-functions in fractional diffusion},
  ⟨\href{http://dx.doi.org/10.1016/j.cam.2004.08.006}{\texttt{DOI: 10.1016/j.cam.2004.08.006}}⟩.

  Selcuk S. Bayin,
  \emph{Fox's $H$-Functions and the Time Fractional Schrödinger Equation},
  ⟨\href{http://arxiv.org/abs/1103.3295}{\texttt{arXiv:1103.3295 [math-ph]}}⟩.

  Yakubovich, S. B. and Luchko, Y. F.,
  \emph{The Hypergeometric Approach to Integral Transforms and Convolutions},
  ⟨\href{http://www.springer.com/mathematics/analysis/book/978-0-7923-2856-8}{\texttt{ISBN: 978-0-7923-2856-8}}⟩.


\bibitem{wolfLCT}
  Kurt Bernardo Wolf,
  \emph{Integral transforms in science and engineering},
  ⟨\href{http://books.google.com/books/about/Integral_transforms_in_science_and_engin.html?id=5Y2nAAAAIAAJ}{\texttt{ISBN: 0306392518, 9780306392511}}⟩.


\bibitem{Ifunction}
  N. S{\"u}dland, G. Baumann, T.~F. Nonnenmacher,
  \emph{Open Problem: Who Knows About the $\aleph$-Functions?},
  ⟨\href{http://www.math.bas.bg/~fcaa/volume1/pp401-402.gif}{Fract. Calc. Appl. Anal., Vol 1, No 4, (1998)}⟩.

  N. S{\"u}dland, G. Baumann, T.~F. Nonnenmacher,
  \emph{Fractional Driftless Fokker-Planck Equations with Power Law Diffusion Coefficients},
  ⟨\href{http://www14.in.tum.de/CASC2001/}{The 4th International Workshop on Computer Algebra in Scientific Computing (CASC, Konstanz, Germany, 2001)}⟩.

  Arjun K. Rathie,
  \emph{A new generalization of generalized hypergeometric functions},
  ⟨\href{http://www.dmi.unict.it/ojs/index.php/lematematiche/article/view/409}{Le Matematiche, Vol 52, No 2 (1997)}⟩.

  Ram K. Saxena, Tibor K. Pog{\'a}ny,
  \emph{Mathieu--type Series for the $\aleph$-function Occurring in Fokker-Planck Equation},
  ⟨\href{http://www.ejpam.com/index.php/ejpam/article/view/792/0}{EJPAM, Vol 3, No 6 (2010)}⟩.


\bibitem{foxHfunction}
  Lev A. Borisov, R. Paul Horja,
  \emph{Mellin--Barnes integrals as Fourier--Mukai transforms},
  ⟨\href{http://arxiv.org/abs/math/0510486}{\texttt{arXiv:math/0510486 [math.AG]}}⟩.

  A.~M. Mathai, R.~K. Saxena, H.~J. Haubold,
  \emph{The $H$-Function: Theory and Applications},
  ⟨\href{http://bit.ly/uiN3Fa}{\texttt{ISBN: 978-1-4419-0915-2}}⟩.


\bibitem{horava}
  Petr Hořava,
  \emph{Quantum Gravity at a Lifshitz Point},
  ⟨\href{http://arxiv.org/abs/0901.3775}{\texttt{arXiv:0901.3775 [hep-th]}}⟩.


\bibitem{reedsimon}
  M.~Reed, B.~Simon,
  \emph{Methods of Modern Mathematical Physics, Vol.~2: Fourier Analysis, Self-Adjointness},
  ⟨\href{http://books.google.com/books?id=Kz7s7bgVe8gC}{\texttt{ISBN: 978-0-1258-5002-5}}⟩.


\bibitem{devilsinvention}
  John P. Boyd,
  \emph{The Devil's Invention: Asymptotic, Superasymptotic and Hyperasymptotic Series},
  ⟨\href{http://dx.doi.org/10.1023/A:1006145903624}{\texttt{DOI: 10.1023/A:1006145903624}}⟩.


\bibitem{stabledistros}
  W. R. Schneider,
  \emph{Stable distributions: Fox function representation and generalization},
  ⟨\href{http://dx.doi.org/10.1007/3540171665_92}{\texttt{DOI: 10.1007/3540171665\_92}}⟩.

  Vladimir V. Uchaikin, V.~M. Zolotarev,
  \emph{Chance and stability: stable distributions and their applications},
  ⟨\href{http://books.google.com/books?id=ExN8DXXrNE8C}{\texttt{ISBN: 9067643017, 9789067643016}}⟩.

  Francesco Mainardi, Yuri Luchko, Gianni Pagnini,
  \emph{The fundamental solution of the space-time fractional diffusion equation},\\
  ⟨\href{http://arxiv.org/abs/cond-mat/0702419}{\texttt{arXiv:cond-mat/0702419 [cond-mat.stat-mech]}}⟩.


\bibitem{jextension}
  Ian Knowles,
  \emph{On $J$-Selfadjoint Extensions of $J$-Symmetric Operators},\\
    ⟨\href{http://dx.doi.org/10.1090/S0002-9939-1980-0560580-X}{\texttt{DOI: 10.1090/S0002-9939-1980-0560580-X}}⟩.

    A. Ibort,
    \emph{Three lectures on global boundary conditions and the theory of self-adjoint extensions of the covariant Laplace--Beltrami and Dirac operators on Riemannian manifolds with boundary},
    ⟨\href{http://arxiv.org/abs/1205.3579}{\texttt{arXiv:1205.3579 [math-ph]}}⟩.


\bibitem{delignesimpson}
  Vladimir Petrov Kostov,
  \emph{The Deligne--Simpson problem --- a survey},\\
  ⟨\href{http://dx.doi.org/10.1016/j.jalgebra.2004.07.013}{\texttt{DOI: 10.1016/j.jalgebra.2004.07.013}}⟩,
  ⟨\href{http://arxiv.org/abs/math/0206298}{\texttt{arXiv:math/0206298 [math.RA]}}⟩.

  Carlos Simpson,
  \emph{Higgs bundles and local systems},
  ⟨\href{http://www.numdam.org/item?id=PMIHES_1992__75__5_0}{\texttt{Publ. Mathématiques de l’IHÉS 75 (1992), p. 5-95}}⟩.

  Kevin J. Costello,
  \emph{Notes on supersymmetric and holomorphic field theories in dimensions 2 and 4},
  ⟨\href{http://arxiv.org/abs/1111.4234}{\texttt{arXiv:1111.4234 [math.QA]}}⟩.


\bibitem{stokeswall}
  Tom Bridgeland, Valerio Toledano-Laredo,
  \emph{Stability conditions and Stokes factors},
  ⟨\href{http://arxiv.org/abs/0801.3974}{\texttt{arXiv:0801.3974 [math.AG]}}⟩.


\bibitem{complexQM}
  Carl M. Bender, Dorje C. Brody, Hugh F. Jones,
  \emph{Complex Extension of Quantum Mechanics},
  ⟨\href{http://arxiv.org/abs/quant-ph/0208076}{\texttt{arXiv:quant-ph/0208076}}⟩.


\bibitem{wieneraskey}
  Dongbin Xiu, George Em Karniadakis,
  \emph{The Wiener--Askey Polynomial Chaos for Stochastic Differential Equations},
  ⟨\href{http://dx.doi.org/10.1137/S1064827501387826}{\texttt{DOI: 10.1137/S1064827501387826}}⟩.


\bibitem{cjt}
  David J. Toms,
  \emph{The Schwinger Action Principle and Effective Action}, \\
  ⟨\href{http://www.cambridge.org/gb/knowledge/isbn/item1174549/}{\texttt{ISBN: 9780521876766}}⟩.

  John M. Cornwall, R. Jackiw and E. Tomboulis,
  \emph{Effective action for composite operators},
  ⟨\href{http://link.aps.org/doi/10.1103/PhysRevD.10.2428}{\texttt{DOI: 10.1103/PhysRevD.10.2428}}⟩.


\bibitem{complexsaddlepoint}
  Frédéric Pham,
  \emph{Vanishing homologies and the $N$ variable saddlepoint method},
  ⟨\href{http://www.ams.org/mathscinet-getitem?mr=85d:32026}{\texttt{MR0713258 (85d:32026)}}⟩.

  P. Deift, X. Zhou,
  \emph{A Steepest Descent Method for Oscillatory Riemann--Hilbert Problems. Asymptotics for the MKdV Equation},
  ⟨\href{http://www.jstor.org/stable/2946540}{\texttt{http://www.jstor.org/stable/2946540}}⟩.

  Robin Pemantle, Marck~C. Wilson,
  \emph{Asymptotic Expansions of Oscillatory Integrals with Complex Phase},
  ⟨\href{http://www.math.upenn.edu/~pemantle/papers/oscillatory.pdf}{\texttt{Contemporary Mathematics Volume 520, 2010}}⟩.

  E. Delabaere and C.~J. Howls,
  \emph{Global asymptotics for multiple integrals with boundaries},
  ⟨\href{http://projecteuclid.org/euclid.dmj/1087575151}{\texttt{doi:10.1215/S0012-9074-02-11221-6}}⟩.

  S. Kamvissis, K.~T.-R. McLaughlin, P.~D. Miller,
  \emph{Semiclassical Soliton Ensembles for the Focusing Nonlinear Schroedinger Equation},
  ⟨\href{http://arxiv.org/abs/nlin/0012034}{\texttt{arXiv:nlin/0012034 [nlin.SI]}}⟩.

  Lars Hörmander,
  \emph{Linear Partial Differential Operators}, (cf. Chapter 7, e.g., ``Paley--Wiener theorem''),
  ⟨\href{http://www.ams.org/mathscinet-getitem?mr=0248435}{\texttt{MR0248435 (40 \#1687)}}⟩.


\bibitem{geometricquantization}
  A. Echeverria--Enriquez, M.C. Munoz--Lecanda, N. Roman--Roy, C. Victoria--Monge,
  \emph{Mathematical Foundations of Geometric Quantization},
  ⟨\href{http://arxiv.org/abs/math-ph/9904008}{\texttt{arXiv:math-ph/9904008}}⟩.

  M. Kontsevich,
  \emph{Deformation quantization of algebraic varieties},
  ⟨\href{http://arxiv.org/abs/math/0106006}{\texttt{arXiv:math/0106006 [math.AG]}}⟩.

  Alberto S. Cattaneo, Giovanni Felder,
  \emph{A path integral approach to the Kontsevich quantization formula},
  ⟨\href{http://arxiv.org/abs/math/9902090}{\texttt{arXiv:math/9902090 [math.QA]}}⟩.

  William Gordon Ritter,
  \emph{Geometric Quantization},
  ⟨\href{http://arxiv.org/abs/math-ph/0208008}{\texttt{arXiv:math-ph/0208008}}⟩.


\bibitem{branesquantization}
  Alexander Polishchuk,
  \emph{Abelian Varieties, Theta Functions and the Fourier Transform},
  ⟨\href{http://dx.doi.org/10.1017/CBO9780511546532}{\texttt{DOI: 10.1017/CBO9780511546532}}⟩.

  Anton Kapustin, Dmitri Orlov,
  \emph{Remarks on A-branes, Mirror Symmetry, and the Fukaya category},
  ⟨\href{http://arxiv.org/abs/hep-th/0109098}{\texttt{arXiv:hep-th/0109098}}⟩.

  P.~Bressler, Y.~Soibelman,
  \emph{Mirror Symmetry and Deformation Quantization}, \\
  ⟨\href{http://arxiv.org/abs/hep-th/0202128}{\texttt{arXiv:hep-th/0202128}}⟩.

  Yong-Geun Oh,
  \emph{Geometry of coisotropic submanifolds in symplectic and Kähler manifolds},
  ⟨\href{http://arxiv.org/abs/math/0310482}{\texttt{arXiv:math/0310482 [math.SG]}}⟩.

  Marco Aldi, Eric Zaslow,
  \emph{Coisotropic Branes, Noncommutativity, and the Mirror Correspondence},
  ⟨\href{http://arxiv.org/abs/hep-th/0501247}{\texttt{arXiv:hep-th/0501247}}⟩.

  M.~Gualtieri,
  \emph{Branes on Poisson Varieties},
  ⟨\href{http://arxiv.org/abs/0710.2719}{\texttt{arXiv:0710.2719 [math.DG]}}⟩.

  Sergei Gukov, Edward Witten,
  \emph{Branes and Quantization},
  ⟨\href{http://arxiv.org/abs/0809.0305}{\texttt{arXiv:0809.0305 [hep-th]}}⟩.

  Sergei Gukov,
  \emph{Quantization via Mirror Symmetry},
  ⟨\href{http://arxiv.org/abs/1011.2218}{\texttt{arXiv:1011.2218 [hep-th]}}⟩.


\bibitem{integraldiscriminants}
  A.~Morozov, Sh.~Shakirov,
  \emph{Introduction to Integral Discriminants},
  ⟨\href{http://arxiv.org/abs/0903.2595}{\texttt{arXiv:0903.2595 [math-ph]}}⟩.

  Kazuyuki Fujii,
  \emph{Beyond Gaussian: A Comment},
  ⟨\href{http://arxiv.org/abs/0905.1363}{\texttt{arXiv:0905.1363 [math-ph]}}⟩.

  Kazuyuki Fujii, Hiroshi Oike,
  \emph{Beyond the Gaussian II: A Mathematical Experiment},
  ⟨\href{http://arxiv.org/abs/1103.4428}{\texttt{arXiv:1103.4428 [math-ph]}}⟩.

  A.~Mironov, A.~Morozov, A.~Popolitov, Sh.~Shakirov,
  \emph{Resolvents and Seiberg-Witten representation for Gaussian $\beta$-ensemble},
  ⟨\href{http://arxiv.org/abs/1103.5470}{\texttt{arXiv:1103.5470 [hep-th]}}⟩.

  Zouhair Mouayn,
  \emph{An index ${}_2F_2$ hypergeometric transform}, \\
  ⟨\href{http://arxiv.org/abs/1105.2164}{\texttt{arXiv:1105.2164 [math-ph]}}⟩.

  Yurii Neretin,
  \emph{Index hypergeometric transform and imitation of analysis of Berezin kernels on hyperbolic spaces},
  ⟨\href{http://arxiv.org/abs/math/0104035}{\texttt{arXiv:math/0104035v1 [math.CA]}}⟩.

  Samuel Friot, David Greynat,
  \emph{Non-Perturbative Asymptotic Improvement of Perturbation Theory and Mellin--Barnes Representation},
  ⟨\href{http://arxiv.org/abs/0907.5593}{\texttt{arXiv:0907.5593v3 [hep-th]}}⟩.

  Samuel Friot, David Greynat,
  \emph{On convergent series representations of Mellin--Barnes integrals},
  ⟨\href{http://arxiv.org/abs/1107.0328}{\texttt{arXiv:1107.0328 [math-ph]}}⟩.


\bibitem{wigner}
  R. F. O'Connell
  \emph{The Wigner Distribution},
  ⟨\href{http://arxiv.org/abs/1009.4431}{\texttt{arXiv:1009.4431 [quant-ph]}}⟩.


\bibitem{conformalblocks}
  Miranda C.~N. Cheng, Robbert Dijkgraaf, Cumrun Vafa,
  \emph{Non-Perturbative Topological Strings And Conformal Blocks},
  ⟨\href{http://arxiv.org/abs/1010.4573}{\texttt{arXiv:1010.4573 [hep-th]}}⟩.


\bibitem{brunoemail}
  Masahito Yamazaki,
  \emph{Crystal Melting and Wall Crossing Phenomena},
  ⟨\href{http://arxiv.org/abs/1002.1709}{\texttt{arXiv:1002.1709 [hep-th]}}⟩.

  Richard~J. Szabo,
  \emph{Crystals, instantons and quantum toric geometry},
  ⟨\href{http://arxiv.org/abs/1102.3861}{\texttt{arXiv:1102.3861 [hep-th]}}⟩.

  Paolo Aluffi, Matilde Marcolli,
  \emph{A motivic approach to phase transitions in Potts models},
  ⟨\href{http://arxiv.org/abs/1102.3462}{\texttt{arXiv:1102.3462 [math-ph]}}⟩.

  Frederik Denef,
  \emph{TASI lectures on complex structures},
  ⟨\href{http://arxiv.org/abs/1104.0254}{\texttt{arXiv:1104.0254 [hep-th]}}⟩.

  L.~Vitagliano,
  \emph{Hamilton-Jacobi Diffieties},
  ⟨\href{http://arxiv.org/abs/1104.0162}{\texttt{arXiv:1104.0162 [math.DG]}}⟩.

  Wolfgang Arendt, Markus Biegert, A.~F.~M.~ter Elst,
  \emph{Diffusion determines the manifold},\\
  ⟨\href{http://arxiv.org/abs/0806.0437}{\texttt{arXiv:0806.0437 [math.AP]}}⟩.

  W. Arendt, A.~F.~M.~ter Elst,
  \emph{Diffusion determines the compact manifold},
  ⟨\href{http://arxiv.org/abs/1104.1012}{\texttt{arXiv:1104.1012 [math.AP]}}⟩.

  Wolfgang Arendt, A.~F.~M.~ter Elst,
  \emph{From forms to semigroups},\\
  ⟨\href{http://arxiv.org/abs/1104.1013}{\texttt{arXiv:1104.1013 [math.AP]}}⟩.

  Pei Hsu,
  \emph{Heat Semigroup on a Complete Riemannian Manifold},\\
  ⟨\href{http://projecteuclid.org/euclid.aop/1176991267}{\texttt{DOI:10.1214/aop/1176991267}}⟩.


\bibitem{coloredtensor}
  Dario Benedetti, Razvan Gurau,
  \emph{Phase Transition in Dually Weighted Colored Tensor Models},
  ⟨\href{http://arxiv.org/abs/1108.5389}{\texttt{arXiv:1108.5389v1 [hep-th]}}⟩.


\bibitem{genfredholm}
  Helmut H. Hofer,
  \emph{A General Fredholm Theory and Applications},
  ⟨\href{http://arxiv.org/abs/math/0509366}{\texttt{arXiv:math/0509366 [math.SG]}}⟩.

  Helmut Hofer, Kris Wysocki, Eduard Zehnder,
  \emph{A General Fredholm Theory I: A Splicing-Based Differential Geometry},
  ⟨\href{http://arxiv.org/abs/math/0612604}{\texttt{arXiv:math/0612604 [math.FA]}}⟩.

  Helmut Hofer, Kris Wysocki, Eduard Zehnder,
  \emph{A General Fredholm Theory II: Implicit Function Theorems},
  ⟨\href{http://arxiv.org/abs/0705.1310}{\texttt{arXiv:0705.1310 [math.FA]}}⟩.

  Helmut Hofer, Kris Wysocki, Eduard Zehnder,
  \emph{A General Fredholm Theory III: Fredholm Functors and Polyfolds},
  ⟨\href{http://arxiv.org/abs/0810.0736}{\texttt{arXiv:0810.0736 [math.FA]}}⟩.

  Helmut Hofer, Kris Wysocki, Eduard Zehnder,
  \emph{Integration Theory for Zero Sets of Polyfold Fredholm Sections},
  ⟨\href{http://arxiv.org/abs/0711.0781}{\texttt{arXiv:0711.0781 [math.FA]}}⟩.

  Helmut Hofer,
  \emph{Polyfolds And A General Fredholm Theory}, \\
  ⟨\href{http://arxiv.org/abs/0809.3753}{\texttt{arXiv:0809.3753 [math.SG]}}⟩.

  Barney Bramham, Helmut Hofer,
  \emph{First Steps Towards a Symplectic Dynamics},\\
  ⟨\href{http://arxiv.org/abs/1102.3723}{\texttt{arXiv:1102.3723 [math.DS]}}⟩.

  Helmut Hofer, Kris Wysocki, Eduard Zehnder,
  \emph{Applications of Polyfold Theory I: The Polyfolds of Gromov--Witten Theory},
  ⟨\href{http://arxiv.org/abs/1107.2097}{\texttt{arXiv:1107.2097 [math.SG]}}⟩.


\bibitem{arnolds}
  V.~I. Arnold,
  \emph{Modes and Quasimodes},
  ⟨\href{http://dx.doi.org/10.1007/BF01077511}{\texttt{DOI: 10.1007/BF01077511}}⟩.

  V.~I. Arnold,
  \emph{Remarks on eigenvalues and eigenvectors of Hermitian matrices, Berry phase, adiabatic connections and quantum Hall effect},
  ⟨\href{http://dx.doi.org/10.1007/BF01614072}{\texttt{DOI: 10.1007/BF01614072}}⟩.

  B.~A. Khesin,
  \emph{Informal Complexification and Poisson Structures on Moduli Spaces},
  ⟨\href{http://citeseerx.ist.psu.edu/viewdoc/summary?doi=10.1.1.54.5034}{\texttt{ISBN: 0821808079, 9780821808078}}⟩.

  A. Agrachev,
  \emph{On the Space of Symmetric Operators with Multiple Ground States},
  ⟨\href{http://arxiv.org/abs/1107.3010}{\texttt{arXiv:1107.3010 [math.AT]}}⟩.


\bibitem{symplecticfieldtheory}
  Yakov Eliashberg, Alexander Givental, Helmut Hofer,
  \emph{Introduction to Symplectic Field Theory},
  ⟨\href{http://arxiv.org/abs/math/0010059}{\texttt{arXiv:math/0010059 [math.SG]}}⟩.

  Yakov Eliashberg,
  \emph{Symplectic field theory and its applications},
  ⟨\href{http://www.icm2006.org/proceedings/Vol_I/14.pdf}{\texttt{Proceedings of
      the International Congress of Mathematicians: Madrid, August 22--30,
      2006, Vol. 1, ISBN 978-3-03719-022-7, 217--246}}⟩.


\bibitem{AtiyahJeffrey}
  M. F. Atiyah and L. C. Jeffrey,
  \emph{Topological Lagrangians and cohomology},\\
  ⟨\href{http://dx.doi.org/10.1016/0393-0440(90)90023-V}{\texttt{DOI:10.1016/0393-0440(90)90023-V}}⟩.

  Andriy Haydys,
  \emph{Fukaya--Seidel category and gauge theory},
  ⟨\href{http://arxiv.org/abs/1010.2353}{\texttt{arXiv:1010.2353 [math.SG]}}⟩.


\bibitem{cosmicclustering}
  Dionysios Anninos, Frederik Denef,
  \emph{Cosmic Clustering},
  ⟨\href{http://arxiv.org/abs/1111.6061}{\texttt{arXiv:1111.6061 [hep-th]}}⟩.


\bibitem{abcdefg}
  Christoph A.~Keller, Noppadol Mekareeya, Jaewon Song, Yuji Tachikawa,
  \emph{The ABCDEFG of Instantons and W-algebras},
  ⟨\href{http://arxiv.org/abs/1111.5624}{\texttt{arXiv:1111.5624 [hep-th]}}⟩.


\bibitem{alexandrov}
  A. Alexandrov,
  \emph{From Hurwitz numbers to Kontsevich--Witten tau-function: a connection by Virasoro operators},
  ⟨\href{http://arxiv.org/abs/1111.5349}{\texttt{arXiv:1111.5349 [hep-th]}}⟩.


\bibitem{airyproperty}
  Maxim Kontsevich,
  \emph{Intersection theory on the moduli space of curves and the matrix Airy function},
  ⟨\href{http://projecteuclid.org/euclid.cmp/1104250524}{\texttt{euclid.cmp/1104250524}}⟩.

  R, N. Fernandez, V. S. Varadarajan,
  \emph{Matrix Airy Functions for Compact Lie Groups},
  ⟨\href{http://dx.doi.org/10.1142/S0129167X09005595}{\texttt{DOI:10.1142/S0129167X09005595}}⟩.

  R. N. Fernandez, V. S. Varadarajan, D. Weisbart,
  \emph{Airy functions over local fields},
  ⟨\href{http://dx.doi.org/10.1007/s11005-009-0311-x}{\texttt{DOI:10.1007/s11005-009-0311-x}}⟩.


\bibitem{lurieTFT}
  Jacob Lurie,
  \emph{On the Classification of Topological Field Theories},
  ⟨\href{http://arxiv.org/abs/0905.0465}{\texttt{arXiv:0905.0465 [math.CT]}}⟩.


\bibitem{multicut}
  Bertrand Eynard, Nicolas Orantin,
  \emph{Algebraic methods in random matrices and enumerative geometry},
  ⟨\href{http://arxiv.org/abs/0811.3531}{\texttt{arXiv:0811.3531 [math-ph]}}⟩.

  C. Crnković, M. Douglas, G. Moore,
  \emph{Loop Equations and the Topological Phase of Multi-Cut Matrix Models},
  ⟨\href{http://arxiv.org/abs/hep-th/9108014}{\texttt{arXiv:hep-th/9108014}}⟩.

  Juan Maldacena, Gregory Moore, Nathan Seiberg, David Shih,
  \emph{Exact vs. Semiclassical Target Space of the Minimal String},
  ⟨\href{http://arxiv.org/abs/hep-th/0408039}{\texttt{arXiv:hep-th/0408039}}⟩.

  Chuan-Tsung Chan, Hirotaka Irie, Sheng-Yu Darren Shih, Chi-Hsien Yeh,
  \emph{Macroscopic loop amplitudes in the multi-cut two-matrix models},
  ⟨\href{http://arxiv.org/abs/0909.1197}{\texttt{arXiv:0909.1197 [hep-th]}}⟩.

  Chuan-Tsung Chan, Hirotaka Irie, Chi-Hsien Yeh,
  \emph{Stokes Phenomena and Quantum Integrability in Non-critical String/M Theory},
  ⟨\href{http://arxiv.org/abs/1109.2598}{\texttt{arXiv:1109.2598 [hep-th]}}⟩.

  Čedomir Crnković, Gregory Moore,
  \emph{Multicritical multi-cut matrix models},
  ⟨\href{http://adsabs.harvard.edu/abs/1991PhLB..257..322C}{\texttt{DOI: 10.1016/0370-2693(91)91900-G}}⟩.


\bibitem{odeim}
  Patrick Dorey, Clare Dunning, Roberto Tateo,
  \emph{Ordinary Differential Equations and Integrable Models},
  ⟨\href{http://arxiv.org/abs/hep-th/0010148}{\texttt{arXiv:hep-th/0010148}}⟩.

  P. Dorey, C. Dunning, A. Millican-Slater, R. Tateo,
  \emph{Differential equations and the Bethe ansatz},
  ⟨\href{http://arxiv.org/abs/hep-th/0309054}{\texttt{arXiv:hep-th/0309054}}⟩.

  Patrick Dorey, Clare Dunning, Roberto Tateo,
  \emph{Aspects of the ODE/IM correspondence},
  ⟨\href{http://arxiv.org/abs/hep-th/0411069}{\texttt{arXiv:hep-th/0411069}}⟩.

  Patrick Dorey, Clare Dunning, Davide Masoero, Junji Suzuki, Roberto Tateo,
  \emph{Pseudo-differential equations, and the Bethe Ansatz for the classical Lie algebras},\\
  ⟨\href{http://arxiv.org/abs/hep-th/0612298}{\texttt{arXiv:hep-th/0612298}}⟩.


\bibitem{ptsymm}
  Carl Bender, Stefan Boettcher, Peter Meisinger,
  \emph{PT-Symmetric Quantum Mechanics},
  ⟨\href{http://arxiv.org/abs/quant-ph/9809072}{\texttt{arXiv:quant-ph/9809072}}⟩.


\bibitem{complexwitten}
  Edward Witten,
  \emph{Analytic Continuation Of Chern--Simons Theory},
  ⟨\href{http://arxiv.org/abs/1001.2933}{\texttt{arXiv:1001.2933 [hep-th]}}⟩.

  Edward Witten,
  \emph{A New Look At The Path Integral Of Quantum Mechanics},
  ⟨\href{http://arxiv.org/abs/1009.6032}{\texttt{arXiv: 1009.6032 [hep-th]}}⟩.

  Edward Witten,
  \emph{Analytic Continuation of Liouville Theory},
  ⟨\href{http://arxiv.org/abs/1108.4417}{\texttt{arXiv: 1108.4417 [hep-th]}}⟩.


\bibitem{holgerscat}
  Keiichi Nagao, Holger Bech Nielsen,
  \emph{Formulation of Complex Action Theory},
  ⟨\href{http://arxiv.org/abs/1104.3381}{\texttt{arXiv: 1104.3381 [quant\--ph]}}⟩.

  Keiichi Nagao, Holger Bech Nielsen,
  \emph{Automatic Hermiticity},
  ⟨\href{http://arxiv.org/abs/1009.0441}{\texttt{arXiv: 1009.0441 [quant-ph]}}⟩.


\bibitem{jaffesneutral}
  Arthur Jaffe, Christian D. Jäkel, Roberto E. Marinez II,
  \emph{Complex Classical Fields: A Framework for Reflection Positivity},
  ⟨\href{http://arxiv.org/abs/1201.6003}{\texttt{arXiv:1201.6003 [math-ph]}}⟩.

\bibitem{jaffespartial}
  Arthur Jaffe, Christian D. Jäkel, Roberto E. Marinez II,
  \emph{Complex Classical Fields and Partial Wick Rotations},
  ⟨\href{http://arxiv.org/abs/1302.5935}{\texttt{arXiv:1302.5935 [math-ph]}}⟩.


\bibitem{donaldson}
  Simon K.~Donaldson,
  \emph{What Is\ldots\/ a Pseudoholomorphic Curve?},
  ⟨\href{http://www.ams.org/notices/200509/what-is.pdf}{Notices of the American Mathematical Society, \textbf{52} (9), pp.1026--1027}⟩.


\bibitem{langlands}
  Robert P.~Langlands,
  \emph{Beyond endoscopy},
  ⟨\href{http://publications.ias.edu/rpl/paper/91}{\texttt{Shalika volume, Johns Hopkins U\-ni\-ver\-si\-ty Press. (2004)}}⟩.

  P.~Edward Herman,
  \emph{The Functional Equation and Beyond Endoscopy},
  ⟨\href{http://arxiv.org/abs/1208.5783}{\texttt{arXiv:1208.5783 [math.NT]}}⟩.

  Robert P.~Langlands,
  \emph{Functoriality and Reciprocity},
  ⟨\href{http://publications.ias.edu/sites/default/files/functoriality.pdf}{\texttt{Lectures at IAS, March, 2011}}⟩.

  Daniel Sternheimer,
  \emph{Quantization: Deformation and/or Functor?},
  ⟨\href{http://dx.doi.org/10.1007/s11005-005-0028-4}{\texttt{DOI: 10.1007/ s11005-005-0028-4}}⟩.

  N.~P.~Landsman,
  \emph{Quantization as a functor},
  ⟨\href{http://arxiv.org/abs/math-ph/0107023}{\texttt{arXiv:math-ph/0107023}}⟩.


\bibitem{mithat}
  Gerald V.~Dunne, Mithat Unsal,
  \emph{Resurgence and Trans-series in Quantum Field Theory: The $\mathbb{CP}^{N-1}$ Model},
  ⟨\href{http://arxiv.org/abs/1210.2423}{arXiv:1210.2423 [hep-th]}⟩.

  Philip C.~Argyres, Mithat Ünsal,
  \emph{The semi-classical expansion and resurgence in gauge theories: new perturbative, instanton, bion, and renormalon effects},
  ⟨\href{http://arxiv.org/abs/1206.1890}{arXiv:1206.1890 [hep-th]}⟩.

  Gerald V.~Dunne, Mithat Unsal,
  \emph{Continuity and Resurgence: towards a continuum definition of the $\mathbb{CP}^{N-1}$ model},
  ⟨\href{http://arxiv.org/abs/1210.3646}{arXiv:1210.3646 [hep-th]}⟩.

  Philip Argyres, Mithat Unsal,
  \emph{A semiclassical realization of infrared renormalons},
  ⟨\href{http://arxiv.org/abs/1204.1661}{arXiv:1204.1661 [hep-th]}⟩.


\bibitem{schiappa}
  Ricardo Schiappa, Ricardo Vaz,
  \emph{The Resurgence of Instantons: Multi-Cuts Stokes Phases and the Painleve II Equation},
  ⟨\href{http://arxiv.org/abs/1302.5138}{arXiv:1302.5138 [hep-th]}⟩.

  Inês Aniceto, Ricardo Schiappa, Marcel Vonk,
  \emph{The Resurgence of Instantons in String Theory},
  ⟨\href{http://arxiv.org/abs/1106.5922}{arXiv:1106.5922 [hep-th]}⟩.

  Marcos Marino, Ricardo Schiappa, Marlene Weiss,
  \emph{Nonperturbative Effects and the Large-Order Behavior of Matrix Models and Topological Strings},
  ⟨\href{http://arxiv.org/abs/0711.1954}{arXiv:0711.1954 [hep-th]}⟩.

  Marcos Marino, Ricardo Schiappa, Marlene Weiss,
  \emph{Multi-Instantons and Multi-Cuts},
  ⟨\href{http://arxiv.org/abs/0809.2619}{arXiv:0809.2619 [hep-th]}⟩.

  Ricardo Schiappa, Niclas Wyllard,
  \emph{An $A_r$ threesome: Matrix models, $2d$ CFTs and $4d\, \mathcal{N}=2$ gauge theories},
  ⟨\href{http://arxiv.org/abs/0911.5337}{arXiv:0911.5337 [hep-th]}⟩.

  Bertrand Eynard, Marcos Marino,
  \emph{A holomorphic and background independent partition function for matrix models and topological strings},
  ⟨\href{http://arxiv.org/abs/0810.4273}{arXiv:0810.4273 [hep-th]}⟩.


%
%
\end{thebibliography}
\end{document}